\documentclass[%
 reprint,
 superscriptaddress,
 amsmath,amssymb,
 aip,
 jap,
]{revtex4-1}

\usepackage{graphicx}% Include figure files
\usepackage{dcolumn}% Align table columns on decimal point
\usepackage{bm}% bold math
\usepackage{hyperref}% add hypertext capabilities
\usepackage[version=4]{mhchem}
\usepackage{amsmath}
\usepackage{xcolor} % for text coloring

\begin{document}

\title{The Physical Significance of Imaginary Phonon Modes in Crystals}

\author{Ioanna Pallikara}
\affiliation{Department of Chemistry, University of Manchester, Oxford Road, Manchester M13 9PL, UK}

\author{Prakriti Kayastha}
\affiliation{Department of Mathematics, Physics and Electrical Engineering, Northumbria University, Newcastle upon Tyne NE1 8QH, UK}

\author{Jonathan M. Skelton}
\email{jonathan.skelton@manchester.ac.uk}
\affiliation{Department of Chemistry, University of Manchester, Oxford Road, Manchester M13 9PL, UK}

\author{Lucy D. Whalley}
\email{l.whalley@northumbria.ac.uk}
\affiliation{Department of Mathematics, Physics and Electrical Engineering, Northumbria University, Newcastle upon Tyne NE1 8QH, UK}

\begin{abstract}

The lattice vibrations (phonon modes) of crystals underpin a large number of material properties.
The harmonic phonon spectrum of a solid is the simplest description of its structural dynamics and can be straightforwardly derived from the Hellman-Feynman forces obtained in a ground-state electronic structure calculation.
The presence of imaginary harmonic modes in the spectrum indicates that a structure is a local maximum on the athermal structural potential-energy surface and can yield important insight into the fundamental nature and physical properties of a material.
In this review article, we discuss the physical significance of imaginary harmonic modes and distinguish between cases where imaginary modes are indicative of such phenomena, and those where they reflect technical problems in the calculations.
We outline basic approaches for exploring and renormalising imaginary modes, and demonstrate their utility through a set of three case studies in the materials sciences.

\end{abstract}

\maketitle

\section{Introduction}

The lattice vibrations (phonon modes) of crystals underpin a large number of material properties.
These include the volumetric and constant-pressure heat capacities, the Helmholtz and Gibbs free energies and the thermal conductivity.\cite{togo2015}
They also provide important spectral signatures (e.g. IR/Raman) that are often used as routine identification and characterisation tools.\cite{skelton2017spectroscopy}
In addition to this, the phonon modes are a key part of describing physical phenomena such as superconductivity,\cite{cea2021coulomb,mankowsky2014nonlinear} phase transitions\cite{steele2020phase,Adams2016,Souvatzis2009,jong2019anharmonic} and ferroelectricity\cite{kamba2021soft,kumar2012coupled} that play a critical role in the performance of established and emerging functional materials.

For predicting material properties from first-principles the most commonly-used tool is density-functional theory (DFT).
Routine DFT calculations however make predictions based on a perfectly static lattice with the atoms in their equilibrium positions.\cite{martin2020electronic,giustino2014materials}
However, this is an approximation: at finite temperature the atoms vibrate around their equilibrium positions, or can undergo more significant anharmonic motion away from their ideal crystallographic positions, and even at $T = 0 \mathrm{K}$ the zero-point atomic motion contributes to the internal energy.\cite{dove1997theory}
To model these vibrational effects it is necessary to combine DFT with other methods such as lattice dynamics or molecular dynamics.
These combinations of techniques then allow prediction, to high accuracy, of a wide variety of thermal properties and temperature effects, albeit at a (sometimes significantly) increased computational cost. 

The base assumption in lattice-dynamics calculations is that the vibrations are harmonic and can be described through a set of normal modes with well-defined wavevectors, frequencies and atomic-displacement patterns. 
Imaginary phonon modes are harmonic vibrations with an imaginary frequency.
While these are often unexpected and regarded as unphysical, as we hope to demonstrate in this review their presence in the phonon spectrum can provide valuable insight into the nature and properties of a material.

This review is organised as follows.
In the first half of the paper, we review the theory underlying imaginary modes.
We start by outlining the harmonic approximation and defining what we mean by an imaginary mode.
We then discuss some possible physical origins of imaginary modes in harmonic phonon spectra and approaches to treating these for subsequent calculations that require real harmonic frequencies.
In the second half of the paper, we then discuss three case studies that utilise imaginary modes for a variety of topical applications in the materials sciences.

\section{Theory}

\subsection{The harmonic approximation}  \label{sec:harmonic}

Atomic motion at small amplitudes is usually well described by the harmonic approximation, where atoms move as if connected by harmonic springs. 
In this case the total energy can be Taylor expanded with respect to the atomic displacement as:
\begin{equation} \label{eq:taylor_expansion}
U(\boldsymbol{u}) \approx U_0+\sum_{j,j^\prime}\sum_{\mu,\nu}\Phi_{jj^\prime}^{\mu\nu} u_j^\mu u_{j^\prime}^\nu 
\end{equation}
where $\boldsymbol{u}$ are displacements of the atoms from their equilibrium positions, $U_0$ is the energy of the static lattice, and $\boldsymbol{\Phi}$ are the second-order force constant matrices given by:
\begin{equation} \label{eq:force_constants}
\Phi_{jj^\prime}^{\mu\nu} = \frac{\delta^2U}{\delta u_j^\mu \delta u_{j^\prime}^\nu}.
\end{equation}
The indices $j$/$j^\prime$ label the atoms, and $\mu$/$\nu$ label the Cartesian directions.
It is assumed that the structure is relaxed to a stationary point on the potential-energy surface (PES) and the forces are equal to zero, so that the linear term in $\boldsymbol{u}$ vanishes.

The $\boldsymbol{\Phi}$ in Equation \eqref{eq:force_constants} are used to construct the dynamical matrix $\boldsymbol{D}(\mathbf{q})$ by performing a mass-reduced Fourier transform, which for solids yields:
\begin{equation}
  \label{eq:dynmat}
  D(\mathbf{q})_{jj^\prime}^{\mu\nu} = \frac{1}{\sqrt{m_j m_{j^\prime}}} \sum_{l^\prime} \Phi_{j0,j^\prime l^\prime}^{\mu\nu} \mathrm{exp}\{i\mathbf{q}\cdot[\boldsymbol{r}_{j^\prime l^\prime}-\boldsymbol{r}_{j0}]\}
\end{equation}
where $\mathbf{q}$ is the phonon wavevector and we have introduced the indices $l$/$l^\prime$ to label the crystallographic unit cells of the atoms.
Diagonalisation of $\boldsymbol{D}(\mathbf{q})$ is equivalent to solving an eigenvalue equation to determine the set of squared frequencies $\omega^2$ and associated mass-weighted displacement vectors $\boldsymbol{W}$ (eigenvectors) for the $3 n_a$ phonon modes at $\mathbf{q}$:
\begin{equation}
  \label{eq:dynmat_diag}
  \boldsymbol{D}(\mathbf{q}) \boldsymbol{W}(\mathbf{q}) = \omega^2(q) \boldsymbol{W}(\mathbf{q})
\end{equation}
For a given mode, the harmonic energy $U(Q)$ as a function of the normal-mode coordinate (displacement amplitude) $Q$ is given by:
\begin{equation}
  \label{eq:e_harm}
  U(Q) = \frac{1}{2} \omega^2 Q^2
\end{equation}
$Q$ is related to the Cartesian displacements of the atoms by:
\begin{equation}
  \label{eq:cart_disp}
    X_j(Q) = \frac{Q}{\sqrt{n_a m_j}} \mathrm{Re}[\mathrm{exp}(i \phi) \times \boldsymbol{W}_j \times \mathrm{exp}(\mathbf{q} \cdot \boldsymbol{r}_j)]
\end{equation}
where $\phi$ is a phase factor.

There are two common approaches to performing a harmonic phonon calculation. The first is to compute the $\boldsymbol{\Phi}$ in real space, typically using numerical differentiation by performing small finite displacements of the atoms.
This is also termed the ``direct'' approach in the literature, and is similar in principle to the ``frozen phonon'' method where atoms are displaced collectively along known eigenvectors (e.g. determined from the crystal symmetry) to compute the corresponding frequencies.
The second method is to compute the $\boldsymbol{D}(\mathbf{q})$ in reciprocal space using density-functional perturbation theory (DFPT, also termed ``linear response'').\cite{togo2015first,baroni2001phonons}
The real-space approach may require a supercell expansion to describe long-wavelength phonons, which can significantly increase the cost of phonon calculations when using methods that scale as a power of the system size.
The approach can be implemented using first-principles methods such as density-functional theory (DFT), i.e. computing the $\boldsymbol{\Phi}$ using the Hellman-Feynman forces,\cite{parlinski1997first} but is ultimately agnostic to the underlying method used to calculate the forces.
This makes the finite-differences approach compatible with the full range of DFT functionals and also with ``beyond DFT'' theories such as dynamical mean field theory.
On the other hand, DFPT can be more efficient, but its implementation in DFT codes is a technical challenge and so it tends to be more restricted in its applications, for example to specific families of DFT functionals.\cite{alyoruk2016piezoelectric,sabatini2016phonons,coutinho20173r,pike2018vibrational}

The truncation of the expansion in Equation \ref{eq:taylor_expansion} to second order in $\boldsymbol{u}$ defines the harmonic approximation.
This leads to a straightforward method of obtaining the $\omega$ and $\boldsymbol{W}$ by diagonalisation of the dynamical matrices, but implicitly yields non-interacting phonon modes with infinite lifetimes and where atoms on average occupy their equilibrium positions regardless of the phonon occupation number (energy level).
To understand processes such as thermal expansion and the creation and annihilation processes that give rise to finite lifetimes and thermal conductivity, one needs to consider the anharmonic interactions described by the third- and higher-order terms in $\boldsymbol{u}$.

Under the harmonic approximation the equilibrium distances between atoms is independent of temperature,
so this model cannot formally predict variations in volume with temperature and dependent properties such as thermal expansion.
The quasi-harmonic approximation (QHA) extends the harmonic approximation to consider such volume-dependent thermal effects by assuming the harmonic approximation remains valid for a series of volumes about the athermal equilibrium and minimising the combined lattice and harmonic phonon energy as a function of volume and temperature.
On a practical level the QHA is a straightforward extension to the harmonic approximation requiring a series of harmonic calculations at a set of compressions and expansions of the unit cell.
The QHA provides access to a number of quantities including the Gibbs free energy, thermal-expansion coefficients, and Gr{\"u}neisen parameters.

Beyond the QHA, higher-order anharmonicity can be incorporated into calculations in a number of ways.
A common approach to determining the phonon lifetimes needed to calculate the lattice thermal conductivity is to apply a perturbation derived from the third-order force constants to the harmonic frequencies and eigenvectors.\cite{togo2015distributions,li2014shengbte,carrete2017almabte}
Alternatively, a variety of approaches exist that aim to build effective harmonic potentials for a given temperature while incorporating the effects of higher-order terms in the Taylor expansion.\cite{hellman2011lattice,hellman2012thermal,hellman2013temperature,carreras2017dynaphopy,monacelli2021stochastic,Souvatzis2009}

Some of these approaches are discussed later in Section \ref{renormalisationBZ}.

\subsection{Imaginary phonon modes}

As shown in Equation \ref{eq:e_harm}, the energy change when atoms are displaced along a phonon mode is quadratic in the amplitude $Q$ with a constant of proportionality $\omega^2$.
A positive $\omega^2$ indicates a positive local curvature on the PES along the mode, and the energy therefore rises as the atoms are displaced from their equilibrium positions.
On the other hand, a negative $\omega^2$ indicates negative local curvature such that the energy decreases as the atoms are displaced.
The harmonic frequencies $\omega$ of modes with $\omega^2<0$ are complex numbers and, as such, are termed ``imaginary modes''% JMS: Pedantic, but...
\footnote{We note that imaginary modes are sometimes indicated with negative frequencies; while strictly speaking this is not mathematically correct, the practice is very common and is often encountered in literature on phonons.}.
\begin{figure}
    \centering
    \includegraphics[width=8cm]{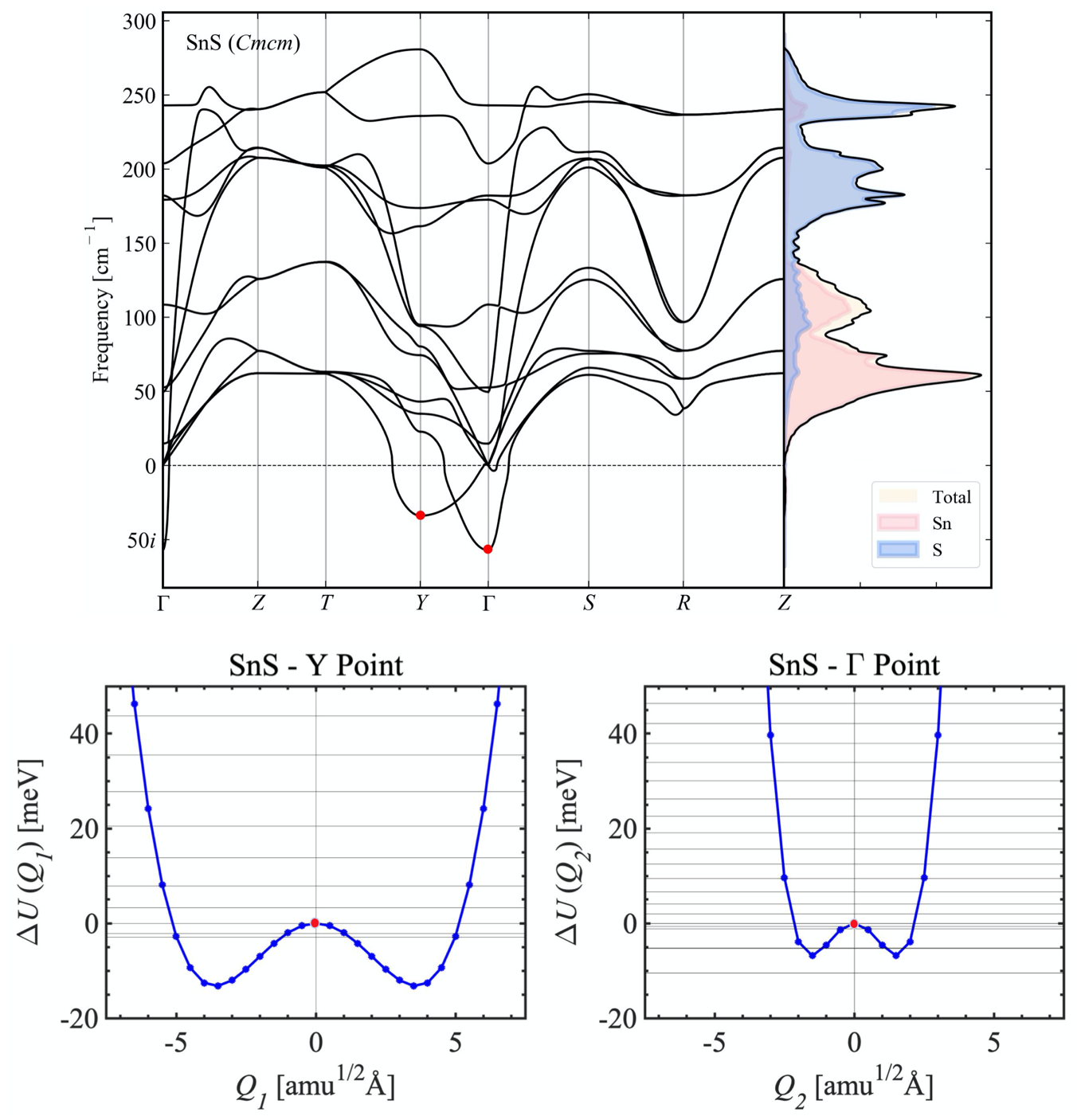}
    \caption{Imaginary modes in the $Cmcm$ phase of SnS. In the phonon dispersion (top) the red circles mark the principal imaginary modes at the $Y$ and $\Gamma$ wavevectors in the $Cmcm$ Brillouin zone. Mapping the potential-energy surfaces by computing the change in energy $\Delta U$ as a function of the displacement amplitude $Q$ (bottom) reveals a double-well potential that shows the $Cmcm$ phase to a maximum on the structural PES. While the magnitude of the imaginary frequency is indicative of the local curvature of the PES at $Q = 0$, it does not necessarily correlate to the size of the stabilisation on descending to the nearest local minimum along the mode: although the frequency of the imaginary mode at $\Gamma$ is larger in magnitude than that at $Y$, the depth of the double well is smaller. Reprinted figure with permission from Ref. \citenum{Pallikara2021}. Copyright 2021 by the Royal Society of Chemistry.}
    \label{figure1}
\end{figure}
Following this, a system for which all the harmonic modes are real is therefore a (local or global) minimum on the structural potential-energy surface (PES), whereas a systems with one or more imaginary modes is an energy maximum (e.g. a transition state/saddle point or hilltop).
As with real modes, imaginary modes have a well-defined eigenvector $\boldsymbol{W}$ that defines the collective atomic displacements (i.e. the the structural distortion) that lowers the energy.
By distorting the crystal structure along a particular phonon eigenvector (or group of eigenvectors), it is possible to map out the PES as a function of mode amplitude (Fig. \ref{figure1}).
This ``mode-mapping'' procedure is discussed further in Section \ref{exploring}. 

For a large class of materials, the imaginary modes take the form of a double-well potential such as that shown in \ref{fig:fig1}.
This double-well PES is characteristic of materials that undergo the symmetry-lowering displacive phase transitions described by Landau theory.\cite{dove1997theory}
As such, the form of the potential is fundamental to modelling ferroelectric and multiferroic materials. 
For example, the perovskite \ce{PbTiO3} transitions from a high-temperature non-ferroelectric $Pm\bar{3}m$ cubic phase to a ferroelectric $P4mm$ tetragonal phase at 763 K, driven by a single zone-centre phonon mode with an associated double-well potential.\cite{yadav2017structural,yuk2017towards}

If the imaginary modes result from dynamical instabilities in the structure, further exploration of the PES offers a number of useful insights including predicting phase transitions,\cite{togo2015first,Skelton2016,Adams2016,beecher2016direct} discovering new crystal polymorphs,\cite{rahim2020polymorph}, identifying dynamic Jahn-Teller instabilities,\cite{yang2017spontaneous} and understanding superionic conduction.\cite{buckeridge2013dynamical,krenzer2021anharmonic}
We will discuss some of these applications as case studies in the second half of this review.

The imaginary modes associated with dynamical instabilities are fundamentally anharmonic in nature and a quantitative description requires moving beyond the harmonic approximation.
Both the real-space finite-differences method and DFPT can also be used to compute the force constants for the higher-order terms in the Taylor expansion in Equation \ref{eq:taylor_expansion}, e.g. the third- and fourth-order force constants $\Phi_{j j^\prime j^{\prime\prime}}$/$\Phi_{j j^\prime j^{\prime\prime} j^{\prime\prime\prime}}$.\cite{togo2015distributions,carrete2017almabte,lazzeri2002first,esfarjani2008method,esfarjani2012erratum,li2014shengbte} 
Using the finite-displacement method to calculate these higher-order terms comes at a considerably higher computational cost than their harmonic equivalents, and perturbation theory is not valid for highly anharmonic materials where the anharmonic corrections to the harmonic energies are significant.\cite{knoop2020anharmonicity}
In these cases, alternative non-perturbative approaches are required to describe the anharmonicity.
For example,
it is possible to extract force constants and phonon frequencies from molecular-dynamics (MD) trajectories, or to fit force constants from stochastic atomic displacements.\cite{hellman2013temperature,tadano2014anharmonic,carreras2017dynaphopy,shulumba2017intrinsic,eriksson2019hiphive}
Although typically more expensive than the perturbation theory-based alternatives, these methods have the advantage of implicitly taking into account higher-order anharmonicity and temperature effects.
Non-perturbative methods for treating anharmonic modes are discussed further in Sections \ref{exploring} and \ref{renormalisationBZ}.

\subsection{The origins of imaginary modes} 

Within the harmonic approximation, the restoring force in response to an atomic displacement is given by minus the derivative of the harmonic energy with respect to the displacement, i.e. it is proportional to minus $\omega^2$.
An imaginary frequency corresponds to a non-restorative force leading to a decrease in potential energy as the atoms are displaced away from their equilibrium positions along the mode (Fig. \ref{figure1}).
The presence of such an instability in a crystal structure can, understandably, lead to to feelings of discomfort.
\footnote{Our wording here is a nod to the enjoyable writing of Marshall Stoneham: A. Stoneham, Rep. Prog. Phys. \textbf{44}, 1251 (1981).}
Being able to distinguish between imaginary modes that are an intrinsic property of the structural dynamics of a system, and those that reflect a technical problem in the calculations, is therefore key. 
As such, we consider three classes of possible origins for imaginary modes (Table \ref{table1}), \textit{viz.} i) numerical issues in the calculations; ii) material-specific physics that are not adequately accounted for when calculating the force constants or dynamical matrices; and iii) dynamical instabilities that are an intrinsic property of the material.

The first class encompasses a range of potential technical problems with calculations.
Since the forces on the atoms in the initial structure are assumed to be zero (c.f. Equation \eqref{eq:taylor_expansion}), a well-converged geometry optimisation is a prerequisite, and the criteria for force convergence may need to be tightened beyond that typically considered sufficient for relaxations.
Accurate forces are essential for computing the $\boldsymbol{\Phi}$ needed in finite-displacement calculations, while determining accurate $\boldsymbol{D}(\mathbf{q})$ using DFPT requires tightly-converged ground-state electronic densities.
Careful consideration must therefore be given to the electron orbital or plane-wave basis set size and the $\mathbf{k}$-point sampling density used to model the electronic structure.
For plane-wave basis sets it is usual to increase the cut-off energy by at least 25\% above the default values used for simple electronic-structure calculations, and larger increases of up to 2$\times$ may be required in some systems.
For numerical atom-centered basis sets, the total energy, density, and force cut-offs may need to be tightened by an order of magnitude.

Auxiliary settings, in particular the sizes of grids used to represent the electron density and to compute quantities such as the exchange-correlation potential, may also need to be considered.
The vibrational Hamiltonian is invariant to a rigid translation of the crystal.
As a consequence, the sum of the acoustic phonon frequencies at the $\Gamma$ wavevector ($\mathbf{q}_{\Gamma}=(0,0,0)$) should sum to zero, which is termed the acoustic-sum rule (ASR).
However, in calculations where the electron density is represented on a spatial grid - commonly referred to as FFT grids in plane-wave codes, or integration grids in codes using numerical atom-centered orbitals - the translational invariance can be broken.\cite{shang2020moving}
Deviation from the ASR results in imaginary modes immediately around the $\Gamma$ wavevector.
To remove such imaginary modes, one solution is to increase the size of the grids.
A computationally cheaper alternative is to enforce the ASR as a post-hoc correction to the computed forces, although this can in some cases leave ``bowing'' artefacts in the dispersion close to the $\Gamma$ wavevector.\cite{da2015phase,zeraati2016highly,zhang2019large}

\begin{figure}[!ht]
    \centering
    \includegraphics[width=8cm]{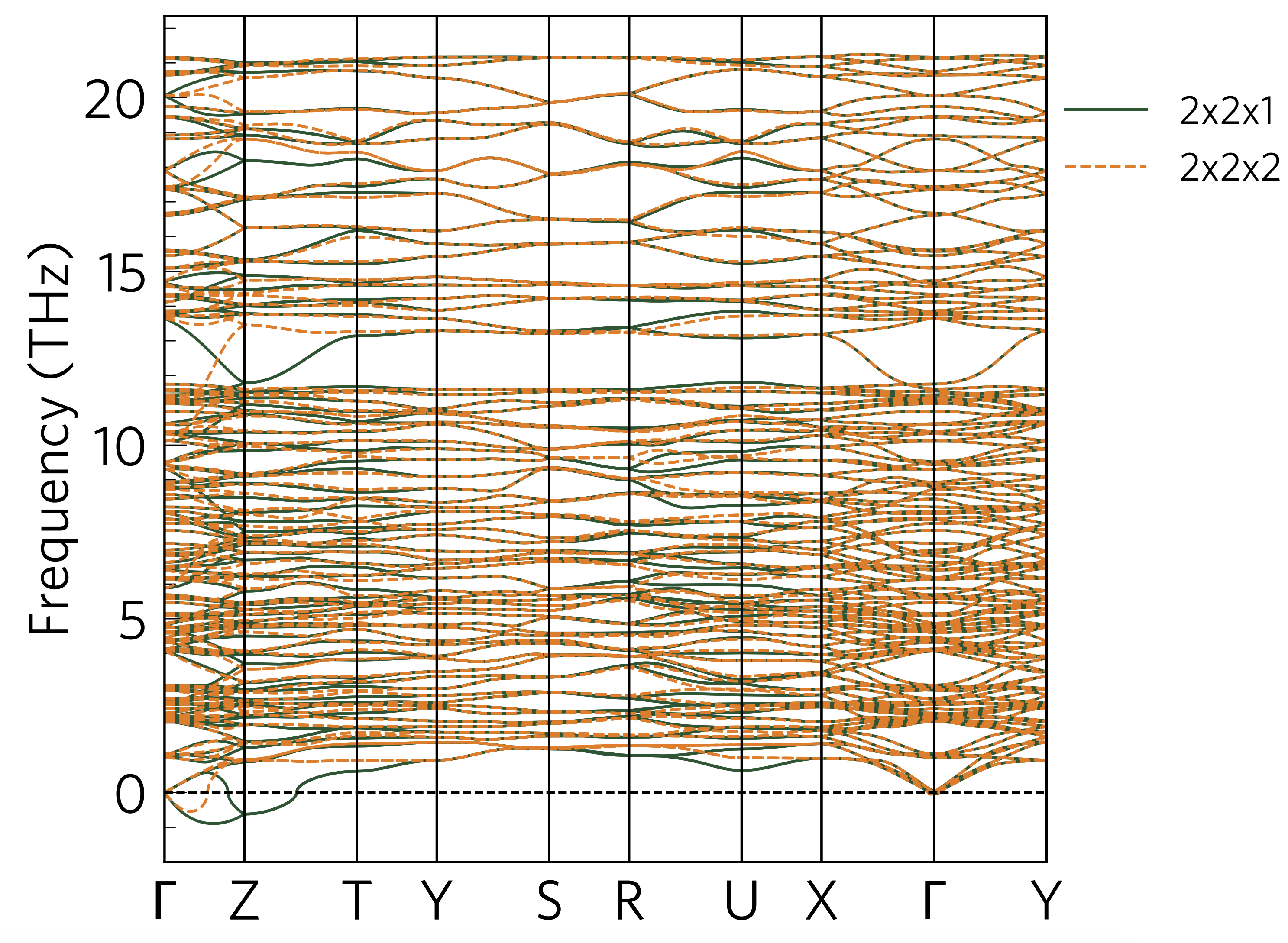}
  \caption{Importance of the choice of suprecell size for calculating phonon dispersion curves. The dispersion of orthorhombic $\delta$-\ce{Bi2Sn2O7} obtained using a $2\times2\times1$ expansion of the unit cell (solid green lines) shows an imaginary mode at the $\mathbf{q}_Z$=(0,0,$\frac{1}{2}$) wavevector. If a $2\times2\times2$ expansion commensurate with $\mathbf{q}_Z$ is used instead (dashed orange lines), the imaginary mode is removed. Reproduced with permission from Ref. \citenum{rahim2020polymorph}. Copyright 2020 by the Royal Society of Chemistry. 
  } 
  \label{fig:fig2a}
\end{figure}

The choice of supercell in a finite-displacement calculation - or, equivalently, the $\mathbf{q}$-point sampling density in a DFPT calculation - is also important.
The sum over the unit-cell index $l^\prime$ in Equation \ref{eq:dynmat} means that construction of the dynamical matrices requires in principle that the supercell used to calculate the force constants is large enough to capture all pairwise interactions for which the corresponding $\boldsymbol{\Phi}$ are non-zero.
For DFPT calculations, the $\mathbf{q}$-point grid must be dense enough to capture the frequency dispersion of the $3n_a$ phonon modes in reciprocal space.
The point at which these criteria are met is material specific and depends on the nature of the bonding. 
For example, silicon is a covalent system which requires a relatively long-ranged force-constant matrix (dense $\mathbf{q}$-point grid) for convergence,\cite{ackland1997practical} whereas in sodium chloride the bonding is more ionic, the structure is more rigid, and the force constants decay at a faster rate, so a smaller supercell (sparser $\mathbf{q}$-point grid) is sufficient.

A useful way to understand the importance of the supercell size is the concept of ``commensurate'' wavevectors.
A given supercell expansion will include the set of $\boldsymbol{\Phi}$ needed to construct the exact $\boldsymbol{D}(\mathbf{q})$ and determine the $\omega$/$\boldsymbol{W}$ at $\mathbf{q}$ for which the cell contains an integer number of wavelengths (modulation periods).
One can evaluate arbitrary dynamical matrices with an incomplete set of $\boldsymbol{\Phi}$ using Equation \eqref{eq:dynmat}, but in this case the resulting frequencies and eigenvectors will not be exact.
This procedure is known as ``Fourier interpolation''.
For a DFPT calculation, a set of $\boldsymbol{D}(\mathbf{q})$ are evaluated exactly by design to obtain phonon frequencies and eigenvectors on a ``coarse'' grid of wavevectors.
The transformation in Equation \eqref{eq:dynmat} can then be reversed to obtain a set of $\boldsymbol{\Phi}$ to an appropriate real-space range, allowing the $\omega$ and $\boldsymbol{W}$ at arbitrary $\textbf{q}$ to be obtained by interpolation in the same way.
Therefore, if an imaginary mode is observed at a given wavevector one should check whether the wavevector is commensurate with the chosen supercell expansion (finite-displacement) or $\textbf{q}$-point grid (DFPT) - if this is not the case, then the mode may be an interpolation artefact.\cite{royo2020using,skelton2020lattice,rahim2020polymorph}
We note that for finite-displacement calculations the zone-center $\mathbf{q} = \Gamma$ wavevector is always commensurate, so an imaginary mode here is either indicative of issues in the calculations, or indicates a distortion of the atom positions within the unit cell that would lower the energy.

As an example we consider the phonon dispersion of $\delta$-\ce{Bi2Sn2O7} (Fig. \ref{fig:fig2a}).\cite{rahim2020polymorph}
With a $2\times2\times1$ supercell expansion, an imaginary mode is obtained at the $\mathbf{q}_{Z} = (0,0,1/2)$ wavevector.
The supercell is however not commensurate with this $\mathbf{q}$ point, and when a $2\times2\times2$ expansion is used instead, which is commensurate with $\mathbf{q}_{Z}$, the imaginary mode is removed, indicating it to be an artefact of the Fourier interpolation.
It is of note that even when using the larger supercell a small imaginary mode remains along the $\Gamma$-$Z$ branch of the dispersion - this is also likely an interpolation artefact, but to correct it would require calculations with an even larger suprecell that would be prohibitively expensive. We also note here that the other acoustic and optical modes of this system have converged, indicating that the overall physics of the system remains invariant of this artefact. 
In cases such as this, it can be useful to turn to non-diagonal supercells, rather than the diagonal ($L\times M\times N$) supercells commonly used in finite-differences calculations, which can reduce costs sometimes by orders of magnitude.\cite{williams2015lattice}

If the theoretical model used in the calculations does not adequately describe the physics of the system, then the resulting force constants will not be correct. We have seen above how chemical bonding determines the supercell size (or $\mathbf{q}$-point grid) required for accurate predictions.
Other material-specific properties to consider when calculating the $\boldsymbol{\Phi}$ and $\boldsymbol{D}(\mathbf{q})$ include the include the extent of electron correlation\cite{Koccer2020efficient,Evarestov2011phonon} and spin-phonon coupling.\cite{kim2020spin,akamatsu2013strong} We address curing of imaginary phonons emerging from these phenomena in Table \ref{table1}. 

Finally, imaginary modes can arise when the crystal structure has intrinsic dynamical instabilities.
Here the structure corresponds to a maximum on the PES and may be accessible as an average over nearby minima at elevated temperature.
The imaginary modes of SnS shown in Fig. \ref{figure1} fall into this category, and it is this type of imaginary mode that forms the focus of this review.

\begin{table*}
\caption{\label{table1} Summary of possible origins of imaginary harmonic modes in calculations and solutions for removing or treating them. Each origin is classified as one of three types, \textit{viz.}: (I) technical (implementation) problems in the calculations; (P) material-specific physics that are not adequately accounted for in the calculations; and (D) a dynamical instability intrinsic to the system being studied. Note that the ``mode-mapping'' technique proposed as a solution to the latter is discussed in detail in Section \ref{exploring}.}
\begin{ruledtabular}
\begin{tabular}{p{6cm} p{10cm}}
Origin & Solution \\
\hline  

(I) Inadequately relaxed structure & Tighten force convergence criteria (e.g. to $\leq0.01$ eV \AA$^{-1}$). \\
(I) Inadequately converged technical parameters & Converge forces with respect to the basis set, $\mathbf{k}$-point sampling density and SCF convergence criteria. \\
(I) Broken translational symmetry & Increase the size of the density grids and/or enforce the acoustic sum rule as a post-hoc correction.\\
(I) Interpolation artefacts & For finite-difference calculations, select supercells commensurate with the wavevector where the imaginary mode is found (consider using non-diagonal supercells if needed). For DFPT calculations, ensure the $\mathbf{q}$-point grid includes the wavevector. \\
(P) Inadequate description of long-range bonding interactions & Increase the size of the supercell (for finite difference calculations) or $\mathbf{q}$-point sampling (for DFPT). \\
(P) Strong electron correlation & Use a suitable exchange-correlation functional for DFT calculations (e.g. DFT+$U$ or a hybrid functional), or a suitable alternative theory (e.g. dynamical mean field theory) to calculate forces.  \\ 
(P) Phonon-Magnetic coupling &  Include the effects of spin and spin-orbit coupling in calculations. \\
(D) Dynamic instability ($\Gamma$ wavevector) & Use phonon mode-mapping to locate lower-energy structure along imaginary mode. \\
(D) Dynamical instability (off-$\Gamma$ wavevector, typically a zone-boundary point) & Ensure point is commensurate with supercell (finite-differences) or included in the $\mathbf{q}$-point sampling grid (DFPT), then use mode-mapping as for $\Gamma$-point imaginary modes. \\
(D) Dynamic instability in a defective structure & Use mode mapping or break crystal symmetry to allow for localised distortions during geometry optimisation.  \\
\end{tabular}
\end{ruledtabular}
\end{table*}

\subsection{Mapping and ``renormalising'' imaginary modes} \label{exploring}

If technical problems in the calculations are ruled out and imaginary modes ascribed to inherent dynamical instabilities in a material, a number of techniques can be used to explore the modes and obtain further physical insight.
In this section we discuss the two most basic approaches: ``mapping'' the structural potential-energy surfaces along the modes, and using the potentials to determine effective (``renormalised'') real frequencies that can be used in harmonic calculations e.g. to determine free energies.\cite{Adams2016,Skelton2016} 

\begin{figure}
    \centering
    \includegraphics[width=8cm]{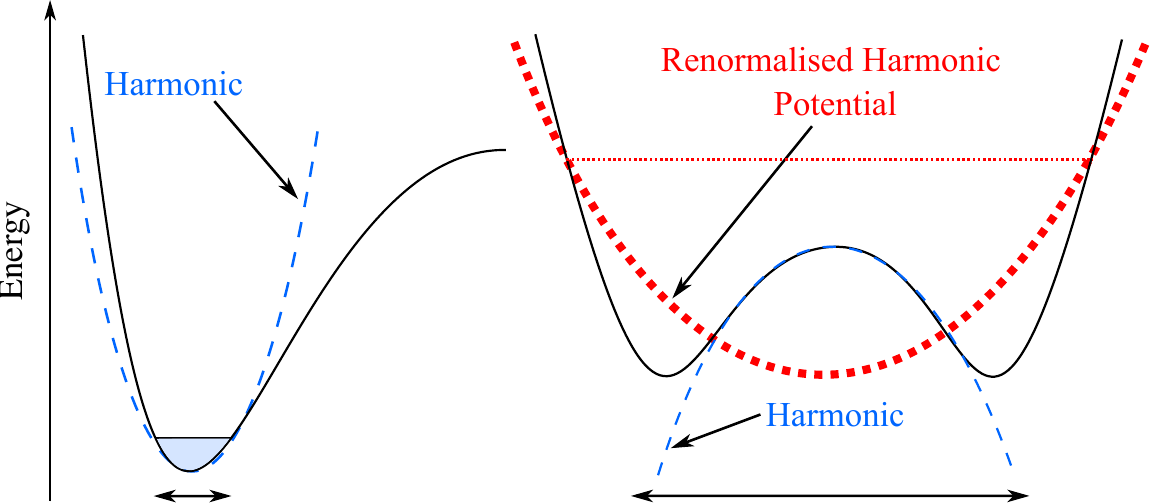}
    \caption{Schematic demonstrating the limitations of the harmonic model and the renormalisation of anharmonic modes. The scheme on the left compares a Morse-like potential for the change in energy as a function of normal-mode amplitude (black solid line), representative of some real materials, with a harmonic model (blue dashed line). The harmonic model provides an excellent description of the potential at small amplitudes, but increasingly deviates at larger amplitudes. The scheme on the right compares the double-well potential typically found along an imaginary harmonic mode, for which there is no restoring force to displacement from equilibrium, with a harmonic model. Here the harmonic model correctly describes the local curvature of the potential near the unstable structure, but does not predict the increase in energy at larger amplitudes. When the form of the potential is known (e.g. by mapping), it is possible to identify the region of the potential that is accessible at a given tempreature $T$ and hence to construct a renormalised harmonic potential (red dotted line) that reproduces the contribution of the anharmonic mode to a physical property of interest such as the thermodynamic partition function if calculating free energies. This schematic is inspired by a similar graphic in Ref. \citenum{hooton1955li}. }
    \label{fig:renormalisationschematic}
\end{figure}

To map the PES along a harmonic mode one can generate a sequence of structures with the atomic positions modulated by the mode eigenvector $\boldsymbol{W}$, with different amplitudes $Q$ using Equation \ref{eq:cart_disp}.
The associated energy changees can then be calculated, using DFT or otherwise, to obtain a $U(Q)$ curve.
At small amplitudes the energy change is quadratic in $Q$ (c.f. Equation \ref{eq:e_harm}), whereas at larger amplitudes the PES becomes increasingly anharmonic and higher-order polynomials are required to describe it (Figure \ref{fig:renormalisationschematic}).
For example, the symmetric double-well potential that is characteristic of many imaginary modes can often be described by the mathematical form $U(Q) = aQ^2 + bQ^4$, as in Landau theory, (where Q is treated as an order parameter) although in principle higher powers of $Q$ may be needed.
While the $U(Q)$ associated with harmonic modes in inorganic materials with symmetric bonding environments tend to be symmetric about $Q = 0$, this need not be the case, and indeed mapping bond-stretching modes in molecular solids, for example, usually yields asymmetric Morse-like potentials.\cite{savin2014thermal}

For a single imaginary mode, mapping allows the nearest local minima along the mode to be located.
In the harmonic approximation, the mode eigenvectors are orthogonal by construction, but this is only valid under the assumption of small displacement amplitudes.
Since the minimum on the imaginary-mode PES often occurs at a $Q$ outside the harmonic regime, if the goal of the mapping is to remove the instability it is usually prudent to perform a geometry optimisation on the minimum in order to relax orthogonal degrees of freedom.

One can also map a multidimensional PES $U(Q_1, ..., Q_N)$ by distorting along multiple harmonic modes simultaneously.
This is technically required for degenerate modes because any linear combination of the $N$ eiegenvectors is a valid solution to the Schr{\"o}dinger equation for the harmonic oscillator. In the case of a $N$-fold degenerate imaginary mode, the minima may be located at an arbitrary combination of displacements along the $N$ eigenvectors $\boldsymbol{W}$.
One can also use this multidimensional mapping procedure to investigate the coupling between specific sets of harmonic modes.

We note in passing that the unit cell used to map the PES along a mode must be commensurate with its wavevector, and for multidimensional maps the unit cell must be commensurate with the $\mathbf{q}$ of all the modes.
This may require that the mapping is performed in a supercell.

An issue with imaginary modes is that many physical properties derived using the harmonic approximation as a base, such as the phonon free energy, are not defined for imaginary modes, and so these modes are often neglected in such calculations.
A convenient way to handle this is to assign the modes a temperature-dependent effective (real) harmonic frequency, a process termed ``renormalisation''.

Consider the typical double-well potential in Figure \ref{fig:renormalisationschematic}.
The structure at $Q = 0$ is a saddle point on the PES between two minima.
In many systems, the saddle point corresponds to a symmetric structure observed at elevated temperature, whereas the two minima are equivalent realisations of a symmetry-lowering distortion that occurs on cooling through the phase-transition temperature.
At $T$ = 0 K, the system will be confined to the minima and will therefore adopt the low-symmetry structure.
As the temperature is raised towards the transition temperature, the system will gain thermal energy and begin to explore the PES in the vicinity of the minima.
Above the transition temperature, sufficient thermal energy will be available to traverse the barrier and the system will then be able to ``shuttle'' between the equivalent minima.
Averaged over time and space, it will appear as though the structure adopts the symmetric $Q = 0$ structure, but locally there will be large-amplitude distortions toward the lower-symmetry structure.
Since the PES is anharmonic, and the region accessible to the system depends strongly on the available thermal energy, the frequency of the mode has a very strong temperature dependence, which can often be used experimentally to track the phase-transition behaviour.
These measured frequencies are often relatively low, particularly close to the phase transition.
This means the modes are heavily populated, and so it is generally undesirable to neglect them when calculating properties.

\begin{figure}[!ht]
    \centering
    \includegraphics[width=8cm]{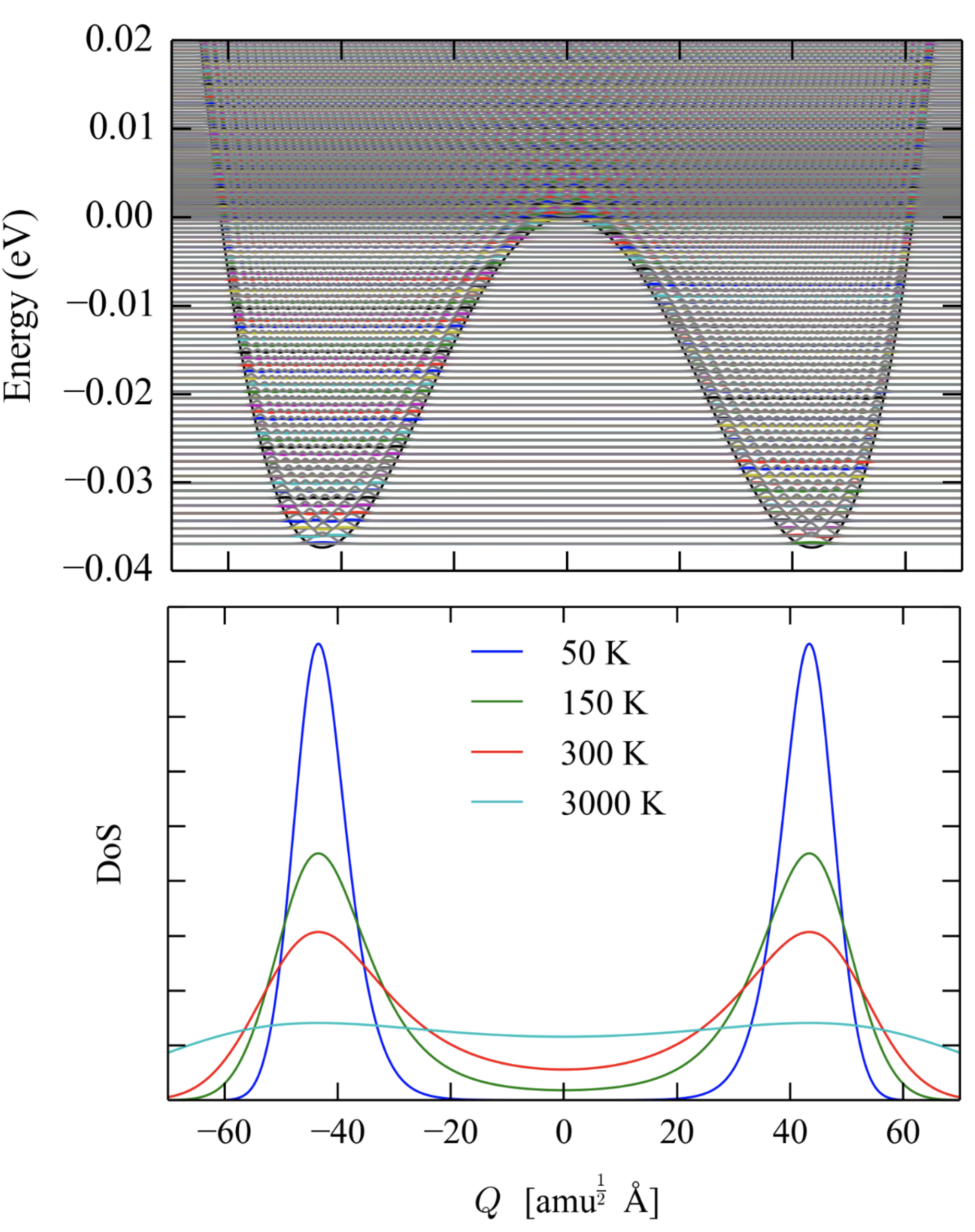}
	\caption{Upper panel: Potential-energy surface obtained by mapping the imaginary phonon mode at the $R$-point in the high-temperature pseudo-cubic phase of \ce{CH3NH3PbI3} as a function of the displacement amplitude (normal-mode coordinate) $Q$. The eigenstates generated by solving the 1D Schr{\"o}dinger equation for this PES are overlaid as coloured lines. Lower panel: Thermalised probability density of states (DoS) modelled using the eigenstate energies. Reprinted with permission from Ref. \citenum{Whalley2016phonon}. Copyright 2016 by the American Physical Society.}
	\label{fig:figMAPI}
\end{figure}

Given that we can obtain the potential by mode mapping, a simple scheme for modelling this temperature-dependent frequency is as follows.
First, the Schr{\"o}dinger equation for the PES is solved to give a set of eigenstates and corresponding energies for the nuclear positions in the anharmonic potential.
This is shown for the $R$-point mode in the hybrid perovskite \ce{(CH3NH3)PbI3} (\ce{MAPbI3}) in Figure \ref{fig:figMAPI}).
The occupation of the eigenstates from a Bose-Einstein distribution at a target temperature $T$ is then used calculate the contribution of the anharmonic mode to the thermodynamic partition function.
Finally, a harmonic frequency $\tilde{\omega}$ is then assigned to the mode such that it reproduces this contribution when used to compute the phonon free energy.
One can also construct a ``density of states'' showing the probability of the system occupying different $Q$ as a function of temperature.
For \ce{MAPbI3}, we see that, as expected, the distribution is heavily skewed toward the minima at low temperatures, but gradually becomes broader on warming and becomes close to homogeneous at a very high temperature of 3000 K.
The $\tilde{\omega}$ can be used to correct the dynamical matrices $\boldsymbol{D}(\mathbf{q}$) at the $\mathbf{q}$ where the imaginary modes are located, which can then be reverse transformed to obtain corrected force constants $\boldsymbol{\Phi}$.
This allows the renormalised frequencies to be used in calculations based on the harmonic frequencies/eigenvectors and force constants with minimal effort.

This free energy-based renormalisation scheme was developed independently by Skelton \textit{et al.} and Adams and Passerone,\cite{Skelton2016,Adams2016} the latter of whom termed it the ``decoupled anharmonic mode approximation'' (DAMA).
Both formalisms provide a convenient means to compute temperature-dependent harmonic frequencies for imaginary modes that can be used to estimate phase-transition temperatures and pressures.\cite{Adams2016,Pallikara2021}
The only substantial difference between them is that in the scheme proposed in Ref. \citenum{Skelton2016} only the imaginary modes are renormalised, whereas in DAMA the renormalisation is applied to all modes (i.e. both real and imaginary).
DAMA was found to provide a good estimate of the phase-transition temperature in cryolite (Na$_3$AlF$_6$), where a high-temperature orthorhombic $Immm$ phases is linked to a lower-temperature monoclinic $P2_1/n$ phase \textit{via} an imaginary mode with a double-well potential.\cite{Adams2016}
The approach in Ref. \citenum{Skelton2016} was also used to explore the impact of imaginary modes on other physical properties including the thermal conductivity of orthorhombic SnSe\cite{Skelton2016} and the electron-phonon coupling in \ce{MAPbI3}.\cite{Whalley2016phonon}
Some notable studies that make use of mode mapping and renormalisation are explored in detail in the case studies in Section \ref{casestudies}.

These simple renormalisation approaches offer a number of advantages.
Firstly, they are computationally relatively inexpensive - particularly in implementations where they are used only to treat a small number of imaginary modes - as they do not require significant numbers of further calculations. 
Secondly, since mapping the PES is agnostic to the code or theoretical method being used, the technique is general and easily adapted to e.g. different DFT functionals or post-DFT methods.
Finally, the methods do not make any assumptions about the form of the PES and therefore do not require the specification of any model parameters. 

On the other hand the main limitation of both approaches is that they retain the orthogonal harmonic eigenvectors, which may be a significant approximation in strongly-anharmonic materials.
The approaches in Refs. \citenum{Skelton2016} and \citenum{Adams2016} are defined for single modes, and as such neglect coupling between modes and may not work for degenerate modes where, as noted above, the multidimensional potential should be considered.
Generalising these renormalisation schemes to multiple modes is in principle possible, but would significantly increase the cost and complexity of both the mapping and subsequent solution of the DFT.
Finally, whereas the method for determining the effective harmonic frequencies is appropriate for thermodynamics, it may not be suitable for other purposes, for example for simulating the outcome of spectroscopic measurements that probe the difference in energy between occupied and virtual eigenstates.

A number of alternative and more sophisticated renormalisation schemes exist, such as the Temperature-Dependent Effective Potential (TDEP)\cite{hellman2013temperature,hellman2011lattice,hellman2012thermal} and the Stochastic Self-Consistent Harmonic Approximation (SSCHA)\cite{monacelli2021stochastic} methods, which are the subject of the following section.

\subsection{Full-spectrum renormalisation}\label{renormalisationBZ}

In systems with very strong intrinsic anharmonicity, or at high temperatures e.g. close to the melting point, the harmonic approximation may break down.
Following the previous section, the consequence of this can be thought of as a substantial renormalisation of the harmonic phonon spectrum.

A useful reference point for approaching this is a molecular-dynamics (MD) simulation.
In a high-temperature MD simulation, the dynamics in principle incorporate anharmonicity to arbitrary order.
A phonon frequency spectrum can be obtained from the Fourier transform of the velocity autocorrelation function and can be thought of as a renormalised harmonic phonon density of states (DoS).
However, MD simulations, particularly \textit{ab initio} MD (AIMD) simulations, often incur a significant computational cost overhead compared to harmonic phonon calculations.
Furthermore, the detail on individual phonon modes that is readily available from the harmonic approximation is lost, which limits the insight available from the simulations, and also the scope for calculating properties that cannot be formulated as integrals over the DoS.

There have therefore been a number of approaches developed that aim to account for anharmonicity within an effective harmonic potential that can then be used in standard harmonic calculations.
While such methods perform a renormalisation of the full phonon spectrum by design, and are thus not specifically targeted at treating imaginary modes, we review them here because intrinsic dynamical instabilities often lead to substantial deviations from the crystallographic average structure and can and do therefore impact upon other modes.
Some full-spectrum renormalisation methods also yield higher-order (e.g. third- and fourth-order) force constants as a useful by-product.
Full-spectrum methods include the self-consistent \textit{ab initio} lattice dynamics (SCAILD),\cite{Souvatzis2009} Temperature-Dependent Effective Potential (TDEP),\cite{hellman2013temperature,hellman2011lattice,hellman2012thermal} and Stochastic Self-Consistent Harmonic Approximation (SSCHA) methods,\cite{monacelli2021stochastic} the Phonon Quasiparticle approach,\cite{PhysRevLett.112.058501} and the Anharmonic LAttice MODEl (ALAMODE) method.\cite{tadano2014anharmonic}

The SCAILD method is based on a self-consistent approach to determining effective harmonic frequencies.\cite{Souvatzis2009}
An initial estimate of the phonon frequencies is made using the harmonic approximation as outlined in Section \ref{sec:harmonic}.
Atomic displacements with random amplitudes are then performed according to Equation \ref{eq:cart_disp} and the resulting forces computed.
These forces are then used to determine new phonon frequencies, and the procedure iterated until the frequencies converge.
SCAILD has been shown accurately to reproduce the phonon spectra of the group IVB elements and the group IIIB elements Sc and Y, and is expected to be applicable to the body-centered cubic (BCC) phases of the lanthanides and actinides, numerous ferroelectric materials, and some transition metals. 
However, as in the methods outlined in Section \ref{exploring} above, SCAILD uses the eigenvectors from the initial harmonic calculation, and, as a consequence, may not give accurate results for highly-anharmonic systems where the displacemens deviate from this.
However, SCAILD is less computationally intensive than alternative renormalisation methods that require MD simulations.\cite{Souvatzis2009}

The TDEP method\cite{hellman2013temperature,hellman2011lattice,hellman2012thermal} uses MD simuations to sample the PES of the material at a finite temperature, and the atomic positions and calculated forces are then mapped onto a model Hamiltonian that describes the lattice dynamics and allows extraction of force constants.\cite{hellman2013temperature,hellman2011lattice,hellman2012thermal,Korotaev2018} 
These can be used within the harmonic approximation to compute properties at finite temperature, and has been shown to yield accurate results for a number of systems including BCC Zr\cite{hellman2013temperature}, PbTe\cite{PhysRevB.91.214310}, MgO\cite{PhysRevB.99.094113} and SnSe\cite{PhysRevLett.117.276601}.
The model Hamiltonian also allows for higher-order (e.g. third-order) force constants to be extracted.
In a typical TDEP calculation, the unit cell volume would be equilibrated at the target renormalisation temperatures.
In contrast to SCAILD, and also to the mode-mapping approach, the method therefore does not assume that the PES is independent of temperature and as such will provide more accurate predictions of thermal properties such as free energies.
The main limitations of TDEP are that the use of MD - typically AIMD with a first-principles method such as DFT - comes at a significant computational cost, as noted above, and also that as a result of using Newtonian dynamics in MD calculations the method does not account for nuclear quantum effects, which can become significant at low temperatures.\cite{hellman2013temperature,hellman2011lattice,hellman2012thermal,Korotaev2018}
The issue of cost can be partially mitigated with the Stochastic TDEP (STDEP) approach, where force-displacement data to fit the force constants is generated by stochastic displacements of the atoms drawn from a Boltzmann distribution corresponding to the temperature of interest, rather than by performing AIMD simulations.\cite{shulumba2017intrinsic}

In the SSCHA, a variational technique is used to minimise the free energy with respect to the density matrix of an auxiliary quadratic Hamiltonian, which allows it to account for anharmonicity at all orders and to incorporate quantum and thermal fluctuations.\cite{monacelli2021stochastic}
The term ``stochastic'' in the name comes from the use of Monte-Carlo to obtain averages of the minimum free energy and anharmonic stress tensor.
The anharmonic force constants are evaluated by minimizing the residual sum of squares of the difference between the calculated forces and those from the anharmonic lattice model.
In principle the SSCHA avoids the limitation of the SCAILD and (S)TDEP models and enables accurate prediction of phonon properties even for highly-anharmonic systems.
For example, it has been shown to effectively predict second-order phase transitions and to accurately reproduce phase diagrams.\cite{bianco2017second} 
It is also a reliable method for structure searching, as it can locate saddle points in complex systems with large numbers of atoms.\cite{errea2020quantum,hou2021quantum}
The scheme also has the capability to determine accurate phonon linewidths e.g. for thermal-conductivity calculations.
The biggest drawback of this method is however its computational cost.\cite{monacelli2021stochastic,Korotaev2018}

It is worth noting that if applied to all the modes in a calculation, as in DAMA, the mapping and renormalisation approaches discussed in the previous section could also be used for full-spectrum renormalisation.\cite{Adams2016}
However, to do this the mapping and subsequent post-processing would invariably represent a substantial workload, at which point the cost may be competitive with some of the other, more sophisticated, methods discussed in this section.

As can be seen, there are a growing number of methods in the literature for going beyond the (quasi-)harmonic approximation by describing anharmonicity in lattice dynamics, including imaginary modes, but most of these come at a substantially higher computational cost over the simpler mode-mapping and renormalisation approaches described in the previous section.
When choosing an appropriate treatment, there is generally therefore a trade-off between computational cost and accuracy, and the balance will be determined by both the system under study, the properties to be calculated, and the required level of accuracy.
For systems with strongly-coupled and highly-anharmonic modes, techniques such as SCAILD, TDEP or the SSCHA will give accurate results but at an increased computational cost.
On the other hand, in systems where the anharmonicity is largely restricted to a small number of imaginary modes, or for studies where cost is a primary concern (e.g. high-throughput modelling), the significantly cheaper mode-mapping approaches may suffice.

\section{Case studies}\label{casestudies}

To demonstrate the utility of imaginary mode we discuss in this section three case studies in materials modelling: i) dynamic disorder in halide perovskites; ii) phase transitions in the tin monochalcogenides SnS and SnSe;  and iii) crystal-structure prediction.

\subsection{Dynamic disorder in halide perovskites} \label{dynamic}

\begin{figure}
    \centering
    \includegraphics[width=8cm]{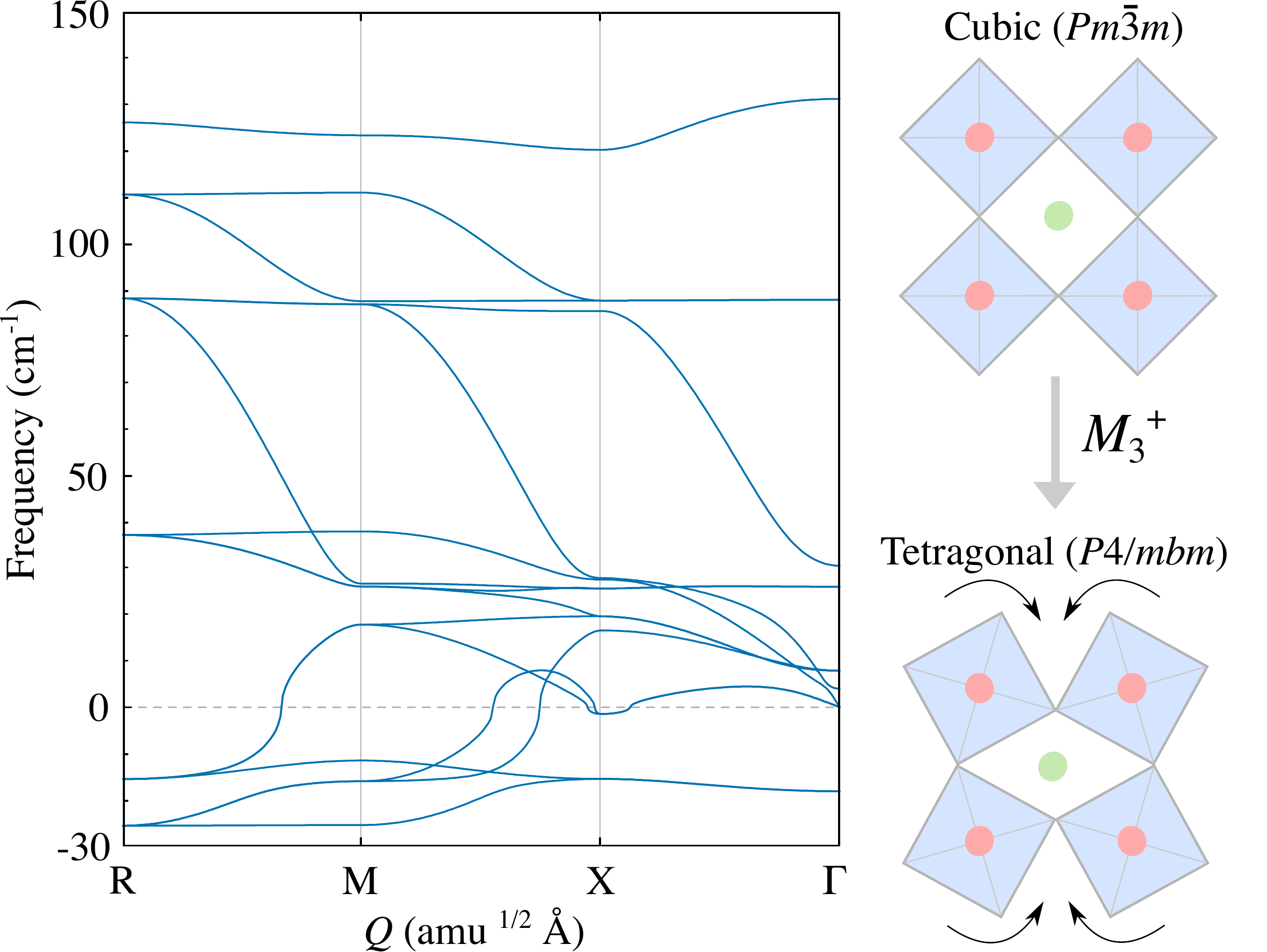}
    \caption{Harmonic phonon dispersion of cubic \ce{CsPbI3}. Above a threshold temperature of $\sim$ 600 K \ce{CsPbI3} adopts the cubic structure as a crystallographic average with large-amplitude local distortions. In the harmonic approximation there are thus multiple imaginary modes. The imaginary modes at the $\Gamma$ and $X$ wavevectors correspond to displacement of the A-site cation (Cs). The imaginary modes at the $M$ and $R$ points correspond to the in-phase ($\mathbf{q}_M=(\frac{1}{2},\frac{1}{2},0)$) and out-of-phase ($\mathbf{q}_R=(\frac{1}{2},\frac{1}{2},\frac{1}{2})$) tilting of the \ce{PbI6} octahedra. The schematic shows the crystal structures of the cubic and tetragonal perovskite structures, the latter of which is obtained by mapping along the $M$-point imaginary mode of the cubic structure.
    We note that some of the imaginary modes, such as the octahedral rotations at the $M$- or $R$- points, are expected to be localised to certain wavevectors but appear extended in reciprocal space due to interpolation.
    Adapted with permission from Ref. \citenum{yang2020assessment}. Copyright 2020 by the American Institute of Physics.}
    \label{fig:CsPbI3}
\end{figure}

It may be surprising that the harmonic phonon dispersion of the high-temperature cubic ($Pm\Bar{3}m$) phase of the halide perovskite \ce{CsPbI3} has several imaginary phonon modes (Fig. \ref{fig:CsPbI3}) when the structure is confirmed by X-ray diffraction (XRD).\cite{trots2008high,liu2019thermal}
This discrepancy however highlights the disconnect between experimental methods that probe a system at finite temperature and average over space and time, and the harmonic model of structures with an intrinsic dynamical instability.

\ce{CsPbI3} is in fact dynamically disordered at room temperature, with large anharmonic displacements of the atoms away from the ideal cubic sites as the crystal explores the anharmonic PES associated with the imaginary modes.\cite{liu2019thermal}
However, standard elastic-scattering techniques---including X-ray and neutron diffraction---probe the mean atomic positions and interatomic distances, averaged over the entire crystal and over the duration of data collection, and will thus identify the average positions albeit with large thermal displacements.
On the other hand, the harmonic phonon calculation does not take account of the higher-order anharmonic terms that would stabilise the cubic structure and produce renormalised (real) frequencies at higher temperatures.
These factors lead to a discrepancy between experiment and harmonic phonon calculations.

Mapping the PES along the imaginary modes in cubic \ce{CsPbI3} demonstrates that this high-symmetry, high-temperature structure is not a local minimum at low temperatures, but a saddle point along multiple double-well modes, making it a ``hilltop'' in multidimensional space.\cite{yang2020assessment}
Displacements along the combination of $M_3^+$ and $R_4^+$ imaginary modes lead to a lower-symmetry and dynamically-stable orthorhombic phase.
The calculated height of the energy barrier between this and the cubic structure is comparable to $k_\mathrm{B} T$, indicating the cubic perovskite structure measured in XRD\cite{trots2008high} is actually an average over the lower-symmetry minima.
Dynamical instabilities in the cubic perovskite structure are present across the inorganic halide perovskite series,\cite{yang2017spontaneous} and are also observed in their hybrid analogues such as methylammonium lead iodide (\ce{(CH3NH3)PbI3}, \ce{MAPbI3}). 
To probe the dynamics directly, inelastic-scattering techniques, such as inelastic neutron scattering (INS)\cite{liu2019thermal} or high energy resolution inelastic X-ray scattering (HERIX)\cite{beecher2016direct} can be used.
Another alternative is to use total-scattering or other methods sensitive to the local atomic structure such as extended X-ray absorption fine structure (EXAFS) or X-ray absorption near edge structure (XANES) combined with atomistic modelling.\cite{liu2019thermal,Bertolotti2017coherent}

Experimental measurements on \ce{MAPbI3} identified a displacive phase transition driven by large-amplitude anharmonic rotations of the \ce{PbI6} octahedra.\cite{beecher2016direct} 
HERIX measurements yielded phonon dispersion curves that showed strong softening at the zone-edge points $M$ and $R$, appointed to the in-phase ($M_\mathrm{TA_1}$) and out-of-phase ($R_\mathrm{TA}$) octahedral tilting away from the high-symmetry cubic orientation.
X-ray atomic pair distribution function (XPDF) measurements further demonstrated that the local structure was best described by the low-temperature orthorhombic phase whereas the longer-range structure was best described by the cubic phase.
Both these outcomes were confirmed by theoretical calculations, where mapping of the imaginary $M_\mathrm{TA_1}$ and $R_\mathrm{TA}$ modes produced two shallow double-well potentials leading to local metastable symmetry-broken minima.
The relative well depths supported the experimentally-observed phase transition sequence of a cubic-to-tetragonal, transition resulting from condensation of the $R_\mathrm{TA}$ mode, followed by a tetragonal-to-orthorhombic transition from the condensation of the $M_\mathrm{TA_1}$ mode.\cite{beecher2016direct}

In the context of applications of (hybrid) halide perovskites to opto-electronic devices, dynamic disorder has several repercussions at the device level.
The connection between dynamic disorder and device performance is mediated by several physical phenomena including band gap broadening,\cite{Whalley2016phonon} spin-splitting,\cite{niesner2018structural} and carrier recombination.\cite{munson2018dynamic}
For example, Niesner \textit{et al.} used resonant excitations of the photocurrent to provide direct experimental evidence for a dynamical Rashba effect (a spin splitting caused by dynamical disorder) in \ce{MAPbI3}.\cite{niesner2018structural}
Munson \textit{et al.} used temperature-dependent and time-resolved infrared spectroscopy to provide evidence for increased dynamic disorder leading to large polaron formation in this system.
The Rashba effect and large polaron formation are significant as they are thought to enhance the lifetimes and diffusion lengths of photocarriers in hybrid perovskite solar cells.

The mode-mapping technique can also be used to investigate how the imaginary phonon modes associated with dynamic disorder affect the electronic structure of a material.
For example, mapping was used to calculate the change in band gap, $\Delta E_\mathrm{g}$, with respect to the imaginary-mode amplitude $Q$ in \ce{MAPbI3}.\cite{Whalley2016phonon}
Summing $\Delta E_\mathrm{g} (Q)$ multiplied by the probability density of states along $Q$ (Fig. \ref{fig:figMAPI}) then yielded a thermally-averaged electron-phonon coupling for each imaginary mode and estimated positive band gap shifts of 35.5 and 27.9 meV for the $R$ and $M$ modes, respectively, at room temperature, which are comparable in magnitude to the photoluminescence broadening measured in experiments.\cite{wright2016electron,bohn2018dephasing,fassl2021revealing}

\subsection{Phase transitions in the tin monochalcogenides SnS and SnSe}

The tin monochalcogenides SnS and SnSe have received significant attention as emerging energy materials made from earth-abundant elements.
SnS has long been studied as a potential thin-film absorber for photovoltaics (PV) applications due to its near-ideal bandgap $E_\mathrm{g}$ and large absorption coefficient $\alpha$.\cite{Sinsermsuksakul-2014-SnSPhotovoltaics,Yun-2019-SnSPhotovoltaics,Cho-2020-SnSPhotovoltaics}
More recently, SnSe has been intensively studied for its potential as a thermoelectric material, as it has the ideal combination of physical properties needed for high heat-to-electricity conversion efficiencies, \textit{viz.} a high Seebeck coefficient $S$, low thermal conductivity $\kappa$, and high electrical conductivity $\sigma$.\cite{Zhao-2014-SnSeThermoelectrics,Zhao-2016-SnSePerspective,Zhou-2021-PolycrystallineSnSe}
SnS and SnSe have been proposed to adopt one of five different crystal structures, \textit{viz.} orthorhombic $Pnma$ and $Cmcm$, cubic rocksalt (RS), zincblende (ZB) and ``$\pi$''  phases.
The $Pnma$ structure is the energetic ground-state of both chalcogenides and transitions reversibly to the $Cmcm$ phase at elevated temperature.\cite{Chattopadhyay-1986-SnSSeNeutronScattering}
Epitaxial growth of RS SnS and SnSe on NaCl was first reported as far back as the 1960s.\cite{Mariano-1967-RocksaltSnX,Bilenkii-1968-RocksaltSnX}
Much more recently, a nanoparticulate form of SnS was assigned to a ZB structure based on X-ray diffraction and the particle morphology, despite theoretical calculations indicating this to be unstable.\cite{Greyson-2006-ZincblendeSnS,Ahmet-2015-ZincblendeSnS}
This debate was eventually settled with the discovery of the $\pi$-cubic phase in the $P2_13$ spacegroup.\cite{Rabkin-2015-PiSnS,Abutbul-2016-PiSnS,Abutbul-2016-PiSnSe}
It is well known that the $Cmcm$ phase has intrinsic dynamical instabilities and, like the cubic perovskite structure, is an average over equivalent $Pnma$ structures separated by a small energetic barrier.\cite{Skelton2016,Pallikara2021}
A comprehensive study of the five structures of SnS analysed the phonon dispersions and further determined that the $\pi$ SnS is dynamically stable, RS SnS has a dynamical instability at the $X$ wavevector, and ZB SnS is highly unstable with $\sim$ 50 \% of the modes being imaginary.\cite{Skelton-2017-ChemLattStability}
It was further found that the imaginary modes in RS SnS harden under compression and eventually become real, providing an explanation for how this phase is stabilised under epitaxial growth.
More recent calculations found that both the $\pi$ and RS phases of SnSe are dynamically stable at equilibrium, a contrast which can be explained by the stereochemical activity of the Sn lone pair.\cite{Pallikara2021}

\begin{figure}[!htp]
    \centering
    \includegraphics[width=7.5cm]{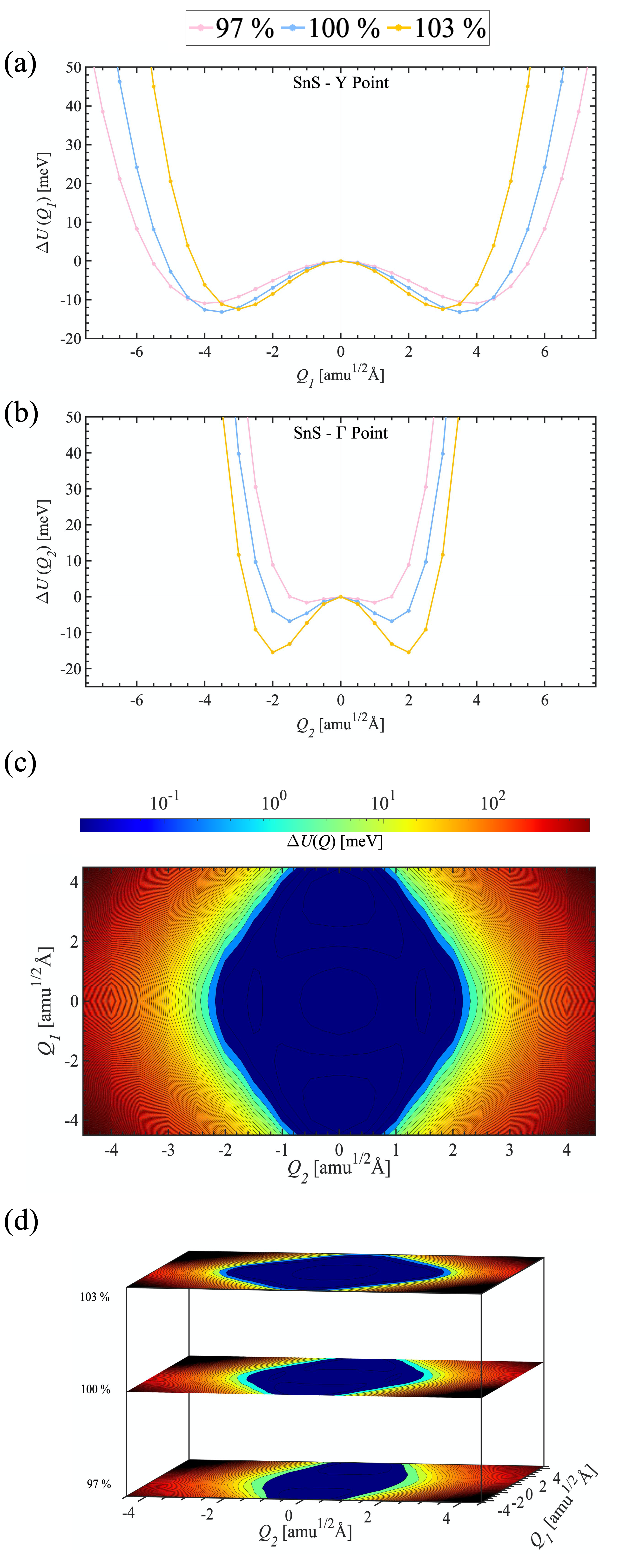}
	\caption{Potential-energy surfaces along the imaginary modes at the $Y$- and $\Gamma$-points in the $Cmcm$ phase of SnS obtained by the mode mapping approach outlined in Section \ref{exploring}. The potentials are shown for the equilibrium volume and 3 \% compressions and expansions to illustrate the effect of pressure. (a) and (b) show the 1D PES along the two modes, while (c) and (d) show the 2D PES spanned by both imaginary modes at the equilibrium volume (c) and at the three volumes in (a) and (b). Reprinted with permission from Ref. \citenum{Pallikara2021}. Copyright 2021 by the Royal Society of Chemistry. }
	\label{fig:fig1}
\end{figure}

The relationship between the low-temperature $Pnma$ and high-temperature $Cmcm$ phases of SnSe was established using the mode-mapping approach outlined in Section \ref{exploring}.\cite{Skelton2016}
The phonon dispersion of $Cmcm$ SnSe has imaginary modes at the $Y$ and $\Gamma$ wavevectors, each of which correspond to two double-well potentials (Figure \ref{fig:fig1}).
Evaluation of the 2D PES spanned by both modes showed that the $Cmcm$ phase is a hilltop on the PES.
The global minima lie along the $Y$-point mode and correspond to the equivalent distorted $Pnma$ phases, whereas the $\Gamma$-point mode leads to a further saddle point.
Interestingly, this is despite the steeper curvature of the PES along this mode at the $Cmcm$ phase.
Again, as for cubic perovskites, at elevated temperature sufficient thermal energy is available for the system to access the $Cmcm$ phase as a hilltop between $Pnma$ minima, and the $Cmcm$ structure is thus observed as a crystallographic average.\cite{Skelton2016}
This finding is consistent with experimental inelastic neturon scattering measurements showing a strong softening of the low-energy optic modes in the $Pnma$ phase towards the phase-transition temperature.\cite{Li2015} 

A more recent theoretical study using the QHA\cite{Pallikara2021} further examined the effect of pressure on the $Pnma$ $\rightarrow$ $Cmcm$ phase transition (Figure \ref{fig:fig1}).
The barrier between the $Pnma$ and $Cmcm$ phase decreases under pressure and eventually disappears, resulting in the $Cmcm$ phase becoming dynamically stable.
This was predicted to occur at pressures around 11.5 and 8 GPa for SnS and SnSe, respectively, and the phase transition temperature, predicted using renormalised harmonic frequencies for the imaginary modes, was also found to decrease with pressure.
The study also examined the impact of soft modes on the Helmholtz and Gibbs free energies, and it was shown that in this case the free energies calculated using renormalised frequencies for the soft modes were only very slightly different to those calculated with them being omitted from the partition functions.

Finally, the study in Ref. \citenum{Pallikara2021} also examined the structural relationship between the RS and $\pi$ phases, which are based on a distorted $2 \times 2 \times 2$ supercell of the eight-atom rocksalt conventional cell.\cite{Abutbul-2016-PiSnS,Abutbul-2016-PiSnSe}
Despite the imaginary mode in RS SnS, an energetic barrier was found between the RS and $\pi$ phases of both monochalcogenides.

\begin{figure}[!ht]
    \centering
    \includegraphics[width=8cm]{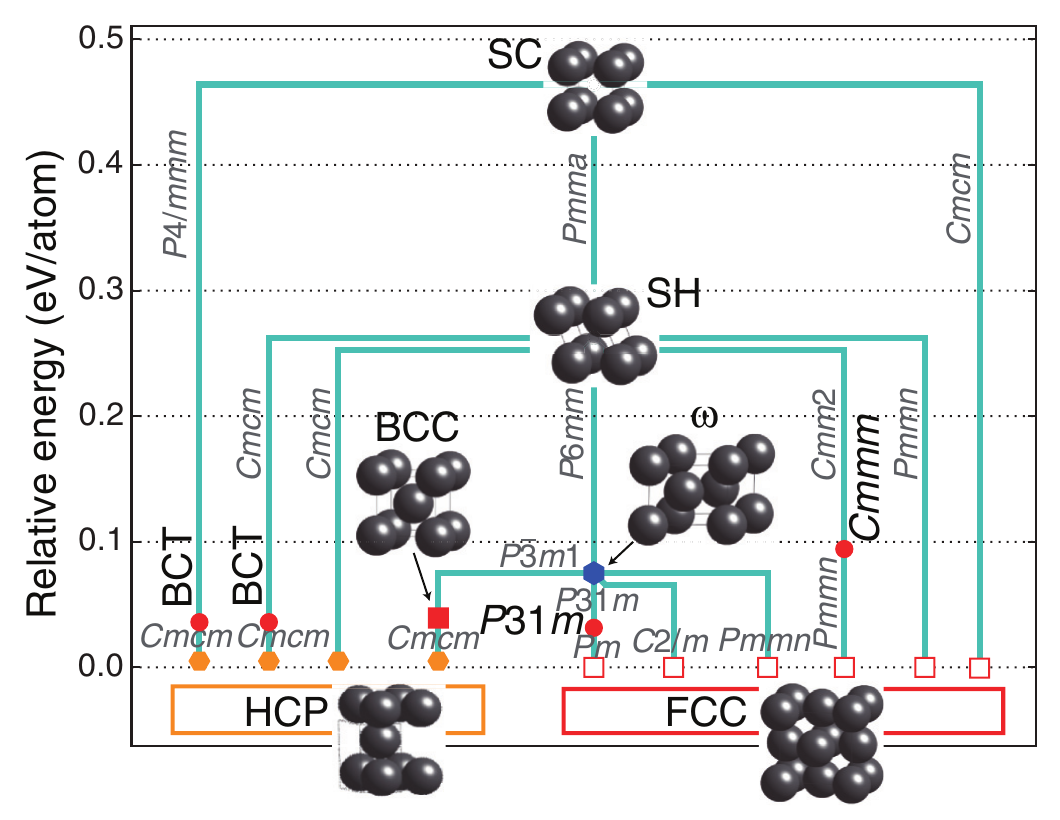}
    \caption{
        ``Tree diagram'' showing the map of the structural potential-energy surface of Cu obtained using imaginary-mode following.
        The branches show the connectivity between a high-energy simple cubic starting structure with multiple phonon instabilities, intermediate lower-energy maxima, and stable hexagonal close-packed and face-centered cubic minima.
        Abbreviations: SC - simple cubic; SH - simple hexagonal; BCC - body-centered cubic; BCT - body-centered tetragonal; HCP - hexagonal close-packed; FCC - face-centered cubic.
        Reprinted with permission from Ref. \citenum{togo2013evolution}.
        Copyright 2013 by the American Physical Society.
        }
    \label{fig:Cu-TreeDiagram}
\end{figure}

These two case studies---dynamic disorder in perovskites and phase transitions in the tin monochalcogenides--- highlight the advantages of combining theory and experiment for materials characterisation.
Returning briefly to \ce{CsPbI3}, as noted by Bertolotti \textit{et al.} the structural similarity of the various \ce{CsPbI3} polymorphs results in very small differences in the XRD patterns, and they are almost indistinguishable by other widely-used characterisation techniques such as transmission electron microscopy.\cite{Bertolotti2017coherent}
In addition to this, single-crystal studies are challenging as the orthorhombic-to-cubic phase transition in \ce{CsPbI3} is accompanied by a large change in volume, which is typical of materials with significant anharmonicity, leading to fracturing.\cite{trots2008high}
In these cases, imaginary modes in the harmonic phonon dispersions calculated for known experimental structures can provide important insight into the complex physics underlying phase transitions and dynamic disorder. 
The extension of this to crystal-structure prediction is explored in our third case study below.

\subsection{Crystal structure prediction} \label{prediction}

The drive to find new materials with improved properties for specific target applications (e.g. efficient absorbers for solar cells) has led to an increasing interest in materials discovery.\cite{butler2016computational}
The scale of this challenge is enormous: the chemical space of multi-component materials is vast, and only a fraction of potential ternary and higher compounds have thus far been made and characterised.\cite{davies2016computational}
While an initial set of candidate materials can be identified using metrics based on the chemical composition alone, later-stage screening requires knowledge of the crystal structure.\cite{davies2018computer}
Efficient methods for crystal-structure prediction are therefore an important challenge in contemporary materials modelling.

Given a composition and possibly a target unit-cell size, the goal of a structure-prediction algorithm is to traverse the structural potential-energy surface (PES) and locate the global energy minimum, a very general problem to which a number of potential approaches have been examined.\cite{woodley2008crystal,reilly2016report}
The simplest conceptually is the so-called \textit{ab initio} random structure searching method,\cite{pickard2011ab} which aims at an unbiased sampling of the PES by generating random structures within some universal chemical constraints such as reasonable minimum bond lengths.
Building on this are a variety of ``accelerated sampling'' methods that attempt to accelerate the search over the PES, with three popular approaches being basin hopping, particle-swarm optimisation and genetic/evolutionary algorithms.\cite{woodley2013knowledge,wang2010crystal,oganov2011evolutionary}
More recently, lower-cost alternatives have been developed based on data-mining of known crystal structures, such as using atomic substitution probabilities to identify known structures that could be adopted by a target composition\cite{hautier2011data} and building new structures from ``modules'' of existing ones.\cite{Collins-2017-MCEMMA}

A relatively common type of structure-prediction sub-problem is that where a high-temperature/high-symmetry structure is known from experiments (or can be guessed) but shows dynamical instabilities, and we wish to find the corresponding stable minima.
Returning again to the \ce{CsPbI3} example discussed above, we might assume that an \ce{ABX3} compound would adopt an analagous, dynamically-unstable cubic structure, and then attempt to find the low-temperature/low-symmetry phases.
A powerful approach to this problem is the following.
We compute the phonon spectrum of the starting structure to identify the imaginary modes at the high-symmetry wavevectors (i.e. the zone-centre and/or the zone-boundary points).
We then map each of the modes sequentially to locate the nearest local minima, perform full geometry optimisations to relax the orthogonal degrees of freedom, and hence obtain a second ``generation'' of structures (one per imaginary mode).
This procedure is then applied to each of the new structures, and iterated until each branch terminates in a dynamically-stable phase with no imaginary modes.

\begin{figure*}[!ht]
    \centering
    \includegraphics[width=16cm]{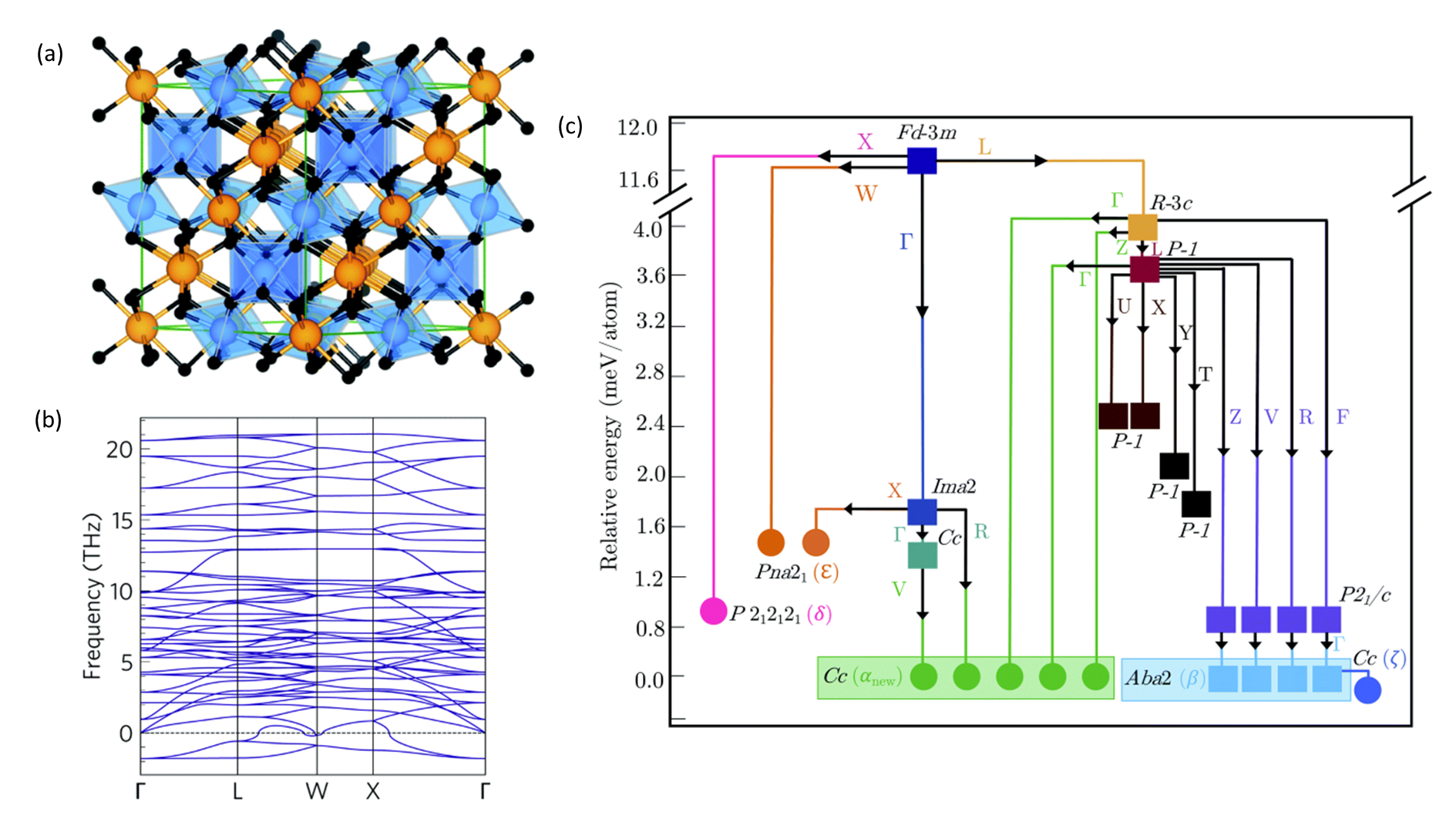}
    \caption{
        Polymorph exploration of the ternary metal oxide \ce{Bi2Sn2O7}.
        \ce{Bi2Sn2O7} adopts a pyrochlore structure ($\gamma$); a) at high temperature, for which the phonon dispersion shows imaginary harmonic modes at multiple wavevectors (b).
        By following the imaginary modes, a route is traced through the structural potential-energy surface (PES; c) that identifies the known $\alpha$ (here labelled $\alpha_\mathrm{new}$) and $\beta$ structures, together with three new polymorphs denoted $\delta$, $\epsilon$ and $\zeta$. Reprinted with permission from Ref. \citenum{rahim2020polymorph}. Copyright 2020 by the Royal Society of Chemistry.
        }
    \label{fig:Bi2Sn2O7-ModeMap}
\end{figure*}

There are a handful of examples where this approach has been applied successfully.
As a proof-of-concept, Togo and Tanaka studied the elemental metals Na, Mg, Al, Ti, Cu, Zr and Hf.\cite{togo2013evolution}
In each case the search was started from a simple cubic (SC) parent structure 0.35--1 eV atom$^{-1}$ higher in energy than the known stable phases and with multiple phonon instabilities.

For Cu, the instabilities in the SC phase led to stable face-centered cubic and hexagonal close-packed (FCC/HCP) minima $\textit{via}$ a set of lower-energy unstable maxima including simple hexagonal (SH), body-centered tetragonal (BCT), body-centred cubic (BCC) and $\omega$ phases (Fig. \ref{fig:Cu-TreeDiagram}).
The resulting ``tree-like diagram'' also enables phase-transition pathways between the stable phases to be identified.
The $\omega$ structure is a ``junction'' between the $\omega$ $\rightarrow$ BCC $\rightarrow$ HCP and $\omega$ $\rightarrow$ FCC pathways and thus serves as a previously unreported low-energy barrier in the FCC $\leftarrow \rightarrow$ BCC transition pathway.
The BCC $\leftarrow \rightarrow$ HCP transition was previously thought to be complex\cite{grimvall2012lattice},
%PK: cited this here: \textcolor{red}{[REF: 10.1103/RevModPhys.84.945]} 
but was found from the PES mapping to proceed \textit{via} an intermediate with a $Cmcm$ space group.

A similar analysis of Mg found HCP, FCC and two $9R$ and $18R$ long-period stacking (LPS) structures as minima together with a metastable $P6\overline{2}m$ structure that has yet to be reported experimentally but is formed by Mg-rich alloys.
Comparison of Ti and Hf reveals an interesting contrast wherein the low-energy $\omega$ structure is linked to a SH phase in Hf but not in Ti, implying the presence of an additional energy barrier in the latter that explains why this phase is not accessible under ambient pressure, and also why the transition under pressure occurs with a large hysteresis.\cite{sikka1982omega}
Searches performed under applied pressure found that pressure stabilises the BCC phase in both Cu and Mg and introduces several LPS minima in Cu, and therefore has a significant effect on the PES.

Across the seven elements examined the FCC and HCP structures were identified as common stable phases together with five LPS structures, but, notably, the $\omega$ phases of Ti and Zr was found to be more stable than the FCC phases despite not being a close-packed structure.
This highlights the level of insight that can potentially be obtained from this structure-prediction approach, even for comparatively simple and well-studied systems.

Rahim \textit{et al.} applied the same approach to the ternary oxide \ce{Bi2Sn2O7} in conjunction with an exhaustive analysis of variable-temperature neutron-diffraction measurements.\cite{rahim2020polymorph}
Experiments indicate that a cubic pyrochlore structure ($\gamma$-\ce{Bi2Sn2O7}; Fig. \ref{fig:Bi2Sn2O7-ModeMap}a) is formed at high temperature, which on cooling converts to an intermediate $\beta$ phase and a low-temperature $\alpha$ phase with two proposed structures (here denoted $\alpha$ and $\alpha_\mathrm{new}$).
The dispersion of $\gamma$-\ce{Bi2Sn2O7} has imaginary modes at four different wavevectors (Fig.  \ref{fig:Bi2Sn2O7-ModeMap}b).
The PES map obtained from the structural search is reproduced in Fig. \ref{fig:Bi2Sn2O7-ModeMap}c.
The structure search recovers the $\beta$ phase and the $\alpha_\mathrm{new}$ structure, providing strong evidence that the latter is the correct model.
Imaginary modes were also found in the $\beta$ dispersion and followed to a new $\zeta$ phase, which proved to be a better match to the experimental diffraction data.
The mapping also identified two new metastable phases, termed $\delta$ and $\epsilon$, that have yet to be identified experimentally.
The parent $\gamma$ phase has 88 atoms in the unit cell, while the intermediate and stable phases have between 44 and 176 atoms, which would make finding these structures with conventional structure-prediction algorithms extremely challenging.

The low-temperature $\alpha_\mathrm{new}$ structure was subsequently predicted to be a high-performance thermoelectric material,\cite{Rahim-2020-aBi2Sn2O7} due to the unusually-low lattice thermal conductivity arising from the complex structure and heavy metals, which further highlights the potential of this approach for materials discovery.

Kayastha \textit{et al.} used a similar mode-mapping approach as part of a high-throughput workflow to predict dynamically stable quasi one-dimensional (Q1D) polymers.\cite{kayastha2021high}
The study considered hypothetical materials produced by bonding one of eleven monovalent metals to one of 109 substituted cyclopentadienyl (Cp) anions in a vertical (stacked) packing arrangement.
The phonon dispersion curves were calculated for the resulting 1,199 Q1D chain polymers along the $\Gamma$-$X$ direction.

\begin{figure}[!ht]
    \centering
    \includegraphics[width=8cm]{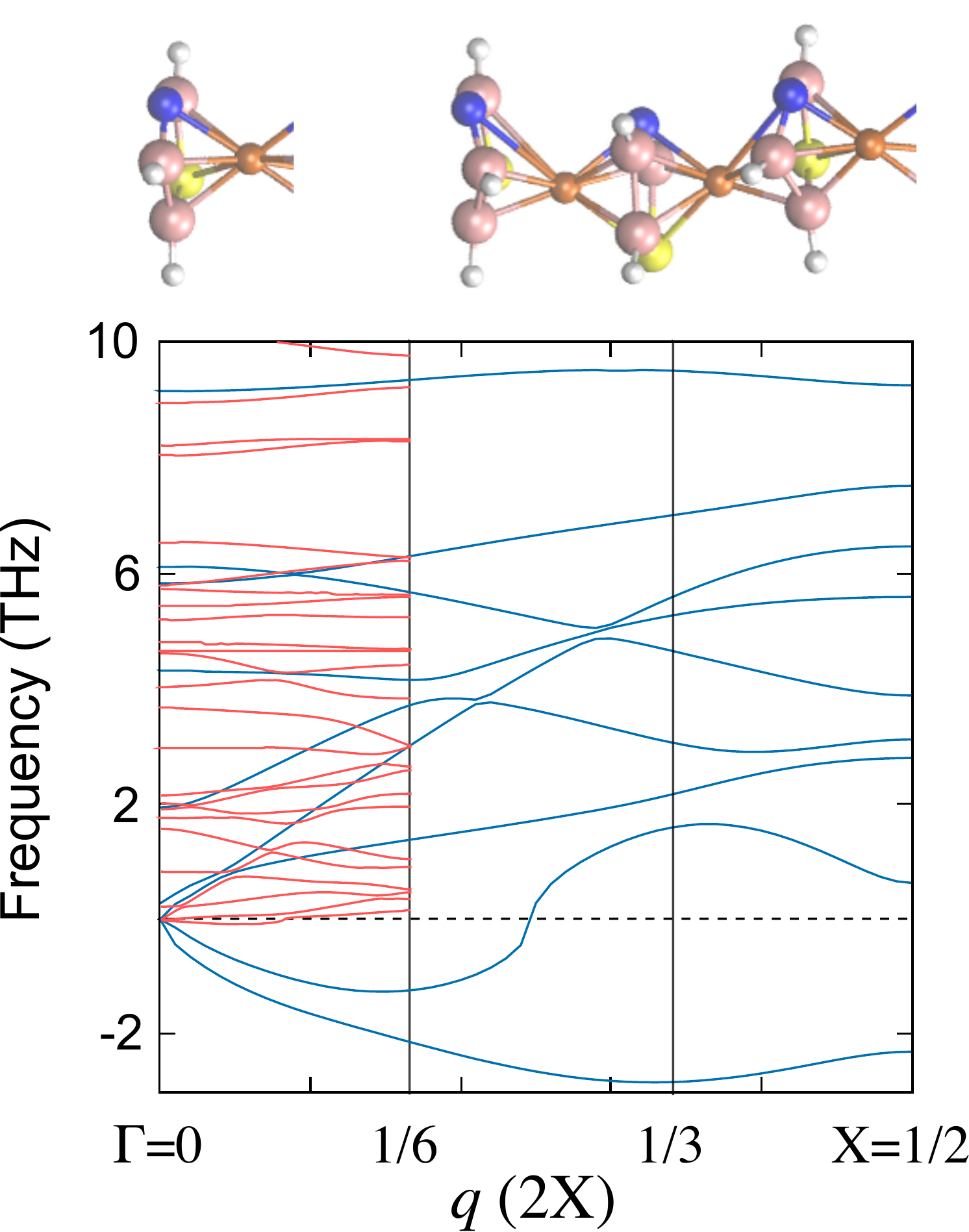}
    \caption{Quasi-1D organometallic materials with a dynamically unstable single unit cell (blue lines) and dynamically stable three-unit supercell (red lines). The imaginary mode in the dispersion of the single unit cell is removed by generating a three-fold expansion along the stacking direction and performing a geometry relaxation with the atomic positions modulated by the imaginary mode. The three-fold expansion in real space results in a three-fold contraction in reciprocal space, with the Brillouin zone edge folding from $\frac{1}{2}$ for the single unit cell to $\frac{1}{6}$ for the supercell. Reproduced with permission from Ref. \citenum{kayastha2021high}. Copyright 2021 by the American Institute of Physics.}
    \label{fig:figQ1D_CDW}
\end{figure}

Based on the form of the dispersion the materials were divided into into three classes, \textit{viz.}: (i) dynamically stable systems with no imaginary harmonic modes; (ii) materials with a single or degenerate instability; and (iii) materials with multiple independent phonon instabilities.
Approximately half of the materials (587) were identified as having maximum instabilities at the Brillouin-zone boundary with $q = \frac{1}{2}$, of which 33 were selected for further analysis to examine trends in the effect of the distortions on the electronic structure.
A further 28 materials were predicted to have charge-density wave like distortions with a instabilities at $q = \frac{1}{3}$.
Fig. \ref{fig:figQ1D_CDW} shows an example where the maximum instability occurs at $q=\frac{1}{3}$.
To remove the imaginary mode and obtain a dynamically-stable structure, the unit cell was expanded threefold in the stacking direction and optimised with the atomic positions modulated by the imaginary mode wavevector and geometry relaxed. Subsequent phonon calculations confirmed the resulting structure to be dynamically stable.

To conclude our discussion of using imaginary phonon modes for crystal structure prediction it is interesting to consider how the mode-following technique compares to other structure-prediction methods.
In general, complex materials are challenging both because structure prediction requires searching over more independent variables, and because the number of minima on the PES grows exponentially with the system size.\cite{stillinger1999exponential}
In the approach adopted in these studies, it can be seen that the search over $3n_a$ atomic positions is reduced to a series of sequential searches over a much smaller number of imaginary modes.
An additional subtlety is that in the mode-mapping approach any changes required to the cell volume are indicated by the wavevectors of the imaginary modes, as was exploited in the example of the Q1D polymer discussed above.\cite{kayastha2021high}
This is in contrast to many prediction algorithms, where the number of formula units is treated as a fixed variable or restricted to a range of values.
As highlighted in the study on metallic elements, the imaginary modes linking structures in each generation can also provide valuable information on the connectivity of the PES (Fig. \ref{fig:Cu-TreeDiagram}).\cite{togo2013evolution}
Finally, the successive freezing in (``condensation'') of imaginary modes mimics how a material naturally explores its PES on cooling, so if the mode-following procedure is started from a high-temperature experimental structure it is not unreasonable to expect that the matching low-temperature structure(s) will be found, as in the case of \ce{Bi2Sn2O7}.\cite{rahim2020polymorph}

On the other hand, the requirement for an initial structure means the mode-mapping approach cannot be used to predict the structures of compositions for which there are no experimental structures or suitable known analogues.
Also, since the stable structures found using this technique must be accessible from the starting structure by following imaginary modes downhill in energy, if the global minimum and/or other local minima are not connected to the starting structure in this way they will not be found.
This is the reason why for example the study in Ref. \citenum{togo2013evolution} does not locate the stable $\omega$ phase of Ti.
If the starting structure is taken from experiment, i.e. it is the phase obtained from the synthesis, this may not be an issue.
However, if the initial structure is guessed then an appropriate exploration of the PES is not guaranteed.
Finally, this technique requires a full phonon calculation on each structure, which can easily impose an overhead of 1-2 orders of magnitude on the geometry-optimisation steps.

If there are other systems like \ce{Bi2Sn2O7}, where a high-temperature structure has been solved but the low-temperature phase(s) are unknown, exploring these with mode-mapping has the potential to yield new materials that should be synthetically accessible.
Alternatively, there may be some scope for using this approach in combination with other structure-prediction algorithms to refine dynamically-unstable structures obtained from a coarse initial sampling of the structural PES.

\section{Conclusions}

Lattice dynamics calculations based on the harmonic approximation are an important tool in computational materials science and can be used to predict a wide range of material properties. 
Imaginary modes in the harmonic phonon spectrum indicate that a structure is dynamically unstable, and to propose suitable solutions it is essential to first identify their origin.
In some cases, the imaginary modes arise because one or more aspects of the technical setup of the calculation are not adequately converged, or because the chosen theoretical method does not accurately describe the electronic structure or chemical bonding of the material under study.
Where imaginary phonon modes are an intrinsic feature of the lattice dynamics of a material their presence can provide insight into structural phase transitions and identify dynamics that are not accessible in traditional experiments using elastic-scattering techniques.
The imaginary modes can also be used as a basis to locate the energy minima on the structural potential energy surface. 

Treatment of imaginary modes usually begins by mapping the associated PES.
If necessary, one may then consider techniques for determining an effective harmonic potential or renormalised frequency that can be used within the harmonic approximation to obtain a corrected phonon spectrum or to compute physical properties.
We make a distinction here between the relatively simple approach of mapping and solving a Schr{\"o}dinger equation for the potential, and more sophisticated methods that aim for a full renormalisation of the harmonic phonon spectrum.
The former is computationally efficient, particularly when used specifically to target the imaginary modes, but retains the harmonic mode eigenvectors, may not account for the coupling of modes, and neglects changes in the PES with temperature.
On the other hand, full-spectrum renormalisation methods address some or all of these limitations, but are computationally more expensive.
We note a recent comparison between renormalisation schemes based on mode mapping and molecular dynamics for FCC aluminium that demonstrates that finite-temperature frequencies can potentially be robustly predicted with the mode-mapping approach,\cite{adams2021anharmonic} but more research is needed to establish the situations for which this method is suitable and whether it holds in materials with strong intrinsic anharmonicity.

Finally, we have demonstrated through a set of case studies some of the contexts in which imaginary modes and their treatment have important applications in the materials sciences.
The presence of imaginary modes in the phonon spectrum is an important indicator of the nature of a material, and can be used together with energetics calculations to assess the relative stabilities of materials.
Further exploration of the imaginary modes can provide insight into phase transitions and the associated structural dynamics.
Given the relatively low computational cost, there is scope for the further automation and integration of this mapping technique into high-throughput calculations, either as part of an optimisation routine or as part of a materials-discovery endeavour to locate new low-temperature phases starting form known high-temperature structures with dynamical instabilities.
We therefore believe that imaginary mode physics will have an increasingly significant role to play in computational materials modelling for materials discovery.

\section*{Data Access Statement}

No new data was created for this study.

\acknowledgments
We thank Jarvist Moore Frost for his valuable insight into phonon renormalisation, Ruo Xi Yang for providing data and insight on the vibrational properties of \ce{CsPbI3}, John Buckeridge for making his 1D Schr{\"o}dinger solver available for our use, and Samantha N. Hood and Aron Walsh for their enthusiasm on this topic.
IP is grateful to the University of Manchester (UoM) for the support of a PhD studentship.
PK acknowledges support from the UK Engineering and Physical Sciences Research Council (EPSRC) CDT in Renewable Energy Northeast Universities (ReNU) for funding through EPSRC grant EP/S023836/1.
JMS is currently supported by a UK Research and Innovation (UKRI) Future Leaders Fellowship (MR/T043121/1), and previously held a UoM Presidential Fellowship.
This work used the Oswald High Performance Computing facility operated by Northumbria University (UK).
\textit{Via} our membership of the UK's HEC Materials Chemistry Consortium, which is funded by EPSRC (EP/R029431), this work used the ARCHER2 UK National Supercomputing Service (http://www.archer2.ac.uk).
A subset of our calculations made use of the UoM Computational Shared Facility (CSF), which is maintained by UoM Research IT.

\bibliography{references}

\begin{thebibliography}{112}%
\makeatletter
\providecommand \@ifxundefined [1]{%
 \@ifx{#1\undefined}
}%
\providecommand \@ifnum [1]{%
 \ifnum #1\expandafter \@firstoftwo
 \else \expandafter \@secondoftwo
 \fi
}%
\providecommand \@ifx [1]{%
 \ifx #1\expandafter \@firstoftwo
 \else \expandafter \@secondoftwo
 \fi
}%
\providecommand \natexlab [1]{#1}%
\providecommand \enquote  [1]{``#1''}%
\providecommand \bibnamefont  [1]{#1}%
\providecommand \bibfnamefont [1]{#1}%
\providecommand \citenamefont [1]{#1}%
\providecommand \href@noop [0]{\@secondoftwo}%
\providecommand \href [0]{\begingroup \@sanitize@url \@href}%
\providecommand \@href[1]{\@@startlink{#1}\@@href}%
\providecommand \@@href[1]{\endgroup#1\@@endlink}%
\providecommand \@sanitize@url [0]{\catcode `\\12\catcode `\$12\catcode
  `\&12\catcode `\#12\catcode `\^12\catcode `\_12\catcode `\%12\relax}%
\providecommand \@@startlink[1]{}%
\providecommand \@@endlink[0]{}%
\providecommand \url  [0]{\begingroup\@sanitize@url \@url }%
\providecommand \@url [1]{\endgroup\@href {#1}{\urlprefix }}%
\providecommand \urlprefix  [0]{URL }%
\providecommand \Eprint [0]{\href }%
\providecommand \doibase [0]{http://dx.doi.org/}%
\providecommand \selectlanguage [0]{\@gobble}%
\providecommand \bibinfo  [0]{\@secondoftwo}%
\providecommand \bibfield  [0]{\@secondoftwo}%
\providecommand \translation [1]{[#1]}%
\providecommand \BibitemOpen [0]{}%
\providecommand \bibitemStop [0]{}%
\providecommand \bibitemNoStop [0]{.\EOS\space}%
\providecommand \EOS [0]{\spacefactor3000\relax}%
\providecommand \BibitemShut  [1]{\csname bibitem#1\endcsname}%
\let\auto@bib@innerbib\@empty
%</preamble>
\bibitem [{\citenamefont {Togo}\ and\ \citenamefont
  {Tanaka}(2015{\natexlab{a}})}]{togo2015}%
  \BibitemOpen
  \bibfield  {author} {\bibinfo {author} {\bibfnamefont {A.}~\bibnamefont
  {Togo}}\ and\ \bibinfo {author} {\bibfnamefont {I.}~\bibnamefont {Tanaka}},\
  }\href {\doibase https://doi.org/10.1016/j.scriptamat.2015.07.021} {\bibfield
   {journal} {\bibinfo  {journal} {Scr. Mater.}\ }\textbf {\bibinfo {volume}
  {108}},\ \bibinfo {pages} {1 } (\bibinfo {year}
  {2015}{\natexlab{a}})}\BibitemShut {NoStop}%
\bibitem [{\citenamefont {Skelton}\ \emph
  {et~al.}(2017{\natexlab{a}})\citenamefont {Skelton}, \citenamefont {Burton},
  \citenamefont {Jackson}, \citenamefont {Oba}, \citenamefont {Parker},\ and\
  \citenamefont {Walsh}}]{skelton2017spectroscopy}%
  \BibitemOpen
  \bibfield  {author} {\bibinfo {author} {\bibfnamefont {J.~M.}\ \bibnamefont
  {Skelton}}, \bibinfo {author} {\bibfnamefont {L.~A.}\ \bibnamefont {Burton}},
  \bibinfo {author} {\bibfnamefont {A.~J.}\ \bibnamefont {Jackson}}, \bibinfo
  {author} {\bibfnamefont {F.}~\bibnamefont {Oba}}, \bibinfo {author}
  {\bibfnamefont {S.~C.}\ \bibnamefont {Parker}}, \ and\ \bibinfo {author}
  {\bibfnamefont {A.}~\bibnamefont {Walsh}},\ }\href {\doibase
  10.1039/C7CP01680H} {\bibfield  {journal} {\bibinfo  {journal} {Phys. Chem.
  Chem. Phys.}\ }\textbf {\bibinfo {volume} {19}},\ \bibinfo {pages} {12452}
  (\bibinfo {year} {2017}{\natexlab{a}})}\BibitemShut {NoStop}%
\bibitem [{\citenamefont {Cea}\ and\ \citenamefont
  {Guinea}(2021)}]{cea2021coulomb}%
  \BibitemOpen
  \bibfield  {author} {\bibinfo {author} {\bibfnamefont {T.}~\bibnamefont
  {Cea}}\ and\ \bibinfo {author} {\bibfnamefont {F.}~\bibnamefont {Guinea}},\
  }\href {https://doi.org/10.1073/pnas.2107874118} {\bibfield  {journal}
  {\bibinfo  {journal} {Proc. Natl. Acad. Sci. U.S.A.}\ }\textbf {\bibinfo
  {volume} {118}} (\bibinfo {year} {2021})}\BibitemShut {NoStop}%
\bibitem [{\citenamefont {Mankowsky}\ \emph {et~al.}(2014)\citenamefont
  {Mankowsky}, \citenamefont {Subedi}, \citenamefont {F{\"o}rst}, \citenamefont
  {Mariager}, \citenamefont {Chollet}, \citenamefont {Lemke}, \citenamefont
  {Robinson}, \citenamefont {Glownia}, \citenamefont {Minitti}, \citenamefont
  {Frano} \emph {et~al.}}]{mankowsky2014nonlinear}%
  \BibitemOpen
  \bibfield  {author} {\bibinfo {author} {\bibfnamefont {R.}~\bibnamefont
  {Mankowsky}}, \bibinfo {author} {\bibfnamefont {A.}~\bibnamefont {Subedi}},
  \bibinfo {author} {\bibfnamefont {M.}~\bibnamefont {F{\"o}rst}}, \bibinfo
  {author} {\bibfnamefont {S.~O.}\ \bibnamefont {Mariager}}, \bibinfo {author}
  {\bibfnamefont {M.}~\bibnamefont {Chollet}}, \bibinfo {author} {\bibfnamefont
  {H.}~\bibnamefont {Lemke}}, \bibinfo {author} {\bibfnamefont {J.~S.}\
  \bibnamefont {Robinson}}, \bibinfo {author} {\bibfnamefont {J.~M.}\
  \bibnamefont {Glownia}}, \bibinfo {author} {\bibfnamefont {M.~P.}\
  \bibnamefont {Minitti}}, \bibinfo {author} {\bibfnamefont {A.}~\bibnamefont
  {Frano}},  \emph {et~al.},\ }\href {https://doi.org/10.1038/nature13875}
  {\bibfield  {journal} {\bibinfo  {journal} {Nature}\ }\textbf {\bibinfo
  {volume} {516}},\ \bibinfo {pages} {71} (\bibinfo {year} {2014})}\BibitemShut
  {NoStop}%
\bibitem [{\citenamefont {Steele}\ \emph {et~al.}(2020)\citenamefont {Steele},
  \citenamefont {Lai}, \citenamefont {Zhang}, \citenamefont {Lin},
  \citenamefont {Hofkens}, \citenamefont {Roeffaers},\ and\ \citenamefont
  {Yang}}]{steele2020phase}%
  \BibitemOpen
  \bibfield  {author} {\bibinfo {author} {\bibfnamefont {J.~A.}\ \bibnamefont
  {Steele}}, \bibinfo {author} {\bibfnamefont {M.}~\bibnamefont {Lai}},
  \bibinfo {author} {\bibfnamefont {Y.}~\bibnamefont {Zhang}}, \bibinfo
  {author} {\bibfnamefont {Z.}~\bibnamefont {Lin}}, \bibinfo {author}
  {\bibfnamefont {J.}~\bibnamefont {Hofkens}}, \bibinfo {author} {\bibfnamefont
  {M.~B.}\ \bibnamefont {Roeffaers}}, \ and\ \bibinfo {author} {\bibfnamefont
  {P.}~\bibnamefont {Yang}},\ }\href
  {https://doi.org/10.1021/accountsmr.0c00009} {\bibfield  {journal} {\bibinfo
  {journal} {Acc. Mater. Res.}\ }\textbf {\bibinfo {volume} {1}},\ \bibinfo
  {pages} {3} (\bibinfo {year} {2020})}\BibitemShut {NoStop}%
\bibitem [{\citenamefont {Adams}\ and\ \citenamefont
  {Passerone}(2016)}]{Adams2016}%
  \BibitemOpen
  \bibfield  {author} {\bibinfo {author} {\bibfnamefont {D.~J.}\ \bibnamefont
  {Adams}}\ and\ \bibinfo {author} {\bibfnamefont {D.}~\bibnamefont
  {Passerone}},\ }\href {https://doi.org/10.1088/0953-8984/28/30/305401}
  {\bibfield  {journal} {\bibinfo  {journal} {J. Condens. Matter Phys.}\
  }\textbf {\bibinfo {volume} {28}} (\bibinfo {year} {2016})}\BibitemShut
  {NoStop}%
\bibitem [{\citenamefont {Souvatzis}\ \emph {et~al.}(2009)\citenamefont
  {Souvatzis}, \citenamefont {Eriksson}, \citenamefont {Katsnelson},\ and\
  \citenamefont {Rudin}}]{Souvatzis2009}%
  \BibitemOpen
  \bibfield  {author} {\bibinfo {author} {\bibfnamefont {P.}~\bibnamefont
  {Souvatzis}}, \bibinfo {author} {\bibfnamefont {O.}~\bibnamefont {Eriksson}},
  \bibinfo {author} {\bibfnamefont {M.~I.}\ \bibnamefont {Katsnelson}}, \ and\
  \bibinfo {author} {\bibfnamefont {S.~P.}\ \bibnamefont {Rudin}},\ }\href
  {\doibase 10.1016/j.commatsci.2008.06.016} {\bibfield  {journal} {\bibinfo
  {journal} {Comput. Mater. Sci.}\ }\textbf {\bibinfo {volume} {44}},\ \bibinfo
  {pages} {888} (\bibinfo {year} {2009})}\BibitemShut {NoStop}%
\bibitem [{\citenamefont {Jong}\ \emph {et~al.}(2019)\citenamefont {Jong},
  \citenamefont {Yu}, \citenamefont {Kye}, \citenamefont {Choe}, \citenamefont
  {Kim},\ and\ \citenamefont {Choe}}]{jong2019anharmonic}%
  \BibitemOpen
  \bibfield  {author} {\bibinfo {author} {\bibfnamefont {U.-G.}\ \bibnamefont
  {Jong}}, \bibinfo {author} {\bibfnamefont {C.-J.}\ \bibnamefont {Yu}},
  \bibinfo {author} {\bibfnamefont {Y.-H.}\ \bibnamefont {Kye}}, \bibinfo
  {author} {\bibfnamefont {S.-H.}\ \bibnamefont {Choe}}, \bibinfo {author}
  {\bibfnamefont {J.-S.}\ \bibnamefont {Kim}}, \ and\ \bibinfo {author}
  {\bibfnamefont {Y.-G.}\ \bibnamefont {Choe}},\ }\href
  {https://doi.org/10.1103/PhysRevB.99.184105} {\bibfield  {journal} {\bibinfo
  {journal} {Phys. Rev. B}\ }\textbf {\bibinfo {volume} {99}},\ \bibinfo
  {pages} {184105} (\bibinfo {year} {2019})}\BibitemShut {NoStop}%
\bibitem [{\citenamefont {Kamba}(2021)}]{kamba2021soft}%
  \BibitemOpen
  \bibfield  {author} {\bibinfo {author} {\bibfnamefont {S.}~\bibnamefont
  {Kamba}},\ }\href {https://doi.org/10.1063/5.0036066} {\bibfield  {journal}
  {\bibinfo  {journal} {APL Mater.}\ }\textbf {\bibinfo {volume} {9}},\
  \bibinfo {pages} {020704} (\bibinfo {year} {2021})}\BibitemShut {NoStop}%
\bibitem [{\citenamefont {Kumar}\ \emph {et~al.}(2012)\citenamefont {Kumar},
  \citenamefont {Bera}, \citenamefont {Muthu}, \citenamefont {Shirodkar},
  \citenamefont {Saha}, \citenamefont {Shireen}, \citenamefont {Sundaresan},
  \citenamefont {Waghmare}, \citenamefont {Sood},\ and\ \citenamefont
  {Rao}}]{kumar2012coupled}%
  \BibitemOpen
  \bibfield  {author} {\bibinfo {author} {\bibfnamefont {P.}~\bibnamefont
  {Kumar}}, \bibinfo {author} {\bibfnamefont {A.}~\bibnamefont {Bera}},
  \bibinfo {author} {\bibfnamefont {D.}~\bibnamefont {Muthu}}, \bibinfo
  {author} {\bibfnamefont {S.~N.}\ \bibnamefont {Shirodkar}}, \bibinfo {author}
  {\bibfnamefont {R.}~\bibnamefont {Saha}}, \bibinfo {author} {\bibfnamefont
  {A.}~\bibnamefont {Shireen}}, \bibinfo {author} {\bibfnamefont
  {A.}~\bibnamefont {Sundaresan}}, \bibinfo {author} {\bibfnamefont {U.~V.}\
  \bibnamefont {Waghmare}}, \bibinfo {author} {\bibfnamefont {A.}~\bibnamefont
  {Sood}}, \ and\ \bibinfo {author} {\bibfnamefont {C.}~\bibnamefont {Rao}},\
  }\href {https://doi.org/10.1103/PhysRevB.85.134449} {\bibfield  {journal}
  {\bibinfo  {journal} {Phys. Rev. B}\ }\textbf {\bibinfo {volume} {85}},\
  \bibinfo {pages} {134449} (\bibinfo {year} {2012})}\BibitemShut {NoStop}%
\bibitem [{\citenamefont {Martin}(2020)}]{martin2020electronic}%
  \BibitemOpen
  \bibfield  {author} {\bibinfo {author} {\bibfnamefont {R.~M.}\ \bibnamefont
  {Martin}},\ }\href@noop {} {\emph {\bibinfo {title} {Electronic structure:
  basic theory and practical methods}}}\ (\bibinfo  {publisher} {Cambridge
  university press},\ \bibinfo {year} {2020})\BibitemShut {NoStop}%
\bibitem [{\citenamefont {Giustino}(2014)}]{giustino2014materials}%
  \BibitemOpen
  \bibfield  {author} {\bibinfo {author} {\bibfnamefont {F.}~\bibnamefont
  {Giustino}},\ }\href@noop {} {\emph {\bibinfo {title} {Materials modelling
  using density functional theory: properties and predictions}}}\ (\bibinfo
  {publisher} {Oxford University Press},\ \bibinfo {year} {2014})\BibitemShut
  {NoStop}%
\bibitem [{\citenamefont {Dove}(1997)}]{dove1997theory}%
  \BibitemOpen
  \bibfield  {author} {\bibinfo {author} {\bibfnamefont {M.~T.}\ \bibnamefont
  {Dove}},\ }\href {https://doi.org/10.2138/am-1997-3-401} {\bibfield
  {journal} {\bibinfo  {journal} {Am. Mineral.}\ }\textbf {\bibinfo {volume}
  {82}},\ \bibinfo {pages} {213} (\bibinfo {year} {1997})}\BibitemShut
  {NoStop}%
\bibitem [{\citenamefont {Togo}\ and\ \citenamefont
  {Tanaka}(2015{\natexlab{b}})}]{togo2015first}%
  \BibitemOpen
  \bibfield  {author} {\bibinfo {author} {\bibfnamefont {A.}~\bibnamefont
  {Togo}}\ and\ \bibinfo {author} {\bibfnamefont {I.}~\bibnamefont {Tanaka}},\
  }\href {https://doi.org/10.1016/j.scriptamat.2015.07.021} {\bibfield
  {journal} {\bibinfo  {journal} {Scr. Mater.}\ }\textbf {\bibinfo {volume}
  {108}},\ \bibinfo {pages} {1} (\bibinfo {year}
  {2015}{\natexlab{b}})}\BibitemShut {NoStop}%
\bibitem [{\citenamefont {Baroni}\ \emph {et~al.}(2001)\citenamefont {Baroni},
  \citenamefont {De~Gironcoli}, \citenamefont {Dal~Corso},\ and\ \citenamefont
  {Giannozzi}}]{baroni2001phonons}%
  \BibitemOpen
  \bibfield  {author} {\bibinfo {author} {\bibfnamefont {S.}~\bibnamefont
  {Baroni}}, \bibinfo {author} {\bibfnamefont {S.}~\bibnamefont
  {De~Gironcoli}}, \bibinfo {author} {\bibfnamefont {A.}~\bibnamefont
  {Dal~Corso}}, \ and\ \bibinfo {author} {\bibfnamefont {P.}~\bibnamefont
  {Giannozzi}},\ }\href {https://doi.org/10.1103/RevModPhys.73.515} {\bibfield
  {journal} {\bibinfo  {journal} {Rev. Mod. Phys.}\ }\textbf {\bibinfo {volume}
  {73}},\ \bibinfo {pages} {515} (\bibinfo {year} {2001})}\BibitemShut
  {NoStop}%
\bibitem [{\citenamefont {Parlinski}, \citenamefont {Li},\ and\ \citenamefont
  {Kawazoe}(1997)}]{parlinski1997first}%
  \BibitemOpen
  \bibfield  {author} {\bibinfo {author} {\bibfnamefont {K.}~\bibnamefont
  {Parlinski}}, \bibinfo {author} {\bibfnamefont {Z.}~\bibnamefont {Li}}, \
  and\ \bibinfo {author} {\bibfnamefont {Y.}~\bibnamefont {Kawazoe}},\ }\href
  {https://doi.org/10.1103/PhysRevLett.78.4063} {\bibfield  {journal} {\bibinfo
   {journal} {Phys. Rev. Lett.}\ }\textbf {\bibinfo {volume} {78}},\ \bibinfo
  {pages} {4063} (\bibinfo {year} {1997})}\BibitemShut {NoStop}%
\bibitem [{\citenamefont {Aly{\"o}r{\"u}k}(2016)}]{alyoruk2016piezoelectric}%
  \BibitemOpen
  \bibfield  {author} {\bibinfo {author} {\bibfnamefont {M.~M.}\ \bibnamefont
  {Aly{\"o}r{\"u}k}},\ }\href {https://doi.org/10.1002/pssb.201600387}
  {\bibfield  {journal} {\bibinfo  {journal} {Phys. Status Solidi B}\ }\textbf
  {\bibinfo {volume} {253}},\ \bibinfo {pages} {2534} (\bibinfo {year}
  {2016})}\BibitemShut {NoStop}%
\bibitem [{\citenamefont {Sabatini}\ \emph {et~al.}(2016)\citenamefont
  {Sabatini}, \citenamefont {K{\"u}{\c{c}}{\"u}kbenli}, \citenamefont {Pham},\
  and\ \citenamefont {de~Gironcoli}}]{sabatini2016phonons}%
  \BibitemOpen
  \bibfield  {author} {\bibinfo {author} {\bibfnamefont {R.}~\bibnamefont
  {Sabatini}}, \bibinfo {author} {\bibfnamefont {E.}~\bibnamefont
  {K{\"u}{\c{c}}{\"u}kbenli}}, \bibinfo {author} {\bibfnamefont {C.~H.}\
  \bibnamefont {Pham}}, \ and\ \bibinfo {author} {\bibfnamefont
  {S.}~\bibnamefont {de~Gironcoli}},\ }\href
  {https://doi.org/10.1103/PhysRevB.93.235120} {\bibfield  {journal} {\bibinfo
  {journal} {Phys. Rev. B}\ }\textbf {\bibinfo {volume} {93}},\ \bibinfo
  {pages} {235120} (\bibinfo {year} {2016})}\BibitemShut {NoStop}%
\bibitem [{\citenamefont {Coutinho}\ \emph {et~al.}(2017)\citenamefont
  {Coutinho}, \citenamefont {Tavares}, \citenamefont {Barboza}, \citenamefont
  {Fraz{\~a}o}, \citenamefont {Moreira},\ and\ \citenamefont
  {Azevedo}}]{coutinho20173r}%
  \BibitemOpen
  \bibfield  {author} {\bibinfo {author} {\bibfnamefont {S.}~\bibnamefont
  {Coutinho}}, \bibinfo {author} {\bibfnamefont {M.}~\bibnamefont {Tavares}},
  \bibinfo {author} {\bibfnamefont {C.}~\bibnamefont {Barboza}}, \bibinfo
  {author} {\bibfnamefont {N.}~\bibnamefont {Fraz{\~a}o}}, \bibinfo {author}
  {\bibfnamefont {E.}~\bibnamefont {Moreira}}, \ and\ \bibinfo {author}
  {\bibfnamefont {D.~L.}\ \bibnamefont {Azevedo}},\ }\href
  {https://doi.org/10.1016/j.jpcs.2017.07.010} {\bibfield  {journal} {\bibinfo
  {journal} {J. Phys. Chem. Solids}\ }\textbf {\bibinfo {volume} {111}},\
  \bibinfo {pages} {25} (\bibinfo {year} {2017})}\BibitemShut {NoStop}%
\bibitem [{\citenamefont {Pike}\ \emph {et~al.}(2018)\citenamefont {Pike},
  \citenamefont {Dewandre}, \citenamefont {Van~Troeye}, \citenamefont {Gonze},\
  and\ \citenamefont {Verstraete}}]{pike2018vibrational}%
  \BibitemOpen
  \bibfield  {author} {\bibinfo {author} {\bibfnamefont {N.~A.}\ \bibnamefont
  {Pike}}, \bibinfo {author} {\bibfnamefont {A.}~\bibnamefont {Dewandre}},
  \bibinfo {author} {\bibfnamefont {B.}~\bibnamefont {Van~Troeye}}, \bibinfo
  {author} {\bibfnamefont {X.}~\bibnamefont {Gonze}}, \ and\ \bibinfo {author}
  {\bibfnamefont {M.~J.}\ \bibnamefont {Verstraete}},\ }\href
  {https://doi.org/10.1103/PhysRevMaterials.2.063608} {\bibfield  {journal}
  {\bibinfo  {journal} {Phys. Rev. Mater.}\ }\textbf {\bibinfo {volume} {2}},\
  \bibinfo {pages} {063608} (\bibinfo {year} {2018})}\BibitemShut {NoStop}%
\bibitem [{\citenamefont {Togo}, \citenamefont {Chaput},\ and\ \citenamefont
  {Tanaka}(2015)}]{togo2015distributions}%
  \BibitemOpen
  \bibfield  {author} {\bibinfo {author} {\bibfnamefont {A.}~\bibnamefont
  {Togo}}, \bibinfo {author} {\bibfnamefont {L.}~\bibnamefont {Chaput}}, \ and\
  \bibinfo {author} {\bibfnamefont {I.}~\bibnamefont {Tanaka}},\ }\href
  {https://doi.org/10.1103/PhysRevB.91.094306} {\bibfield  {journal} {\bibinfo
  {journal} {Phys. Rev. B}\ }\textbf {\bibinfo {volume} {91}},\ \bibinfo
  {pages} {094306} (\bibinfo {year} {2015})}\BibitemShut {NoStop}%
\bibitem [{\citenamefont {Li}\ \emph {et~al.}(2014)\citenamefont {Li},
  \citenamefont {Carrete}, \citenamefont {Katcho},\ and\ \citenamefont
  {Mingo}}]{li2014shengbte}%
  \BibitemOpen
  \bibfield  {author} {\bibinfo {author} {\bibfnamefont {W.}~\bibnamefont
  {Li}}, \bibinfo {author} {\bibfnamefont {J.}~\bibnamefont {Carrete}},
  \bibinfo {author} {\bibfnamefont {N.~A.}\ \bibnamefont {Katcho}}, \ and\
  \bibinfo {author} {\bibfnamefont {N.}~\bibnamefont {Mingo}},\ }\href
  {https://doi.org/10.1016/j.cpc.2014.02.015} {\bibfield  {journal} {\bibinfo
  {journal} {Comput. Phys. Commun.}\ }\textbf {\bibinfo {volume} {185}},\
  \bibinfo {pages} {1747} (\bibinfo {year} {2014})}\BibitemShut {NoStop}%
\bibitem [{\citenamefont {Carrete}\ \emph {et~al.}(2017)\citenamefont
  {Carrete}, \citenamefont {Vermeersch}, \citenamefont {Katre}, \citenamefont
  {van Roekeghem}, \citenamefont {Wang}, \citenamefont {Madsen},\ and\
  \citenamefont {Mingo}}]{carrete2017almabte}%
  \BibitemOpen
  \bibfield  {author} {\bibinfo {author} {\bibfnamefont {J.}~\bibnamefont
  {Carrete}}, \bibinfo {author} {\bibfnamefont {B.}~\bibnamefont {Vermeersch}},
  \bibinfo {author} {\bibfnamefont {A.}~\bibnamefont {Katre}}, \bibinfo
  {author} {\bibfnamefont {A.}~\bibnamefont {van Roekeghem}}, \bibinfo {author}
  {\bibfnamefont {T.}~\bibnamefont {Wang}}, \bibinfo {author} {\bibfnamefont
  {G.~K.}\ \bibnamefont {Madsen}}, \ and\ \bibinfo {author} {\bibfnamefont
  {N.}~\bibnamefont {Mingo}},\ }\href
  {https://doi.org/10.1016/j.cpc.2017.06.023} {\bibfield  {journal} {\bibinfo
  {journal} {Comput. Phys. Commun.}\ }\textbf {\bibinfo {volume} {220}},\
  \bibinfo {pages} {351} (\bibinfo {year} {2017})}\BibitemShut {NoStop}%
\bibitem [{\citenamefont {Hellman}, \citenamefont {Abrikosov},\ and\
  \citenamefont {Simak}(2011)}]{hellman2011lattice}%
  \BibitemOpen
  \bibfield  {author} {\bibinfo {author} {\bibfnamefont {O.}~\bibnamefont
  {Hellman}}, \bibinfo {author} {\bibfnamefont {I.}~\bibnamefont {Abrikosov}},
  \ and\ \bibinfo {author} {\bibfnamefont {S.}~\bibnamefont {Simak}},\ }\href
  {https://doi.org/10.1103/PhysRevB.84.180301} {\bibfield  {journal} {\bibinfo
  {journal} {Phys. Rev. B}\ }\textbf {\bibinfo {volume} {84}},\ \bibinfo
  {pages} {180301} (\bibinfo {year} {2011})}\BibitemShut {NoStop}%
\bibitem [{\citenamefont {Hellman}(2012)}]{hellman2012thermal}%
  \BibitemOpen
  \bibfield  {author} {\bibinfo {author} {\bibfnamefont {O.}~\bibnamefont
  {Hellman}},\ }\emph {\bibinfo {title} {Thermal properties of materials from
  first principles}},\ \href@noop {} {Ph.D. thesis},\ \bibinfo  {school}
  {Link{\"o}ping University Electronic Press} (\bibinfo {year}
  {2012})\BibitemShut {NoStop}%
\bibitem [{\citenamefont {Hellman}\ \emph {et~al.}(2013)\citenamefont
  {Hellman}, \citenamefont {Steneteg}, \citenamefont {Abrikosov},\ and\
  \citenamefont {Simak}}]{hellman2013temperature}%
  \BibitemOpen
  \bibfield  {author} {\bibinfo {author} {\bibfnamefont {O.}~\bibnamefont
  {Hellman}}, \bibinfo {author} {\bibfnamefont {P.}~\bibnamefont {Steneteg}},
  \bibinfo {author} {\bibfnamefont {I.~A.}\ \bibnamefont {Abrikosov}}, \ and\
  \bibinfo {author} {\bibfnamefont {S.~I.}\ \bibnamefont {Simak}},\ }\href
  {https://doi.org/10.1103/PhysRevB.87.104111} {\bibfield  {journal} {\bibinfo
  {journal} {Phys. Rev. B}\ }\textbf {\bibinfo {volume} {87}},\ \bibinfo
  {pages} {104111} (\bibinfo {year} {2013})}\BibitemShut {NoStop}%
\bibitem [{\citenamefont {Carreras}, \citenamefont {Togo},\ and\ \citenamefont
  {Tanaka}(2017)}]{carreras2017dynaphopy}%
  \BibitemOpen
  \bibfield  {author} {\bibinfo {author} {\bibfnamefont {A.}~\bibnamefont
  {Carreras}}, \bibinfo {author} {\bibfnamefont {A.}~\bibnamefont {Togo}}, \
  and\ \bibinfo {author} {\bibfnamefont {I.}~\bibnamefont {Tanaka}},\ }\href
  {https://doi.org/10.1016/j.cpc.2017.08.017} {\bibfield  {journal} {\bibinfo
  {journal} {Comput. Phys. Commun.}\ }\textbf {\bibinfo {volume} {221}},\
  \bibinfo {pages} {221} (\bibinfo {year} {2017})}\BibitemShut {NoStop}%
\bibitem [{\citenamefont {Monacelli}\ \emph {et~al.}(2021)\citenamefont
  {Monacelli}, \citenamefont {Bianco}, \citenamefont {Cherubini}, \citenamefont
  {Calandra}, \citenamefont {Errea},\ and\ \citenamefont
  {Mauri}}]{monacelli2021stochastic}%
  \BibitemOpen
  \bibfield  {author} {\bibinfo {author} {\bibfnamefont {L.}~\bibnamefont
  {Monacelli}}, \bibinfo {author} {\bibfnamefont {R.}~\bibnamefont {Bianco}},
  \bibinfo {author} {\bibfnamefont {M.}~\bibnamefont {Cherubini}}, \bibinfo
  {author} {\bibfnamefont {M.}~\bibnamefont {Calandra}}, \bibinfo {author}
  {\bibfnamefont {I.}~\bibnamefont {Errea}}, \ and\ \bibinfo {author}
  {\bibfnamefont {F.}~\bibnamefont {Mauri}},\ }\href
  {https://doi.org/10.1088/1361-648X/ac066b} {\bibfield  {journal} {\bibinfo
  {journal} {J. Condens. Matter Phys.}\ }\textbf {\bibinfo {volume} {33}},\
  \bibinfo {pages} {363001} (\bibinfo {year} {2021})}\BibitemShut {NoStop}%
\bibitem [{Note1()}]{Note1}%
  \BibitemOpen
  \bibinfo {note} {We note that imaginary modes are sometimes indicated with
  negative frequencies; while strictly speaking this is not mathematically
  correct, the practice is very common and is often encountered in literature
  on phonons.}\BibitemShut {Stop}%
\bibitem [{\citenamefont {Pallikara}\ and\ \citenamefont
  {Skelton}(2021)}]{Pallikara2021}%
  \BibitemOpen
  \bibfield  {author} {\bibinfo {author} {\bibfnamefont {I.}~\bibnamefont
  {Pallikara}}\ and\ \bibinfo {author} {\bibfnamefont {J.~M.}\ \bibnamefont
  {Skelton}},\ }\href {\doibase 10.1039/d1cp02597j} {\bibfield  {journal}
  {\bibinfo  {journal} {Phys. Chem. Chem. Phys.}\ }\textbf {\bibinfo {volume}
  {23}},\ \bibinfo {pages} {19219} (\bibinfo {year} {2021})}\BibitemShut
  {NoStop}%
\bibitem [{\citenamefont {Yadav}\ \emph {et~al.}(2017)\citenamefont {Yadav},
  \citenamefont {Anita}, \citenamefont {Kumar}, \citenamefont {Panchwanee},
  \citenamefont {Reddy}, \citenamefont {Shirage}, \citenamefont {Biring},\ and\
  \citenamefont {Sen}}]{yadav2017structural}%
  \BibitemOpen
  \bibfield  {author} {\bibinfo {author} {\bibfnamefont {A.~K.}\ \bibnamefont
  {Yadav}}, \bibinfo {author} {\bibnamefont {Anita}}, \bibinfo {author}
  {\bibfnamefont {S.}~\bibnamefont {Kumar}}, \bibinfo {author} {\bibfnamefont
  {A.}~\bibnamefont {Panchwanee}}, \bibinfo {author} {\bibfnamefont {V.~R.}\
  \bibnamefont {Reddy}}, \bibinfo {author} {\bibfnamefont {P.~M.}\ \bibnamefont
  {Shirage}}, \bibinfo {author} {\bibfnamefont {S.}~\bibnamefont {Biring}}, \
  and\ \bibinfo {author} {\bibfnamefont {S.}~\bibnamefont {Sen}},\ }\href
  {\doibase 10.1039/C7RA07130B} {\bibfield  {journal} {\bibinfo  {journal} {RSC
  Adv.}\ }\textbf {\bibinfo {volume} {7}},\ \bibinfo {pages} {39434} (\bibinfo
  {year} {2017})}\BibitemShut {NoStop}%
\bibitem [{\citenamefont {Yuk}\ \emph {et~al.}(2017)\citenamefont {Yuk},
  \citenamefont {Pitike}, \citenamefont {Nakhmanson}, \citenamefont
  {Eisenbach}, \citenamefont {Li},\ and\ \citenamefont
  {Cooper}}]{yuk2017towards}%
  \BibitemOpen
  \bibfield  {author} {\bibinfo {author} {\bibfnamefont {S.~F.}\ \bibnamefont
  {Yuk}}, \bibinfo {author} {\bibfnamefont {K.~C.}\ \bibnamefont {Pitike}},
  \bibinfo {author} {\bibfnamefont {S.~M.}\ \bibnamefont {Nakhmanson}},
  \bibinfo {author} {\bibfnamefont {M.}~\bibnamefont {Eisenbach}}, \bibinfo
  {author} {\bibfnamefont {Y.~W.}\ \bibnamefont {Li}}, \ and\ \bibinfo {author}
  {\bibfnamefont {V.~R.}\ \bibnamefont {Cooper}},\ }\href {\doibase
  10.1038/srep43482} {\bibfield  {journal} {\bibinfo  {journal} {Sci. Rep.}\
  }\textbf {\bibinfo {volume} {7}},\ \bibinfo {pages} {43482} (\bibinfo {year}
  {2017})}\BibitemShut {NoStop}%
\bibitem [{\citenamefont {Skelton}\ \emph {et~al.}(2016)\citenamefont
  {Skelton}, \citenamefont {Burton}, \citenamefont {Parker}, \citenamefont
  {Walsh}, \citenamefont {Kim}, \citenamefont {Soon}, \citenamefont
  {Buckeridge}, \citenamefont {Sokol}, \citenamefont {Catlow}, \citenamefont
  {Togo},\ and\ \citenamefont {Tanaka}}]{Skelton2016}%
  \BibitemOpen
  \bibfield  {author} {\bibinfo {author} {\bibfnamefont {J.~M.}\ \bibnamefont
  {Skelton}}, \bibinfo {author} {\bibfnamefont {L.~A.}\ \bibnamefont {Burton}},
  \bibinfo {author} {\bibfnamefont {S.~C.}\ \bibnamefont {Parker}}, \bibinfo
  {author} {\bibfnamefont {A.}~\bibnamefont {Walsh}}, \bibinfo {author}
  {\bibfnamefont {C.-E.}\ \bibnamefont {Kim}}, \bibinfo {author} {\bibfnamefont
  {A.}~\bibnamefont {Soon}}, \bibinfo {author} {\bibfnamefont {J.}~\bibnamefont
  {Buckeridge}}, \bibinfo {author} {\bibfnamefont {A.~A.}\ \bibnamefont
  {Sokol}}, \bibinfo {author} {\bibfnamefont {C.~R.~A.}\ \bibnamefont
  {Catlow}}, \bibinfo {author} {\bibfnamefont {A.}~\bibnamefont {Togo}}, \ and\
  \bibinfo {author} {\bibfnamefont {I.}~\bibnamefont {Tanaka}},\ }\href
  {https://link.aps.org/doi/10.1103/PhysRevLett.117.075502} {\bibfield
  {journal} {\bibinfo  {journal} {Phys. Rev. Lett.}\ }\textbf {\bibinfo
  {volume} {117}} (\bibinfo {year} {2016})}\BibitemShut {NoStop}%
\bibitem [{\citenamefont {Beecher}\ \emph {et~al.}(2016)\citenamefont
  {Beecher}, \citenamefont {Semonin}, \citenamefont {Skelton}, \citenamefont
  {Frost}, \citenamefont {Terban}, \citenamefont {Zhai}, \citenamefont
  {Alatas}, \citenamefont {Owen}, \citenamefont {Walsh},\ and\ \citenamefont
  {Billinge}}]{beecher2016direct}%
  \BibitemOpen
  \bibfield  {author} {\bibinfo {author} {\bibfnamefont {A.~N.}\ \bibnamefont
  {Beecher}}, \bibinfo {author} {\bibfnamefont {O.~E.}\ \bibnamefont
  {Semonin}}, \bibinfo {author} {\bibfnamefont {J.~M.}\ \bibnamefont
  {Skelton}}, \bibinfo {author} {\bibfnamefont {J.~M.}\ \bibnamefont {Frost}},
  \bibinfo {author} {\bibfnamefont {M.~W.}\ \bibnamefont {Terban}}, \bibinfo
  {author} {\bibfnamefont {H.}~\bibnamefont {Zhai}}, \bibinfo {author}
  {\bibfnamefont {A.}~\bibnamefont {Alatas}}, \bibinfo {author} {\bibfnamefont
  {J.~S.}\ \bibnamefont {Owen}}, \bibinfo {author} {\bibfnamefont
  {A.}~\bibnamefont {Walsh}}, \ and\ \bibinfo {author} {\bibfnamefont {S.~J.}\
  \bibnamefont {Billinge}},\ }\href
  {https://doi.org/10.1021/acsenergylett.6b00381} {\bibfield  {journal}
  {\bibinfo  {journal} {ACS Energy Lett.}\ }\textbf {\bibinfo {volume} {1}},\
  \bibinfo {pages} {880} (\bibinfo {year} {2016})}\BibitemShut {NoStop}%
\bibitem [{\citenamefont {Rahim}\ \emph {et~al.}(2020)\citenamefont {Rahim},
  \citenamefont {Skelton}, \citenamefont {Savory}, \citenamefont {Evans},
  \citenamefont {Evans}, \citenamefont {Walsh},\ and\ \citenamefont
  {Scanlon}}]{rahim2020polymorph}%
  \BibitemOpen
  \bibfield  {author} {\bibinfo {author} {\bibfnamefont {W.}~\bibnamefont
  {Rahim}}, \bibinfo {author} {\bibfnamefont {J.~M.}\ \bibnamefont {Skelton}},
  \bibinfo {author} {\bibfnamefont {C.~N.}\ \bibnamefont {Savory}}, \bibinfo
  {author} {\bibfnamefont {I.~R.}\ \bibnamefont {Evans}}, \bibinfo {author}
  {\bibfnamefont {J.~S.}\ \bibnamefont {Evans}}, \bibinfo {author}
  {\bibfnamefont {A.}~\bibnamefont {Walsh}}, \ and\ \bibinfo {author}
  {\bibfnamefont {D.~O.}\ \bibnamefont {Scanlon}},\ }\href
  {https://doi.org/10.1039/D0SC02995E} {\bibfield  {journal} {\bibinfo
  {journal} {Chem. Sci.}\ }\textbf {\bibinfo {volume} {11}},\ \bibinfo {pages}
  {7904} (\bibinfo {year} {2020})}\BibitemShut {NoStop}%
\bibitem [{\citenamefont {Yang}\ \emph {et~al.}(2017)\citenamefont {Yang},
  \citenamefont {Skelton}, \citenamefont {Da~Silva}, \citenamefont {Frost},\
  and\ \citenamefont {Walsh}}]{yang2017spontaneous}%
  \BibitemOpen
  \bibfield  {author} {\bibinfo {author} {\bibfnamefont {R.~X.}\ \bibnamefont
  {Yang}}, \bibinfo {author} {\bibfnamefont {J.~M.}\ \bibnamefont {Skelton}},
  \bibinfo {author} {\bibfnamefont {E.~L.}\ \bibnamefont {Da~Silva}}, \bibinfo
  {author} {\bibfnamefont {J.~M.}\ \bibnamefont {Frost}}, \ and\ \bibinfo
  {author} {\bibfnamefont {A.}~\bibnamefont {Walsh}},\ }\href
  {https://doi.org/10.1021/acs.jpclett.7b02423} {\bibfield  {journal} {\bibinfo
   {journal} {J. Phys. Chem. Lett.}\ }\textbf {\bibinfo {volume} {8}},\
  \bibinfo {pages} {4720} (\bibinfo {year} {2017})}\BibitemShut {NoStop}%
\bibitem [{\citenamefont {Buckeridge}\ \emph {et~al.}(2013)\citenamefont
  {Buckeridge}, \citenamefont {Scanlon}, \citenamefont {Walsh}, \citenamefont
  {Catlow},\ and\ \citenamefont {Sokol}}]{buckeridge2013dynamical}%
  \BibitemOpen
  \bibfield  {author} {\bibinfo {author} {\bibfnamefont {J.}~\bibnamefont
  {Buckeridge}}, \bibinfo {author} {\bibfnamefont {D.}~\bibnamefont {Scanlon}},
  \bibinfo {author} {\bibfnamefont {A.}~\bibnamefont {Walsh}}, \bibinfo
  {author} {\bibfnamefont {C.}~\bibnamefont {Catlow}}, \ and\ \bibinfo {author}
  {\bibfnamefont {A.}~\bibnamefont {Sokol}},\ }\href
  {https://doi.org/10.1103/PhysRevB.87.214304} {\bibfield  {journal} {\bibinfo
  {journal} {Phys. Rev. B}\ }\textbf {\bibinfo {volume} {87}},\ \bibinfo
  {pages} {214304} (\bibinfo {year} {2013})}\BibitemShut {NoStop}%
\bibitem [{\citenamefont {Krenzer}\ \emph {et~al.}(2021)\citenamefont
  {Krenzer}, \citenamefont {Kim}, \citenamefont {Tolborg}, \citenamefont
  {Morgan},\ and\ \citenamefont {Walsh}}]{krenzer2021anharmonic}%
  \BibitemOpen
  \bibfield  {author} {\bibinfo {author} {\bibfnamefont {G.}~\bibnamefont
  {Krenzer}}, \bibinfo {author} {\bibfnamefont {C.~E.}\ \bibnamefont {Kim}},
  \bibinfo {author} {\bibfnamefont {K.}~\bibnamefont {Tolborg}}, \bibinfo
  {author} {\bibfnamefont {B.}~\bibnamefont {Morgan}}, \ and\ \bibinfo {author}
  {\bibfnamefont {A.}~\bibnamefont {Walsh}},\ }\href
  {https://doi.org/10.1039/D1TA07631K} {\bibfield  {journal} {\bibinfo
  {journal} {J. Mater. Chem. A}\ } (\bibinfo {year} {2021})}\BibitemShut
  {NoStop}%
\bibitem [{\citenamefont {Lazzeri}\ and\ \citenamefont
  {de~Gironcoli}(2002)}]{lazzeri2002first}%
  \BibitemOpen
  \bibfield  {author} {\bibinfo {author} {\bibfnamefont {M.}~\bibnamefont
  {Lazzeri}}\ and\ \bibinfo {author} {\bibfnamefont {S.}~\bibnamefont
  {de~Gironcoli}},\ }\href {https://doi.org/10.1103/PhysRevB.65.245402}
  {\bibfield  {journal} {\bibinfo  {journal} {Phys. Rev. B}\ }\textbf {\bibinfo
  {volume} {65}},\ \bibinfo {pages} {245402} (\bibinfo {year}
  {2002})}\BibitemShut {NoStop}%
\bibitem [{\citenamefont {Esfarjani}\ and\ \citenamefont
  {Stokes}(2008)}]{esfarjani2008method}%
  \BibitemOpen
  \bibfield  {author} {\bibinfo {author} {\bibfnamefont {K.}~\bibnamefont
  {Esfarjani}}\ and\ \bibinfo {author} {\bibfnamefont {H.~T.}\ \bibnamefont
  {Stokes}},\ }\href {https://doi.org/10.1103/PhysRevB.77.144112} {\bibfield
  {journal} {\bibinfo  {journal} {Phys. Rev. B}\ }\textbf {\bibinfo {volume}
  {77}},\ \bibinfo {pages} {144112} (\bibinfo {year} {2008})}\BibitemShut
  {NoStop}%
\bibitem [{\citenamefont {Esfarjani}\ and\ \citenamefont
  {Stokes}(2012)}]{esfarjani2012erratum}%
  \BibitemOpen
  \bibfield  {author} {\bibinfo {author} {\bibfnamefont {K.}~\bibnamefont
  {Esfarjani}}\ and\ \bibinfo {author} {\bibfnamefont {H.~T.}\ \bibnamefont
  {Stokes}},\ }\href {https://doi.org/10.1103/PhysRevB.86.019904} {\bibfield
  {journal} {\bibinfo  {journal} {Phys. Rev. B}\ }\textbf {\bibinfo {volume}
  {86}},\ \bibinfo {pages} {019904} (\bibinfo {year} {2012})}\BibitemShut
  {NoStop}%
\bibitem [{\citenamefont {Knoop}\ \emph {et~al.}(2020)\citenamefont {Knoop},
  \citenamefont {Purcell}, \citenamefont {Scheffler},\ and\ \citenamefont
  {Carbogno}}]{knoop2020anharmonicity}%
  \BibitemOpen
  \bibfield  {author} {\bibinfo {author} {\bibfnamefont {F.}~\bibnamefont
  {Knoop}}, \bibinfo {author} {\bibfnamefont {T.~A.~R.}\ \bibnamefont
  {Purcell}}, \bibinfo {author} {\bibfnamefont {M.}~\bibnamefont {Scheffler}},
  \ and\ \bibinfo {author} {\bibfnamefont {C.}~\bibnamefont {Carbogno}},\
  }\href {\doibase 10.1103/PhysRevMaterials.4.083809} {\bibfield  {journal}
  {\bibinfo  {journal} {Phys. Rev. Materials}\ }\textbf {\bibinfo {volume}
  {4}},\ \bibinfo {pages} {083809} (\bibinfo {year} {2020})}\BibitemShut
  {NoStop}%
\bibitem [{\citenamefont {Tadano}, \citenamefont {Gohda},\ and\ \citenamefont
  {Tsuneyuki}(2014)}]{tadano2014anharmonic}%
  \BibitemOpen
  \bibfield  {author} {\bibinfo {author} {\bibfnamefont {T.}~\bibnamefont
  {Tadano}}, \bibinfo {author} {\bibfnamefont {Y.}~\bibnamefont {Gohda}}, \
  and\ \bibinfo {author} {\bibfnamefont {S.}~\bibnamefont {Tsuneyuki}},\ }\href
  {https://doi.org/10.1088/0953-8984/26/22/225402} {\bibfield  {journal}
  {\bibinfo  {journal} {J. Condens. Matter Phys.}\ }\textbf {\bibinfo {volume}
  {26}},\ \bibinfo {pages} {225402} (\bibinfo {year} {2014})}\BibitemShut
  {NoStop}%
\bibitem [{\citenamefont {Shulumba}, \citenamefont {Hellman},\ and\
  \citenamefont {Minnich}(2017)}]{shulumba2017intrinsic}%
  \BibitemOpen
  \bibfield  {author} {\bibinfo {author} {\bibfnamefont {N.}~\bibnamefont
  {Shulumba}}, \bibinfo {author} {\bibfnamefont {O.}~\bibnamefont {Hellman}}, \
  and\ \bibinfo {author} {\bibfnamefont {A.~J.}\ \bibnamefont {Minnich}},\
  }\href {https://doi.org/10.1103/PhysRevB.95.014302} {\bibfield  {journal}
  {\bibinfo  {journal} {Phys. Rev. B}\ }\textbf {\bibinfo {volume} {95}},\
  \bibinfo {pages} {014302} (\bibinfo {year} {2017})}\BibitemShut {NoStop}%
\bibitem [{\citenamefont {Eriksson}, \citenamefont {Fransson},\ and\
  \citenamefont {Erhart}(2019)}]{eriksson2019hiphive}%
  \BibitemOpen
  \bibfield  {author} {\bibinfo {author} {\bibfnamefont {F.}~\bibnamefont
  {Eriksson}}, \bibinfo {author} {\bibfnamefont {E.}~\bibnamefont {Fransson}},
  \ and\ \bibinfo {author} {\bibfnamefont {P.}~\bibnamefont {Erhart}},\ }\href
  {\doibase https://doi.org/10.1002/adts.201800184} {\bibfield  {journal}
  {\bibinfo  {journal} {Advanced Theory and Simulations}\ }\textbf {\bibinfo
  {volume} {2}},\ \bibinfo {pages} {1800184} (\bibinfo {year} {2019})},\
  \Eprint
  {http://arxiv.org/abs/https://onlinelibrary.wiley.com/doi/pdf/10.1002/adts.201800184}
  {https://onlinelibrary.wiley.com/doi/pdf/10.1002/adts.201800184} \BibitemShut
  {NoStop}%
\bibitem [{Note2()}]{Note2}%
  \BibitemOpen
  \bibinfo {note} {Our wording here is a nod to the enjoyable writing of
  Marshall Stoneham: A. Stoneham, Rep. Prog. Phys. \protect \textbf {44}, 1251
  (1981).}\BibitemShut {Stop}%
\bibitem [{\citenamefont {Shang}\ and\ \citenamefont
  {Yang}(2020)}]{shang2020moving}%
  \BibitemOpen
  \bibfield  {author} {\bibinfo {author} {\bibfnamefont {H.}~\bibnamefont
  {Shang}}\ and\ \bibinfo {author} {\bibfnamefont {J.}~\bibnamefont {Yang}},\
  }\href {https://doi.org/10.1021/acs.jpca.0c01453} {\bibfield  {journal}
  {\bibinfo  {journal} {J. Phys. Chem. A}\ }\textbf {\bibinfo {volume} {124}},\
  \bibinfo {pages} {2897} (\bibinfo {year} {2020})}\BibitemShut {NoStop}%
\bibitem [{\citenamefont {da~Silva}\ \emph {et~al.}(2015)\citenamefont
  {da~Silva}, \citenamefont {Skelton}, \citenamefont {Parker},\ and\
  \citenamefont {Walsh}}]{da2015phase}%
  \BibitemOpen
  \bibfield  {author} {\bibinfo {author} {\bibfnamefont {E.~L.}\ \bibnamefont
  {da~Silva}}, \bibinfo {author} {\bibfnamefont {J.~M.}\ \bibnamefont
  {Skelton}}, \bibinfo {author} {\bibfnamefont {S.~C.}\ \bibnamefont {Parker}},
  \ and\ \bibinfo {author} {\bibfnamefont {A.}~\bibnamefont {Walsh}},\ }\href
  {https://doi.org/10.1103/PhysRevB.91.144107} {\bibfield  {journal} {\bibinfo
  {journal} {Phys. Rev. B}\ }\textbf {\bibinfo {volume} {91}},\ \bibinfo
  {pages} {144107} (\bibinfo {year} {2015})}\BibitemShut {NoStop}%
\bibitem [{\citenamefont {Zeraati}\ \emph {et~al.}(2016)\citenamefont
  {Zeraati}, \citenamefont {Allaei}, \citenamefont {Sarsari}, \citenamefont
  {Pourfath},\ and\ \citenamefont {Donadio}}]{zeraati2016highly}%
  \BibitemOpen
  \bibfield  {author} {\bibinfo {author} {\bibfnamefont {M.}~\bibnamefont
  {Zeraati}}, \bibinfo {author} {\bibfnamefont {S.~M.~V.}\ \bibnamefont
  {Allaei}}, \bibinfo {author} {\bibfnamefont {I.~A.}\ \bibnamefont {Sarsari}},
  \bibinfo {author} {\bibfnamefont {M.}~\bibnamefont {Pourfath}}, \ and\
  \bibinfo {author} {\bibfnamefont {D.}~\bibnamefont {Donadio}},\ }\href
  {https://doi.org/10.1103/PhysRevB.93.085424} {\bibfield  {journal} {\bibinfo
  {journal} {Phys. Rev. B}\ }\textbf {\bibinfo {volume} {93}},\ \bibinfo
  {pages} {085424} (\bibinfo {year} {2016})}\BibitemShut {NoStop}%
\bibitem [{\citenamefont {Zhang}\ \emph {et~al.}(2019)\citenamefont {Zhang},
  \citenamefont {Cheng}, \citenamefont {Lu}, \citenamefont {Briggs},
  \citenamefont {Ramirez-Cuesta},\ and\ \citenamefont
  {Bernholc}}]{zhang2019large}%
  \BibitemOpen
  \bibfield  {author} {\bibinfo {author} {\bibfnamefont {J.}~\bibnamefont
  {Zhang}}, \bibinfo {author} {\bibfnamefont {Y.}~\bibnamefont {Cheng}},
  \bibinfo {author} {\bibfnamefont {W.}~\bibnamefont {Lu}}, \bibinfo {author}
  {\bibfnamefont {E.}~\bibnamefont {Briggs}}, \bibinfo {author} {\bibfnamefont
  {A.~J.}\ \bibnamefont {Ramirez-Cuesta}}, \ and\ \bibinfo {author}
  {\bibfnamefont {J.}~\bibnamefont {Bernholc}},\ }\href
  {https://doi.org/10.1021/acs.jctc.9b00802} {\bibfield  {journal} {\bibinfo
  {journal} {J. Chem. Theory Comput.}\ }\textbf {\bibinfo {volume} {15}},\
  \bibinfo {pages} {6859} (\bibinfo {year} {2019})}\BibitemShut {NoStop}%
\bibitem [{\citenamefont {Ackland}, \citenamefont {Warren},\ and\ \citenamefont
  {Clark}(1997)}]{ackland1997practical}%
  \BibitemOpen
  \bibfield  {author} {\bibinfo {author} {\bibfnamefont {G.~J.}\ \bibnamefont
  {Ackland}}, \bibinfo {author} {\bibfnamefont {M.~C.}\ \bibnamefont {Warren}},
  \ and\ \bibinfo {author} {\bibfnamefont {S.~J.}\ \bibnamefont {Clark}},\
  }\href {\doibase 10.1088/0953-8984/9/37/017} {\bibfield  {journal} {\bibinfo
  {journal} {J. Condens. Matter Phys.}\ }\textbf {\bibinfo {volume} {9}},\
  \bibinfo {pages} {7861} (\bibinfo {year} {1997})}\BibitemShut {NoStop}%
\bibitem [{\citenamefont {Royo}, \citenamefont {Hahn},\ and\ \citenamefont
  {Stengel}(2020)}]{royo2020using}%
  \BibitemOpen
  \bibfield  {author} {\bibinfo {author} {\bibfnamefont {M.}~\bibnamefont
  {Royo}}, \bibinfo {author} {\bibfnamefont {K.~R.}\ \bibnamefont {Hahn}}, \
  and\ \bibinfo {author} {\bibfnamefont {M.}~\bibnamefont {Stengel}},\ }\href
  {https://doi.org/10.1103/PhysRevLett.125.217602} {\bibfield  {journal}
  {\bibinfo  {journal} {Phys. Rev. Lett.}\ }\textbf {\bibinfo {volume} {125}},\
  \bibinfo {pages} {217602} (\bibinfo {year} {2020})}\BibitemShut {NoStop}%
\bibitem [{\citenamefont {Skelton}(2020)}]{skelton2020lattice}%
  \BibitemOpen
  \bibfield  {author} {\bibinfo {author} {\bibfnamefont {J.~M.}\ \bibnamefont
  {Skelton}},\ }\href {https://doi.org/10.1088/2515-7655/ab7839} {\bibfield
  {journal} {\bibinfo  {journal} {J. Phys. Energy}\ }\textbf {\bibinfo {volume}
  {2}},\ \bibinfo {pages} {025006} (\bibinfo {year} {2020})}\BibitemShut
  {NoStop}%
\bibitem [{\citenamefont {Lloyd-Williams}\ and\ \citenamefont
  {Monserrat}(2015)}]{williams2015lattice}%
  \BibitemOpen
  \bibfield  {author} {\bibinfo {author} {\bibfnamefont {J.~H.}\ \bibnamefont
  {Lloyd-Williams}}\ and\ \bibinfo {author} {\bibfnamefont {B.}~\bibnamefont
  {Monserrat}},\ }\href {\doibase 10.1103/PhysRevB.92.184301} {\bibfield
  {journal} {\bibinfo  {journal} {Phys. Rev. B}\ }\textbf {\bibinfo {volume}
  {92}},\ \bibinfo {pages} {184301} (\bibinfo {year} {2015})}\BibitemShut
  {NoStop}%
\bibitem [{\citenamefont {Ko\ifmmode~\mbox{\c{c}}\else \c{c}\fi{}er}\ \emph
  {et~al.}(2020)\citenamefont {Ko\ifmmode~\mbox{\c{c}}\else \c{c}\fi{}er},
  \citenamefont {Haule}, \citenamefont {Pascut},\ and\ \citenamefont
  {Monserrat}}]{Koccer2020efficient}%
  \BibitemOpen
  \bibfield  {author} {\bibinfo {author} {\bibfnamefont {C.~P.}\ \bibnamefont
  {Ko\ifmmode~\mbox{\c{c}}\else \c{c}\fi{}er}}, \bibinfo {author}
  {\bibfnamefont {K.}~\bibnamefont {Haule}}, \bibinfo {author} {\bibfnamefont
  {G.~L.}\ \bibnamefont {Pascut}}, \ and\ \bibinfo {author} {\bibfnamefont
  {B.}~\bibnamefont {Monserrat}},\ }\href {\doibase
  10.1103/PhysRevB.102.245104} {\bibfield  {journal} {\bibinfo  {journal}
  {Phys. Rev. B}\ }\textbf {\bibinfo {volume} {102}},\ \bibinfo {pages}
  {245104} (\bibinfo {year} {2020})}\BibitemShut {NoStop}%
\bibitem [{\citenamefont {Evarestov}\ \emph {et~al.}(2011)\citenamefont
  {Evarestov}, \citenamefont {Blokhin}, \citenamefont {Gryaznov}, \citenamefont
  {Kotomin},\ and\ \citenamefont {Maier}}]{Evarestov2011phonon}%
  \BibitemOpen
  \bibfield  {author} {\bibinfo {author} {\bibfnamefont {R.~A.}\ \bibnamefont
  {Evarestov}}, \bibinfo {author} {\bibfnamefont {E.}~\bibnamefont {Blokhin}},
  \bibinfo {author} {\bibfnamefont {D.}~\bibnamefont {Gryaznov}}, \bibinfo
  {author} {\bibfnamefont {E.~A.}\ \bibnamefont {Kotomin}}, \ and\ \bibinfo
  {author} {\bibfnamefont {J.}~\bibnamefont {Maier}},\ }\href {\doibase
  10.1103/PhysRevB.83.134108} {\bibfield  {journal} {\bibinfo  {journal} {Phys.
  Rev. B}\ }\textbf {\bibinfo {volume} {83}},\ \bibinfo {pages} {134108}
  (\bibinfo {year} {2011})}\BibitemShut {NoStop}%
\bibitem [{\citenamefont {Kim}\ \emph {et~al.}(2020)\citenamefont {Kim},
  \citenamefont {Kim}, \citenamefont {Jeong}, \citenamefont {Park},
  \citenamefont {Park}, \citenamefont {Lee}, \citenamefont {Leiner},
  \citenamefont {Ishikawa}, \citenamefont {Baron}, \citenamefont {Hiroi},\ and\
  \citenamefont {Park}}]{kim2020spin}%
  \BibitemOpen
  \bibfield  {author} {\bibinfo {author} {\bibfnamefont {T.}~\bibnamefont
  {Kim}}, \bibinfo {author} {\bibfnamefont {C.~H.}\ \bibnamefont {Kim}},
  \bibinfo {author} {\bibfnamefont {J.}~\bibnamefont {Jeong}}, \bibinfo
  {author} {\bibfnamefont {P.}~\bibnamefont {Park}}, \bibinfo {author}
  {\bibfnamefont {K.}~\bibnamefont {Park}}, \bibinfo {author} {\bibfnamefont
  {K.~H.}\ \bibnamefont {Lee}}, \bibinfo {author} {\bibfnamefont {J.~C.}\
  \bibnamefont {Leiner}}, \bibinfo {author} {\bibfnamefont {D.}~\bibnamefont
  {Ishikawa}}, \bibinfo {author} {\bibfnamefont {A.~Q.~R.}\ \bibnamefont
  {Baron}}, \bibinfo {author} {\bibfnamefont {Z.}~\bibnamefont {Hiroi}}, \ and\
  \bibinfo {author} {\bibfnamefont {J.-G.}\ \bibnamefont {Park}},\ }\href
  {\doibase 10.1103/PhysRevB.102.201101} {\bibfield  {journal} {\bibinfo
  {journal} {Phys. Rev. B}\ }\textbf {\bibinfo {volume} {102}},\ \bibinfo
  {pages} {201101} (\bibinfo {year} {2020})}\BibitemShut {NoStop}%
\bibitem [{\citenamefont {Akamatsu}\ \emph {et~al.}(2013)\citenamefont
  {Akamatsu}, \citenamefont {Kumagai}, \citenamefont {Oba}, \citenamefont
  {Fujita}, \citenamefont {Tanaka},\ and\ \citenamefont
  {Tanaka}}]{akamatsu2013strong}%
  \BibitemOpen
  \bibfield  {author} {\bibinfo {author} {\bibfnamefont {H.}~\bibnamefont
  {Akamatsu}}, \bibinfo {author} {\bibfnamefont {Y.}~\bibnamefont {Kumagai}},
  \bibinfo {author} {\bibfnamefont {F.}~\bibnamefont {Oba}}, \bibinfo {author}
  {\bibfnamefont {K.}~\bibnamefont {Fujita}}, \bibinfo {author} {\bibfnamefont
  {K.}~\bibnamefont {Tanaka}}, \ and\ \bibinfo {author} {\bibfnamefont
  {I.}~\bibnamefont {Tanaka}},\ }\href {https://doi.org/10.1002/adfm.201202477}
  {\bibfield  {journal} {\bibinfo  {journal} {Adv. Funct. Mater.}\ }\textbf
  {\bibinfo {volume} {23}},\ \bibinfo {pages} {1864} (\bibinfo {year}
  {2013})}\BibitemShut {NoStop}%
\bibitem [{\citenamefont {Savin}\ and\ \citenamefont
  {Kosevich}(2014)}]{savin2014thermal}%
  \BibitemOpen
  \bibfield  {author} {\bibinfo {author} {\bibfnamefont {A.~V.}\ \bibnamefont
  {Savin}}\ and\ \bibinfo {author} {\bibfnamefont {Y.~A.}\ \bibnamefont
  {Kosevich}},\ }\href {https://doi.org/10.1103/PhysRevE.89.032102} {\bibfield
  {journal} {\bibinfo  {journal} {Phys. Rev. E}\ }\textbf {\bibinfo {volume}
  {89}},\ \bibinfo {pages} {032102} (\bibinfo {year} {2014})}\BibitemShut
  {NoStop}%
\bibitem [{\citenamefont {Hooton}(1955)}]{hooton1955li}%
  \BibitemOpen
  \bibfield  {author} {\bibinfo {author} {\bibfnamefont {D.}~\bibnamefont
  {Hooton}},\ }\href {https://doi.org/10.1080/14786440408520575} {\bibfield
  {journal} {\bibinfo  {journal} {Lond. Edinb. Dublin philos. mag. j. sci.}\
  }\textbf {\bibinfo {volume} {46}},\ \bibinfo {pages} {422} (\bibinfo {year}
  {1955})}\BibitemShut {NoStop}%
\bibitem [{\citenamefont {Whalley}\ \emph {et~al.}(2016)\citenamefont
  {Whalley}, \citenamefont {Skelton}, \citenamefont {Frost},\ and\
  \citenamefont {Walsh}}]{Whalley2016phonon}%
  \BibitemOpen
  \bibfield  {author} {\bibinfo {author} {\bibfnamefont {L.~D.}\ \bibnamefont
  {Whalley}}, \bibinfo {author} {\bibfnamefont {J.~M.}\ \bibnamefont
  {Skelton}}, \bibinfo {author} {\bibfnamefont {J.~M.}\ \bibnamefont {Frost}},
  \ and\ \bibinfo {author} {\bibfnamefont {A.}~\bibnamefont {Walsh}},\ }\href
  {\doibase 10.1103/PhysRevB.94.220301} {\bibfield  {journal} {\bibinfo
  {journal} {Phys. Rev. B}\ }\textbf {\bibinfo {volume} {94}},\ \bibinfo
  {pages} {220301} (\bibinfo {year} {2016})}\BibitemShut {NoStop}%
\bibitem [{\citenamefont {Zhang}, \citenamefont {Sun},\ and\ \citenamefont
  {Wentzcovitch}(2014)}]{PhysRevLett.112.058501}%
  \BibitemOpen
  \bibfield  {author} {\bibinfo {author} {\bibfnamefont {D.-B.}\ \bibnamefont
  {Zhang}}, \bibinfo {author} {\bibfnamefont {T.}~\bibnamefont {Sun}}, \ and\
  \bibinfo {author} {\bibfnamefont {R.~M.}\ \bibnamefont {Wentzcovitch}},\
  }\href {\doibase 10.1103/PhysRevLett.112.058501} {\bibfield  {journal}
  {\bibinfo  {journal} {Phys. Rev. Lett.}\ }\textbf {\bibinfo {volume} {112}},\
  \bibinfo {pages} {058501} (\bibinfo {year} {2014})}\BibitemShut {NoStop}%
\bibitem [{\citenamefont {Korotaev}, \citenamefont {Belov},\ and\ \citenamefont
  {Yanilkin}(2018)}]{Korotaev2018}%
  \BibitemOpen
  \bibfield  {author} {\bibinfo {author} {\bibfnamefont {P.}~\bibnamefont
  {Korotaev}}, \bibinfo {author} {\bibfnamefont {M.}~\bibnamefont {Belov}}, \
  and\ \bibinfo {author} {\bibfnamefont {A.}~\bibnamefont {Yanilkin}},\ }\href
  {\doibase 10.1016/j.commatsci.2018.03.057} {\bibfield  {journal} {\bibinfo
  {journal} {Comput. Mater. Sci.}\ }\textbf {\bibinfo {volume} {150}},\
  \bibinfo {pages} {47} (\bibinfo {year} {2018})}\BibitemShut {NoStop}%
\bibitem [{\citenamefont {Romero}\ \emph {et~al.}(2015)\citenamefont {Romero},
  \citenamefont {Gross}, \citenamefont {Verstraete},\ and\ \citenamefont
  {Hellman}}]{PhysRevB.91.214310}%
  \BibitemOpen
  \bibfield  {author} {\bibinfo {author} {\bibfnamefont {A.~H.}\ \bibnamefont
  {Romero}}, \bibinfo {author} {\bibfnamefont {E.~K.~U.}\ \bibnamefont
  {Gross}}, \bibinfo {author} {\bibfnamefont {M.~J.}\ \bibnamefont
  {Verstraete}}, \ and\ \bibinfo {author} {\bibfnamefont {O.}~\bibnamefont
  {Hellman}},\ }\href {\doibase 10.1103/PhysRevB.91.214310} {\bibfield
  {journal} {\bibinfo  {journal} {Phys. Rev. B}\ }\textbf {\bibinfo {volume}
  {91}},\ \bibinfo {pages} {214310} (\bibinfo {year} {2015})}\BibitemShut
  {NoStop}%
\bibitem [{\citenamefont {Bouchet}\ \emph {et~al.}(2019)\citenamefont
  {Bouchet}, \citenamefont {Bottin}, \citenamefont {Recoules}, \citenamefont
  {Remus}, \citenamefont {Morard}, \citenamefont {Bolis},\ and\ \citenamefont
  {Benuzzi-Mounaix}}]{PhysRevB.99.094113}%
  \BibitemOpen
  \bibfield  {author} {\bibinfo {author} {\bibfnamefont {J.}~\bibnamefont
  {Bouchet}}, \bibinfo {author} {\bibfnamefont {F.}~\bibnamefont {Bottin}},
  \bibinfo {author} {\bibfnamefont {V.}~\bibnamefont {Recoules}}, \bibinfo
  {author} {\bibfnamefont {F.}~\bibnamefont {Remus}}, \bibinfo {author}
  {\bibfnamefont {G.}~\bibnamefont {Morard}}, \bibinfo {author} {\bibfnamefont
  {R.~M.}\ \bibnamefont {Bolis}}, \ and\ \bibinfo {author} {\bibfnamefont
  {A.}~\bibnamefont {Benuzzi-Mounaix}},\ }\href {\doibase
  10.1103/PhysRevB.99.094113} {\bibfield  {journal} {\bibinfo  {journal} {Phys.
  Rev. B}\ }\textbf {\bibinfo {volume} {99}},\ \bibinfo {pages} {094113}
  (\bibinfo {year} {2019})}\BibitemShut {NoStop}%
\bibitem [{\citenamefont {Dewandre}\ \emph {et~al.}(2016)\citenamefont
  {Dewandre}, \citenamefont {Hellman}, \citenamefont {Bhattacharya},
  \citenamefont {Romero}, \citenamefont {Madsen},\ and\ \citenamefont
  {Verstraete}}]{PhysRevLett.117.276601}%
  \BibitemOpen
  \bibfield  {author} {\bibinfo {author} {\bibfnamefont {A.}~\bibnamefont
  {Dewandre}}, \bibinfo {author} {\bibfnamefont {O.}~\bibnamefont {Hellman}},
  \bibinfo {author} {\bibfnamefont {S.}~\bibnamefont {Bhattacharya}}, \bibinfo
  {author} {\bibfnamefont {A.~H.}\ \bibnamefont {Romero}}, \bibinfo {author}
  {\bibfnamefont {G.~K.~H.}\ \bibnamefont {Madsen}}, \ and\ \bibinfo {author}
  {\bibfnamefont {M.~J.}\ \bibnamefont {Verstraete}},\ }\href {\doibase
  10.1103/PhysRevLett.117.276601} {\bibfield  {journal} {\bibinfo  {journal}
  {Phys. Rev. Lett.}\ }\textbf {\bibinfo {volume} {117}},\ \bibinfo {pages}
  {276601} (\bibinfo {year} {2016})}\BibitemShut {NoStop}%
\bibitem [{\citenamefont {Bianco}\ \emph {et~al.}(2017)\citenamefont {Bianco},
  \citenamefont {Errea}, \citenamefont {Paulatto}, \citenamefont {Calandra},\
  and\ \citenamefont {Mauri}}]{bianco2017second}%
  \BibitemOpen
  \bibfield  {author} {\bibinfo {author} {\bibfnamefont {R.}~\bibnamefont
  {Bianco}}, \bibinfo {author} {\bibfnamefont {I.}~\bibnamefont {Errea}},
  \bibinfo {author} {\bibfnamefont {L.}~\bibnamefont {Paulatto}}, \bibinfo
  {author} {\bibfnamefont {M.}~\bibnamefont {Calandra}}, \ and\ \bibinfo
  {author} {\bibfnamefont {F.}~\bibnamefont {Mauri}},\ }\href
  {https://doi.org/10.1103/PhysRevB.96.014111} {\bibfield  {journal} {\bibinfo
  {journal} {Phys. Rev. B}\ }\textbf {\bibinfo {volume} {96}},\ \bibinfo
  {pages} {014111} (\bibinfo {year} {2017})}\BibitemShut {NoStop}%
\bibitem [{\citenamefont {Errea}\ \emph {et~al.}(2020)\citenamefont {Errea},
  \citenamefont {Belli}, \citenamefont {Monacelli}, \citenamefont {Sanna},
  \citenamefont {Koretsune}, \citenamefont {Tadano}, \citenamefont {Bianco},
  \citenamefont {Calandra}, \citenamefont {Arita}, \citenamefont {Mauri} \emph
  {et~al.}}]{errea2020quantum}%
  \BibitemOpen
  \bibfield  {author} {\bibinfo {author} {\bibfnamefont {I.}~\bibnamefont
  {Errea}}, \bibinfo {author} {\bibfnamefont {F.}~\bibnamefont {Belli}},
  \bibinfo {author} {\bibfnamefont {L.}~\bibnamefont {Monacelli}}, \bibinfo
  {author} {\bibfnamefont {A.}~\bibnamefont {Sanna}}, \bibinfo {author}
  {\bibfnamefont {T.}~\bibnamefont {Koretsune}}, \bibinfo {author}
  {\bibfnamefont {T.}~\bibnamefont {Tadano}}, \bibinfo {author} {\bibfnamefont
  {R.}~\bibnamefont {Bianco}}, \bibinfo {author} {\bibfnamefont
  {M.}~\bibnamefont {Calandra}}, \bibinfo {author} {\bibfnamefont
  {R.}~\bibnamefont {Arita}}, \bibinfo {author} {\bibfnamefont
  {F.}~\bibnamefont {Mauri}},  \emph {et~al.},\ }\href
  {https://doi.org/10.1038/s41586-020-1955-z} {\bibfield  {journal} {\bibinfo
  {journal} {Nature}\ }\textbf {\bibinfo {volume} {578}},\ \bibinfo {pages}
  {66} (\bibinfo {year} {2020})}\BibitemShut {NoStop}%
\bibitem [{\citenamefont {Hou}\ \emph {et~al.}(2021)\citenamefont {Hou},
  \citenamefont {Belli}, \citenamefont {Bianco},\ and\ \citenamefont
  {Errea}}]{hou2021quantum}%
  \BibitemOpen
  \bibfield  {author} {\bibinfo {author} {\bibfnamefont {P.}~\bibnamefont
  {Hou}}, \bibinfo {author} {\bibfnamefont {F.}~\bibnamefont {Belli}}, \bibinfo
  {author} {\bibfnamefont {R.}~\bibnamefont {Bianco}}, \ and\ \bibinfo {author}
  {\bibfnamefont {I.}~\bibnamefont {Errea}},\ }\href
  {https://doi.org/10.1063/5.0063968} {\bibfield  {journal} {\bibinfo
  {journal} {J. Appl. Phys.}\ }\textbf {\bibinfo {volume} {130}},\ \bibinfo
  {pages} {175902} (\bibinfo {year} {2021})}\BibitemShut {NoStop}%
\bibitem [{\citenamefont {Yang}\ \emph {et~al.}(2020)\citenamefont {Yang},
  \citenamefont {Skelton}, \citenamefont {Da~Silva}, \citenamefont {Frost},\
  and\ \citenamefont {Walsh}}]{yang2020assessment}%
  \BibitemOpen
  \bibfield  {author} {\bibinfo {author} {\bibfnamefont {R.~X.}\ \bibnamefont
  {Yang}}, \bibinfo {author} {\bibfnamefont {J.~M.}\ \bibnamefont {Skelton}},
  \bibinfo {author} {\bibfnamefont {E.~L.}\ \bibnamefont {Da~Silva}}, \bibinfo
  {author} {\bibfnamefont {J.~M.}\ \bibnamefont {Frost}}, \ and\ \bibinfo
  {author} {\bibfnamefont {A.}~\bibnamefont {Walsh}},\ }\href
  {https://doi.org/10.1063/1.5131575} {\bibfield  {journal} {\bibinfo
  {journal} {J. Chem. Phys.}\ }\textbf {\bibinfo {volume} {152}},\ \bibinfo
  {pages} {024703} (\bibinfo {year} {2020})}\BibitemShut {NoStop}%
\bibitem [{\citenamefont {Trots}\ and\ \citenamefont
  {Myagkota}(2008)}]{trots2008high}%
  \BibitemOpen
  \bibfield  {author} {\bibinfo {author} {\bibfnamefont {D.}~\bibnamefont
  {Trots}}\ and\ \bibinfo {author} {\bibfnamefont {S.}~\bibnamefont
  {Myagkota}},\ }\href {https://doi.org/10.1016/j.jpcs.2008.05.007} {\bibfield
  {journal} {\bibinfo  {journal} {J. Phys. Chem. Solids}\ }\textbf {\bibinfo
  {volume} {69}},\ \bibinfo {pages} {2520} (\bibinfo {year}
  {2008})}\BibitemShut {NoStop}%
\bibitem [{\citenamefont {Liu}\ \emph {et~al.}(2019)\citenamefont {Liu},
  \citenamefont {Phillips}, \citenamefont {Keen},\ and\ \citenamefont
  {Dove}}]{liu2019thermal}%
  \BibitemOpen
  \bibfield  {author} {\bibinfo {author} {\bibfnamefont {J.}~\bibnamefont
  {Liu}}, \bibinfo {author} {\bibfnamefont {A.~E.}\ \bibnamefont {Phillips}},
  \bibinfo {author} {\bibfnamefont {D.~A.}\ \bibnamefont {Keen}}, \ and\
  \bibinfo {author} {\bibfnamefont {M.~T.}\ \bibnamefont {Dove}},\ }\href
  {https://doi.org/10.1021/acs.jpcc.9b02936} {\bibfield  {journal} {\bibinfo
  {journal} {J. Phys. Chem. C}\ }\textbf {\bibinfo {volume} {123}},\ \bibinfo
  {pages} {14934} (\bibinfo {year} {2019})}\BibitemShut {NoStop}%
\bibitem [{\citenamefont {Bertolotti}\ \emph {et~al.}(2017)\citenamefont
  {Bertolotti}, \citenamefont {Protesescu}, \citenamefont {Kovalenko},
  \citenamefont {Yakunin}, \citenamefont {Cervellino}, \citenamefont
  {Billinge}, \citenamefont {Terban}, \citenamefont {Pedersen}, \citenamefont
  {Masciocchi},\ and\ \citenamefont {Guagliardi}}]{Bertolotti2017coherent}%
  \BibitemOpen
  \bibfield  {author} {\bibinfo {author} {\bibfnamefont {F.}~\bibnamefont
  {Bertolotti}}, \bibinfo {author} {\bibfnamefont {L.}~\bibnamefont
  {Protesescu}}, \bibinfo {author} {\bibfnamefont {M.~V.}\ \bibnamefont
  {Kovalenko}}, \bibinfo {author} {\bibfnamefont {S.}~\bibnamefont {Yakunin}},
  \bibinfo {author} {\bibfnamefont {A.}~\bibnamefont {Cervellino}}, \bibinfo
  {author} {\bibfnamefont {S.~J.~L.}\ \bibnamefont {Billinge}}, \bibinfo
  {author} {\bibfnamefont {M.~W.}\ \bibnamefont {Terban}}, \bibinfo {author}
  {\bibfnamefont {J.~S.}\ \bibnamefont {Pedersen}}, \bibinfo {author}
  {\bibfnamefont {N.}~\bibnamefont {Masciocchi}}, \ and\ \bibinfo {author}
  {\bibfnamefont {A.}~\bibnamefont {Guagliardi}},\ }\href
  {https://doi.org/10.1021/acsnano.7b00017} {\bibfield  {journal} {\bibinfo
  {journal} {ACS Nano.}\ }\textbf {\bibinfo {volume} {11}},\ \bibinfo {pages}
  {3819} (\bibinfo {year} {2017})}\BibitemShut {NoStop}%
\bibitem [{\citenamefont {Niesner}\ \emph {et~al.}(2018)\citenamefont
  {Niesner}, \citenamefont {Hauck}, \citenamefont {Shrestha}, \citenamefont
  {Levchuk}, \citenamefont {Matt}, \citenamefont {Osvet}, \citenamefont
  {Batentschuk}, \citenamefont {Brabec}, \citenamefont {Weber},\ and\
  \citenamefont {Fauster}}]{niesner2018structural}%
  \BibitemOpen
  \bibfield  {author} {\bibinfo {author} {\bibfnamefont {D.}~\bibnamefont
  {Niesner}}, \bibinfo {author} {\bibfnamefont {M.}~\bibnamefont {Hauck}},
  \bibinfo {author} {\bibfnamefont {S.}~\bibnamefont {Shrestha}}, \bibinfo
  {author} {\bibfnamefont {I.}~\bibnamefont {Levchuk}}, \bibinfo {author}
  {\bibfnamefont {G.~J.}\ \bibnamefont {Matt}}, \bibinfo {author}
  {\bibfnamefont {A.}~\bibnamefont {Osvet}}, \bibinfo {author} {\bibfnamefont
  {M.}~\bibnamefont {Batentschuk}}, \bibinfo {author} {\bibfnamefont
  {C.}~\bibnamefont {Brabec}}, \bibinfo {author} {\bibfnamefont {H.~B.}\
  \bibnamefont {Weber}}, \ and\ \bibinfo {author} {\bibfnamefont
  {T.}~\bibnamefont {Fauster}},\ }\href
  {https://doi.org/10.1073/pnas.1805422115} {\bibfield  {journal} {\bibinfo
  {journal} {Proc. Natl. Acad. Sci. U.S.A.}\ }\textbf {\bibinfo {volume}
  {115}},\ \bibinfo {pages} {9509} (\bibinfo {year} {2018})}\BibitemShut
  {NoStop}%
\bibitem [{\citenamefont {Munson}\ \emph {et~al.}(2018)\citenamefont {Munson},
  \citenamefont {Kennehan}, \citenamefont {Doucette},\ and\ \citenamefont
  {Asbury}}]{munson2018dynamic}%
  \BibitemOpen
  \bibfield  {author} {\bibinfo {author} {\bibfnamefont {K.~T.}\ \bibnamefont
  {Munson}}, \bibinfo {author} {\bibfnamefont {E.~R.}\ \bibnamefont
  {Kennehan}}, \bibinfo {author} {\bibfnamefont {G.~S.}\ \bibnamefont
  {Doucette}}, \ and\ \bibinfo {author} {\bibfnamefont {J.~B.}\ \bibnamefont
  {Asbury}},\ }\href {https://doi.org/10.1016/j.chempr.2018.09.001} {\bibfield
  {journal} {\bibinfo  {journal} {Chem}\ }\textbf {\bibinfo {volume} {4}},\
  \bibinfo {pages} {2826} (\bibinfo {year} {2018})}\BibitemShut {NoStop}%
\bibitem [{\citenamefont {Wright}\ \emph {et~al.}(2016)\citenamefont {Wright},
  \citenamefont {Verdi}, \citenamefont {Milot}, \citenamefont {Eperon},
  \citenamefont {P{\'e}rez-Osorio}, \citenamefont {Snaith}, \citenamefont
  {Giustino}, \citenamefont {Johnston},\ and\ \citenamefont
  {Herz}}]{wright2016electron}%
  \BibitemOpen
  \bibfield  {author} {\bibinfo {author} {\bibfnamefont {A.~D.}\ \bibnamefont
  {Wright}}, \bibinfo {author} {\bibfnamefont {C.}~\bibnamefont {Verdi}},
  \bibinfo {author} {\bibfnamefont {R.~L.}\ \bibnamefont {Milot}}, \bibinfo
  {author} {\bibfnamefont {G.~E.}\ \bibnamefont {Eperon}}, \bibinfo {author}
  {\bibfnamefont {M.~A.}\ \bibnamefont {P{\'e}rez-Osorio}}, \bibinfo {author}
  {\bibfnamefont {H.~J.}\ \bibnamefont {Snaith}}, \bibinfo {author}
  {\bibfnamefont {F.}~\bibnamefont {Giustino}}, \bibinfo {author}
  {\bibfnamefont {M.~B.}\ \bibnamefont {Johnston}}, \ and\ \bibinfo {author}
  {\bibfnamefont {L.~M.}\ \bibnamefont {Herz}},\ }\href
  {https://doi.org/10.1038/ncomms11755} {\bibfield  {journal} {\bibinfo
  {journal} {Nat. Commun.}\ }\textbf {\bibinfo {volume} {7}},\ \bibinfo {pages}
  {1} (\bibinfo {year} {2016})}\BibitemShut {NoStop}%
\bibitem [{\citenamefont {Bohn}\ \emph {et~al.}(2018)\citenamefont {Bohn},
  \citenamefont {Simon}, \citenamefont {Gramlich}, \citenamefont {Richter},
  \citenamefont {Polavarapu}, \citenamefont {Urban},\ and\ \citenamefont
  {Feldmann}}]{bohn2018dephasing}%
  \BibitemOpen
  \bibfield  {author} {\bibinfo {author} {\bibfnamefont {B.~J.}\ \bibnamefont
  {Bohn}}, \bibinfo {author} {\bibfnamefont {T.}~\bibnamefont {Simon}},
  \bibinfo {author} {\bibfnamefont {M.}~\bibnamefont {Gramlich}}, \bibinfo
  {author} {\bibfnamefont {A.~F.}\ \bibnamefont {Richter}}, \bibinfo {author}
  {\bibfnamefont {L.}~\bibnamefont {Polavarapu}}, \bibinfo {author}
  {\bibfnamefont {A.~S.}\ \bibnamefont {Urban}}, \ and\ \bibinfo {author}
  {\bibfnamefont {J.}~\bibnamefont {Feldmann}},\ }\href
  {https://doi.org/10.1021/acsphotonics.7b01292} {\bibfield  {journal}
  {\bibinfo  {journal} {ACS Photonics}\ }\textbf {\bibinfo {volume} {5}},\
  \bibinfo {pages} {648} (\bibinfo {year} {2018})}\BibitemShut {NoStop}%
\bibitem [{\citenamefont {Fassl}\ \emph {et~al.}(2021)\citenamefont {Fassl},
  \citenamefont {Lami}, \citenamefont {Berger}, \citenamefont {Falk},
  \citenamefont {Zaumseil}, \citenamefont {Richards}, \citenamefont {Howard},
  \citenamefont {Vaynzof},\ and\ \citenamefont
  {Paetzold}}]{fassl2021revealing}%
  \BibitemOpen
  \bibfield  {author} {\bibinfo {author} {\bibfnamefont {P.}~\bibnamefont
  {Fassl}}, \bibinfo {author} {\bibfnamefont {V.}~\bibnamefont {Lami}},
  \bibinfo {author} {\bibfnamefont {F.~J.}\ \bibnamefont {Berger}}, \bibinfo
  {author} {\bibfnamefont {L.~M.}\ \bibnamefont {Falk}}, \bibinfo {author}
  {\bibfnamefont {J.}~\bibnamefont {Zaumseil}}, \bibinfo {author}
  {\bibfnamefont {B.~S.}\ \bibnamefont {Richards}}, \bibinfo {author}
  {\bibfnamefont {I.~A.}\ \bibnamefont {Howard}}, \bibinfo {author}
  {\bibfnamefont {Y.}~\bibnamefont {Vaynzof}}, \ and\ \bibinfo {author}
  {\bibfnamefont {U.~W.}\ \bibnamefont {Paetzold}},\ }\href
  {https://doi.org/10.1016/j.matt.2021.01.019} {\bibfield  {journal} {\bibinfo
  {journal} {Matter}\ }\textbf {\bibinfo {volume} {4}},\ \bibinfo {pages}
  {1391} (\bibinfo {year} {2021})}\BibitemShut {NoStop}%
\bibitem [{\citenamefont {Sinsermsuksakul}\ \emph {et~al.}(2014)\citenamefont
  {Sinsermsuksakul}, \citenamefont {Sun}, \citenamefont {Lee}, \citenamefont
  {Park}, \citenamefont {Kim}, \citenamefont {Yang},\ and\ \citenamefont
  {Gordon}}]{Sinsermsuksakul-2014-SnSPhotovoltaics}%
  \BibitemOpen
  \bibfield  {author} {\bibinfo {author} {\bibfnamefont {P.}~\bibnamefont
  {Sinsermsuksakul}}, \bibinfo {author} {\bibfnamefont {L.}~\bibnamefont
  {Sun}}, \bibinfo {author} {\bibfnamefont {S.~W.}\ \bibnamefont {Lee}},
  \bibinfo {author} {\bibfnamefont {H.~H.}\ \bibnamefont {Park}}, \bibinfo
  {author} {\bibfnamefont {S.~B.}\ \bibnamefont {Kim}}, \bibinfo {author}
  {\bibfnamefont {C.}~\bibnamefont {Yang}}, \ and\ \bibinfo {author}
  {\bibfnamefont {R.~G.}\ \bibnamefont {Gordon}},\ }\href {\doibase
  10.1002/aenm.201400496} {\bibfield  {journal} {\bibinfo  {journal} {Adv.
  Energy Mater.}\ }\textbf {\bibinfo {volume} {4}},\ \bibinfo {pages} {1400496}
  (\bibinfo {year} {2014})}\BibitemShut {NoStop}%
\bibitem [{\citenamefont {Yun}\ \emph {et~al.}(2019)\citenamefont {Yun},
  \citenamefont {Park}, \citenamefont {Choi}, \citenamefont {Im}, \citenamefont
  {Shin},\ and\ \citenamefont {Seok}}]{Yun-2019-SnSPhotovoltaics}%
  \BibitemOpen
  \bibfield  {author} {\bibinfo {author} {\bibfnamefont {H.-S.}\ \bibnamefont
  {Yun}}, \bibinfo {author} {\bibfnamefont {B.-w.}\ \bibnamefont {Park}},
  \bibinfo {author} {\bibfnamefont {Y.~C.}\ \bibnamefont {Choi}}, \bibinfo
  {author} {\bibfnamefont {J.}~\bibnamefont {Im}}, \bibinfo {author}
  {\bibfnamefont {T.~J.}\ \bibnamefont {Shin}}, \ and\ \bibinfo {author}
  {\bibfnamefont {S.~I.}\ \bibnamefont {Seok}},\ }\href {\doibase
  10.1002/aenm.201901343} {\bibfield  {journal} {\bibinfo  {journal} {Advanced
  Energy Materials}\ }\textbf {\bibinfo {volume} {9}},\ \bibinfo {pages}
  {1901343} (\bibinfo {year} {2019})}\BibitemShut {NoStop}%
\bibitem [{\citenamefont {Cho}\ \emph {et~al.}(2020)\citenamefont {Cho},
  \citenamefont {Kim}, \citenamefont {Nandi}, \citenamefont {Jang},
  \citenamefont {Yun}, \citenamefont {Enkhbayar}, \citenamefont {Kim},
  \citenamefont {Lee}, \citenamefont {Chung}, \citenamefont {Kim},\ and\
  \citenamefont {Heo}}]{Cho-2020-SnSPhotovoltaics}%
  \BibitemOpen
  \bibfield  {author} {\bibinfo {author} {\bibfnamefont {J.~Y.}\ \bibnamefont
  {Cho}}, \bibinfo {author} {\bibfnamefont {S.}~\bibnamefont {Kim}}, \bibinfo
  {author} {\bibfnamefont {R.}~\bibnamefont {Nandi}}, \bibinfo {author}
  {\bibfnamefont {J.}~\bibnamefont {Jang}}, \bibinfo {author} {\bibfnamefont
  {H.-S.}\ \bibnamefont {Yun}}, \bibinfo {author} {\bibfnamefont
  {E.}~\bibnamefont {Enkhbayar}}, \bibinfo {author} {\bibfnamefont {J.~H.}\
  \bibnamefont {Kim}}, \bibinfo {author} {\bibfnamefont {D.-K.}\ \bibnamefont
  {Lee}}, \bibinfo {author} {\bibfnamefont {C.-H.}\ \bibnamefont {Chung}},
  \bibinfo {author} {\bibfnamefont {J.}~\bibnamefont {Kim}}, \ and\ \bibinfo
  {author} {\bibfnamefont {J.}~\bibnamefont {Heo}},\ }\href {\doibase
  10.1039/D0TA06937J} {\bibfield  {journal} {\bibinfo  {journal} {J. Mater.
  Chem. A}\ }\textbf {\bibinfo {volume} {8}},\ \bibinfo {pages} {20658}
  (\bibinfo {year} {2020})}\BibitemShut {NoStop}%
\bibitem [{\citenamefont {Zhao}\ \emph {et~al.}(2014)\citenamefont {Zhao},
  \citenamefont {Lo}, \citenamefont {Zhang}, \citenamefont {Sun}, \citenamefont
  {Tan}, \citenamefont {Uher}, \citenamefont {Wolverton}, \citenamefont
  {Dravid},\ and\ \citenamefont {Kanatzidis}}]{Zhao-2014-SnSeThermoelectrics}%
  \BibitemOpen
  \bibfield  {author} {\bibinfo {author} {\bibfnamefont {L.-D.}\ \bibnamefont
  {Zhao}}, \bibinfo {author} {\bibfnamefont {S.-H.}\ \bibnamefont {Lo}},
  \bibinfo {author} {\bibfnamefont {Y.}~\bibnamefont {Zhang}}, \bibinfo
  {author} {\bibfnamefont {H.}~\bibnamefont {Sun}}, \bibinfo {author}
  {\bibfnamefont {G.}~\bibnamefont {Tan}}, \bibinfo {author} {\bibfnamefont
  {C.}~\bibnamefont {Uher}}, \bibinfo {author} {\bibfnamefont {C.}~\bibnamefont
  {Wolverton}}, \bibinfo {author} {\bibfnamefont {V.~P.}\ \bibnamefont
  {Dravid}}, \ and\ \bibinfo {author} {\bibfnamefont {M.~G.}\ \bibnamefont
  {Kanatzidis}},\ }\href {\doibase 10.1038/nature13184} {\bibfield  {journal}
  {\bibinfo  {journal} {Nature}\ }\textbf {\bibinfo {volume} {508}},\ \bibinfo
  {pages} {373} (\bibinfo {year} {2014})}\BibitemShut {NoStop}%
\bibitem [{\citenamefont {Zhao}\ \emph {et~al.}(2016)\citenamefont {Zhao},
  \citenamefont {Chang}, \citenamefont {Tan},\ and\ \citenamefont
  {Kanatzidis}}]{Zhao-2016-SnSePerspective}%
  \BibitemOpen
  \bibfield  {author} {\bibinfo {author} {\bibfnamefont {L.-D.}\ \bibnamefont
  {Zhao}}, \bibinfo {author} {\bibfnamefont {C.}~\bibnamefont {Chang}},
  \bibinfo {author} {\bibfnamefont {G.}~\bibnamefont {Tan}}, \ and\ \bibinfo
  {author} {\bibfnamefont {M.~G.}\ \bibnamefont {Kanatzidis}},\ }\href
  {\doibase 10.1039/C6EE01755J} {\bibfield  {journal} {\bibinfo  {journal}
  {Energy Environ. Sci.}\ }\textbf {\bibinfo {volume} {9}},\ \bibinfo {pages}
  {3044} (\bibinfo {year} {2016})}\BibitemShut {NoStop}%
\bibitem [{\citenamefont {Zhou}\ \emph {et~al.}(2021)\citenamefont {Zhou},
  \citenamefont {Lee}, \citenamefont {Yu}, \citenamefont {Byun}, \citenamefont
  {Luo}, \citenamefont {Lee}, \citenamefont {Ge}, \citenamefont {Lee},
  \citenamefont {Chen}, \citenamefont {Lee}, \citenamefont
  {Cojocaru-Mir{\'e}din}, \citenamefont {Chang}, \citenamefont {Im},
  \citenamefont {Cho}, \citenamefont {Wuttig}, \citenamefont {Dravid},
  \citenamefont {Kanatzidis},\ and\ \citenamefont
  {Chung}}]{Zhou-2021-PolycrystallineSnSe}%
  \BibitemOpen
  \bibfield  {author} {\bibinfo {author} {\bibfnamefont {C.}~\bibnamefont
  {Zhou}}, \bibinfo {author} {\bibfnamefont {Y.~K.}\ \bibnamefont {Lee}},
  \bibinfo {author} {\bibfnamefont {Y.}~\bibnamefont {Yu}}, \bibinfo {author}
  {\bibfnamefont {S.}~\bibnamefont {Byun}}, \bibinfo {author} {\bibfnamefont
  {Z.-Z.}\ \bibnamefont {Luo}}, \bibinfo {author} {\bibfnamefont
  {H.}~\bibnamefont {Lee}}, \bibinfo {author} {\bibfnamefont {B.}~\bibnamefont
  {Ge}}, \bibinfo {author} {\bibfnamefont {Y.-L.}\ \bibnamefont {Lee}},
  \bibinfo {author} {\bibfnamefont {X.}~\bibnamefont {Chen}}, \bibinfo {author}
  {\bibfnamefont {J.~Y.}\ \bibnamefont {Lee}}, \bibinfo {author} {\bibfnamefont
  {O.}~\bibnamefont {Cojocaru-Mir{\'e}din}}, \bibinfo {author} {\bibfnamefont
  {H.}~\bibnamefont {Chang}}, \bibinfo {author} {\bibfnamefont
  {J.}~\bibnamefont {Im}}, \bibinfo {author} {\bibfnamefont {S.-P.}\
  \bibnamefont {Cho}}, \bibinfo {author} {\bibfnamefont {M.}~\bibnamefont
  {Wuttig}}, \bibinfo {author} {\bibfnamefont {V.~P.}\ \bibnamefont {Dravid}},
  \bibinfo {author} {\bibfnamefont {M.~G.}\ \bibnamefont {Kanatzidis}}, \ and\
  \bibinfo {author} {\bibfnamefont {I.}~\bibnamefont {Chung}},\ }\href
  {\doibase 10.1038/s41563-021-01064-6} {\bibfield  {journal} {\bibinfo
  {journal} {Nat. Mater.}\ }\textbf {\bibinfo {volume} {20}},\ \bibinfo {pages}
  {1378} (\bibinfo {year} {2021})}\BibitemShut {NoStop}%
\bibitem [{\citenamefont {{T. Chattopadhyay and J. Pannetier and H.G. {Von
  Schnering}}}(1986)}]{Chattopadhyay-1986-SnSSeNeutronScattering}%
  \BibitemOpen
  \bibfield  {author} {\bibinfo {author} {\bibnamefont {{T. Chattopadhyay and
  J. Pannetier and H.G. {Von Schnering}}}},\ }\href {\doibase
  10.1016/0022-3697(86)90059-4} {\bibfield  {journal} {\bibinfo  {journal} {J.
  Phys. Chem. Solids}\ }\textbf {\bibinfo {volume} {47}},\ \bibinfo {pages}
  {879} (\bibinfo {year} {1986})}\BibitemShut {NoStop}%
\bibitem [{\citenamefont {Mariano}\ and\ \citenamefont
  {Chopra}(1967)}]{Mariano-1967-RocksaltSnX}%
  \BibitemOpen
  \bibfield  {author} {\bibinfo {author} {\bibfnamefont {A.~N.}\ \bibnamefont
  {Mariano}}\ and\ \bibinfo {author} {\bibfnamefont {K.~L.}\ \bibnamefont
  {Chopra}},\ }\href {\doibase 10.1063/1.1754812} {\bibfield  {journal}
  {\bibinfo  {journal} {Appl. Phys. Lett.}\ }\textbf {\bibinfo {volume} {10}},\
  \bibinfo {pages} {282} (\bibinfo {year} {1967})}\BibitemShut {NoStop}%
\bibitem [{\citenamefont {Bilenkii}, \citenamefont {Mikolaichuk},\ and\
  \citenamefont {Freik}(1968)}]{Bilenkii-1968-RocksaltSnX}%
  \BibitemOpen
  \bibfield  {author} {\bibinfo {author} {\bibfnamefont {B.~F.}\ \bibnamefont
  {Bilenkii}}, \bibinfo {author} {\bibfnamefont {A.~G.}\ \bibnamefont
  {Mikolaichuk}}, \ and\ \bibinfo {author} {\bibfnamefont {D.~M.}\ \bibnamefont
  {Freik}},\ }\href {\doibase 10.1002/pssb.19680280146} {\bibfield  {journal}
  {\bibinfo  {journal} {Phys. Status Solidi B}\ }\textbf {\bibinfo {volume}
  {28}},\ \bibinfo {pages} {K5} (\bibinfo {year} {1968})}\BibitemShut {NoStop}%
\bibitem [{\citenamefont {Greyson}, \citenamefont {Barton},\ and\ \citenamefont
  {Odom}(2006)}]{Greyson-2006-ZincblendeSnS}%
  \BibitemOpen
  \bibfield  {author} {\bibinfo {author} {\bibfnamefont {E.}~\bibnamefont
  {Greyson}}, \bibinfo {author} {\bibfnamefont {J.}~\bibnamefont {Barton}}, \
  and\ \bibinfo {author} {\bibfnamefont {T.}~\bibnamefont {Odom}},\ }\href
  {\doibase 10.1002/smll.200500460} {\bibfield  {journal} {\bibinfo  {journal}
  {Small}\ }\textbf {\bibinfo {volume} {2}},\ \bibinfo {pages} {368} (\bibinfo
  {year} {2006})}\BibitemShut {NoStop}%
\bibitem [{\citenamefont {Ahmet}\ \emph {et~al.}(2015)\citenamefont {Ahmet},
  \citenamefont {Hill}, \citenamefont {Johnson},\ and\ \citenamefont
  {Peter}}]{Ahmet-2015-ZincblendeSnS}%
  \BibitemOpen
  \bibfield  {author} {\bibinfo {author} {\bibfnamefont {I.~Y.}\ \bibnamefont
  {Ahmet}}, \bibinfo {author} {\bibfnamefont {M.~S.}\ \bibnamefont {Hill}},
  \bibinfo {author} {\bibfnamefont {A.~L.}\ \bibnamefont {Johnson}}, \ and\
  \bibinfo {author} {\bibfnamefont {L.~M.}\ \bibnamefont {Peter}},\ }\href
  {\doibase 10.1021/acs.chemmater.5b03220} {\bibfield  {journal} {\bibinfo
  {journal} {Chem. Mater.}\ }\textbf {\bibinfo {volume} {27}},\ \bibinfo
  {pages} {7680} (\bibinfo {year} {2015})}\BibitemShut {NoStop}%
\bibitem [{\citenamefont {Rabkin}\ \emph {et~al.}(2015)\citenamefont {Rabkin},
  \citenamefont {Samuha}, \citenamefont {Abutbul}, \citenamefont {Ezersky},
  \citenamefont {Meshi},\ and\ \citenamefont {Golan}}]{Rabkin-2015-PiSnS}%
  \BibitemOpen
  \bibfield  {author} {\bibinfo {author} {\bibfnamefont {A.}~\bibnamefont
  {Rabkin}}, \bibinfo {author} {\bibfnamefont {S.}~\bibnamefont {Samuha}},
  \bibinfo {author} {\bibfnamefont {R.~E.}\ \bibnamefont {Abutbul}}, \bibinfo
  {author} {\bibfnamefont {V.}~\bibnamefont {Ezersky}}, \bibinfo {author}
  {\bibfnamefont {L.}~\bibnamefont {Meshi}}, \ and\ \bibinfo {author}
  {\bibfnamefont {Y.}~\bibnamefont {Golan}},\ }\href {\doibase
  10.1021/acs.nanolett.5b00209} {\bibfield  {journal} {\bibinfo  {journal}
  {Nano Lett.}\ }\textbf {\bibinfo {volume} {15}},\ \bibinfo {pages} {2174}
  (\bibinfo {year} {2015})}\BibitemShut {NoStop}%
\bibitem [{\citenamefont {Abutbul}\ \emph
  {et~al.}(2016{\natexlab{a}})\citenamefont {Abutbul}, \citenamefont
  {Garcia-Angelmo}, \citenamefont {Burshtein}, \citenamefont {Nair},
  \citenamefont {Nair},\ and\ \citenamefont {Golan}}]{Abutbul-2016-PiSnS}%
  \BibitemOpen
  \bibfield  {author} {\bibinfo {author} {\bibfnamefont {R.~E.}\ \bibnamefont
  {Abutbul}}, \bibinfo {author} {\bibfnamefont {A.~R.}\ \bibnamefont
  {Garcia-Angelmo}}, \bibinfo {author} {\bibfnamefont {Z.}~\bibnamefont
  {Burshtein}}, \bibinfo {author} {\bibfnamefont {M.~T.~S.}\ \bibnamefont
  {Nair}}, \bibinfo {author} {\bibfnamefont {P.~K.}\ \bibnamefont {Nair}}, \
  and\ \bibinfo {author} {\bibfnamefont {Y.}~\bibnamefont {Golan}},\ }\href
  {\doibase 10.1039/C6CE00647G} {\bibfield  {journal} {\bibinfo  {journal}
  {Cryst. Eng. Comm.}\ }\textbf {\bibinfo {volume} {18}},\ \bibinfo {pages}
  {5188} (\bibinfo {year} {2016}{\natexlab{a}})}\BibitemShut {NoStop}%
\bibitem [{\citenamefont {Abutbul}\ \emph
  {et~al.}(2016{\natexlab{b}})\citenamefont {Abutbul}, \citenamefont {Segev},
  \citenamefont {Samuha}, \citenamefont {Zeiri}, \citenamefont {Ezersky},
  \citenamefont {Makov},\ and\ \citenamefont {Golan}}]{Abutbul-2016-PiSnSe}%
  \BibitemOpen
  \bibfield  {author} {\bibinfo {author} {\bibfnamefont {R.~E.}\ \bibnamefont
  {Abutbul}}, \bibinfo {author} {\bibfnamefont {E.}~\bibnamefont {Segev}},
  \bibinfo {author} {\bibfnamefont {S.}~\bibnamefont {Samuha}}, \bibinfo
  {author} {\bibfnamefont {L.}~\bibnamefont {Zeiri}}, \bibinfo {author}
  {\bibfnamefont {V.}~\bibnamefont {Ezersky}}, \bibinfo {author} {\bibfnamefont
  {G.}~\bibnamefont {Makov}}, \ and\ \bibinfo {author} {\bibfnamefont
  {Y.}~\bibnamefont {Golan}},\ }\href {\doibase 10.1039/C5CE02437D} {\bibfield
  {journal} {\bibinfo  {journal} {Cryst. Eng. Comm.}\ }\textbf {\bibinfo
  {volume} {18}},\ \bibinfo {pages} {1918} (\bibinfo {year}
  {2016}{\natexlab{b}})}\BibitemShut {NoStop}%
\bibitem [{\citenamefont {Skelton}\ \emph
  {et~al.}(2017{\natexlab{b}})\citenamefont {Skelton}, \citenamefont {Burton},
  \citenamefont {Oba},\ and\ \citenamefont
  {Walsh}}]{Skelton-2017-ChemLattStability}%
  \BibitemOpen
  \bibfield  {author} {\bibinfo {author} {\bibfnamefont {J.~M.}\ \bibnamefont
  {Skelton}}, \bibinfo {author} {\bibfnamefont {L.~A.}\ \bibnamefont {Burton}},
  \bibinfo {author} {\bibfnamefont {F.}~\bibnamefont {Oba}}, \ and\ \bibinfo
  {author} {\bibfnamefont {A.}~\bibnamefont {Walsh}},\ }\href {\doibase
  10.1021/acs.jpcc.6b12581} {\bibfield  {journal} {\bibinfo  {journal} {J.
  Phys. Chem. C}\ }\textbf {\bibinfo {volume} {121}},\ \bibinfo {pages} {6446}
  (\bibinfo {year} {2017}{\natexlab{b}})}\BibitemShut {NoStop}%
\bibitem [{\citenamefont {Li}\ \emph {et~al.}(2015)\citenamefont {Li},
  \citenamefont {Hong}, \citenamefont {May}, \citenamefont {Bansal},
  \citenamefont {Chi}, \citenamefont {Hong}, \citenamefont {Ehlers},\ and\
  \citenamefont {Delaire}}]{Li2015}%
  \BibitemOpen
  \bibfield  {author} {\bibinfo {author} {\bibfnamefont {C.~W.}\ \bibnamefont
  {Li}}, \bibinfo {author} {\bibfnamefont {J.}~\bibnamefont {Hong}}, \bibinfo
  {author} {\bibfnamefont {A.~F.}\ \bibnamefont {May}}, \bibinfo {author}
  {\bibfnamefont {D.}~\bibnamefont {Bansal}}, \bibinfo {author} {\bibfnamefont
  {S.}~\bibnamefont {Chi}}, \bibinfo {author} {\bibfnamefont {T.}~\bibnamefont
  {Hong}}, \bibinfo {author} {\bibfnamefont {G.}~\bibnamefont {Ehlers}}, \ and\
  \bibinfo {author} {\bibfnamefont {O.}~\bibnamefont {Delaire}},\ }\href
  {\doibase 10.1038/nphys3492} {\bibfield  {journal} {\bibinfo  {journal} {Nat.
  Phys.}\ }\textbf {\bibinfo {volume} {11}},\ \bibinfo {pages} {1063} (\bibinfo
  {year} {2015})}\BibitemShut {NoStop}%
\bibitem [{\citenamefont {Togo}\ and\ \citenamefont
  {Tanaka}(2013)}]{togo2013evolution}%
  \BibitemOpen
  \bibfield  {author} {\bibinfo {author} {\bibfnamefont {A.}~\bibnamefont
  {Togo}}\ and\ \bibinfo {author} {\bibfnamefont {I.}~\bibnamefont {Tanaka}},\
  }\href {https://doi.org/10.1103/PhysRevB.87.184104} {\bibfield  {journal}
  {\bibinfo  {journal} {Phys. Rev. B}\ }\textbf {\bibinfo {volume} {87}},\
  \bibinfo {pages} {184104} (\bibinfo {year} {2013})}\BibitemShut {NoStop}%
\bibitem [{\citenamefont {Butler}\ \emph {et~al.}(2016)\citenamefont {Butler},
  \citenamefont {Frost}, \citenamefont {Skelton}, \citenamefont {Svane},\ and\
  \citenamefont {Walsh}}]{butler2016computational}%
  \BibitemOpen
  \bibfield  {author} {\bibinfo {author} {\bibfnamefont {K.~T.}\ \bibnamefont
  {Butler}}, \bibinfo {author} {\bibfnamefont {J.~M.}\ \bibnamefont {Frost}},
  \bibinfo {author} {\bibfnamefont {J.~M.}\ \bibnamefont {Skelton}}, \bibinfo
  {author} {\bibfnamefont {K.~L.}\ \bibnamefont {Svane}}, \ and\ \bibinfo
  {author} {\bibfnamefont {A.}~\bibnamefont {Walsh}},\ }\href
  {https://doi.org/10.1039/C5CS00841G} {\bibfield  {journal} {\bibinfo
  {journal} {Chem. Soc. Rev.}\ }\textbf {\bibinfo {volume} {45}},\ \bibinfo
  {pages} {6138} (\bibinfo {year} {2016})}\BibitemShut {NoStop}%
\bibitem [{\citenamefont {Davies}\ \emph {et~al.}(2016)\citenamefont {Davies},
  \citenamefont {Butler}, \citenamefont {Jackson}, \citenamefont {Morris},
  \citenamefont {Frost}, \citenamefont {Skelton},\ and\ \citenamefont
  {Walsh}}]{davies2016computational}%
  \BibitemOpen
  \bibfield  {author} {\bibinfo {author} {\bibfnamefont {D.~W.}\ \bibnamefont
  {Davies}}, \bibinfo {author} {\bibfnamefont {K.~T.}\ \bibnamefont {Butler}},
  \bibinfo {author} {\bibfnamefont {A.~J.}\ \bibnamefont {Jackson}}, \bibinfo
  {author} {\bibfnamefont {A.}~\bibnamefont {Morris}}, \bibinfo {author}
  {\bibfnamefont {J.~M.}\ \bibnamefont {Frost}}, \bibinfo {author}
  {\bibfnamefont {J.~M.}\ \bibnamefont {Skelton}}, \ and\ \bibinfo {author}
  {\bibfnamefont {A.}~\bibnamefont {Walsh}},\ }\href
  {https://doi.org/10.1016/j.chempr.2016.09.010} {\bibfield  {journal}
  {\bibinfo  {journal} {Chem}\ }\textbf {\bibinfo {volume} {1}},\ \bibinfo
  {pages} {617} (\bibinfo {year} {2016})}\BibitemShut {NoStop}%
\bibitem [{\citenamefont {Davies}\ \emph {et~al.}(2018)\citenamefont {Davies},
  \citenamefont {Butler}, \citenamefont {Skelton}, \citenamefont {Xie},
  \citenamefont {Oganov},\ and\ \citenamefont {Walsh}}]{davies2018computer}%
  \BibitemOpen
  \bibfield  {author} {\bibinfo {author} {\bibfnamefont {D.~W.}\ \bibnamefont
  {Davies}}, \bibinfo {author} {\bibfnamefont {K.~T.}\ \bibnamefont {Butler}},
  \bibinfo {author} {\bibfnamefont {J.~M.}\ \bibnamefont {Skelton}}, \bibinfo
  {author} {\bibfnamefont {C.}~\bibnamefont {Xie}}, \bibinfo {author}
  {\bibfnamefont {A.~R.}\ \bibnamefont {Oganov}}, \ and\ \bibinfo {author}
  {\bibfnamefont {A.}~\bibnamefont {Walsh}},\ }\href
  {https://doi.org/10.1039/C7SC03961A} {\bibfield  {journal} {\bibinfo
  {journal} {Chem. Sci.}\ }\textbf {\bibinfo {volume} {9}},\ \bibinfo {pages}
  {1022} (\bibinfo {year} {2018})}\BibitemShut {NoStop}%
\bibitem [{\citenamefont {Woodley}\ and\ \citenamefont
  {Catlow}(2008)}]{woodley2008crystal}%
  \BibitemOpen
  \bibfield  {author} {\bibinfo {author} {\bibfnamefont {S.~M.}\ \bibnamefont
  {Woodley}}\ and\ \bibinfo {author} {\bibfnamefont {R.}~\bibnamefont
  {Catlow}},\ }\href {https://doi.org/10.1038/nmat2321} {\bibfield  {journal}
  {\bibinfo  {journal} {Nat. Mater.}\ }\textbf {\bibinfo {volume} {7}},\
  \bibinfo {pages} {937} (\bibinfo {year} {2008})}\BibitemShut {NoStop}%
\bibitem [{\citenamefont {Reilly}\ \emph {et~al.}(2016)\citenamefont {Reilly},
  \citenamefont {Cooper}, \citenamefont {Adjiman}, \citenamefont
  {Bhattacharya}, \citenamefont {Boese}, \citenamefont {Brandenburg},
  \citenamefont {Bygrave}, \citenamefont {Bylsma}, \citenamefont {Campbell},
  \citenamefont {Car} \emph {et~al.}}]{reilly2016report}%
  \BibitemOpen
  \bibfield  {author} {\bibinfo {author} {\bibfnamefont {A.~M.}\ \bibnamefont
  {Reilly}}, \bibinfo {author} {\bibfnamefont {R.~I.}\ \bibnamefont {Cooper}},
  \bibinfo {author} {\bibfnamefont {C.~S.}\ \bibnamefont {Adjiman}}, \bibinfo
  {author} {\bibfnamefont {S.}~\bibnamefont {Bhattacharya}}, \bibinfo {author}
  {\bibfnamefont {A.~D.}\ \bibnamefont {Boese}}, \bibinfo {author}
  {\bibfnamefont {J.~G.}\ \bibnamefont {Brandenburg}}, \bibinfo {author}
  {\bibfnamefont {P.~J.}\ \bibnamefont {Bygrave}}, \bibinfo {author}
  {\bibfnamefont {R.}~\bibnamefont {Bylsma}}, \bibinfo {author} {\bibfnamefont
  {J.~E.}\ \bibnamefont {Campbell}}, \bibinfo {author} {\bibfnamefont
  {R.}~\bibnamefont {Car}},  \emph {et~al.},\ }\href
  {https://doi.org/10.1107/S2052520616007447} {\bibfield  {journal} {\bibinfo
  {journal} {Acta. Crystallogr. B. Struct. Sci. Cryst. Eng. Mater.}\ }\textbf
  {\bibinfo {volume} {72}},\ \bibinfo {pages} {439} (\bibinfo {year}
  {2016})}\BibitemShut {NoStop}%
\bibitem [{\citenamefont {Pickard}\ and\ \citenamefont
  {Needs}(2011)}]{pickard2011ab}%
  \BibitemOpen
  \bibfield  {author} {\bibinfo {author} {\bibfnamefont {C.~J.}\ \bibnamefont
  {Pickard}}\ and\ \bibinfo {author} {\bibfnamefont {R.}~\bibnamefont
  {Needs}},\ }\href {https://doi.org/10.1088/0953-8984/23/5/053201} {\bibfield
  {journal} {\bibinfo  {journal} {J. Condens. Matter Phys.}\ }\textbf {\bibinfo
  {volume} {23}},\ \bibinfo {pages} {053201} (\bibinfo {year}
  {2011})}\BibitemShut {NoStop}%
\bibitem [{\citenamefont {Woodley}(2013)}]{woodley2013knowledge}%
  \BibitemOpen
  \bibfield  {author} {\bibinfo {author} {\bibfnamefont {S.~M.}\ \bibnamefont
  {Woodley}},\ }\href {https://doi.org/10.1021/jp406854j} {\bibfield  {journal}
  {\bibinfo  {journal} {J. Phys. Chem. C}\ }\textbf {\bibinfo {volume} {117}},\
  \bibinfo {pages} {24003} (\bibinfo {year} {2013})}\BibitemShut {NoStop}%
\bibitem [{\citenamefont {Wang}\ \emph {et~al.}(2010)\citenamefont {Wang},
  \citenamefont {Lv}, \citenamefont {Zhu},\ and\ \citenamefont
  {Ma}}]{wang2010crystal}%
  \BibitemOpen
  \bibfield  {author} {\bibinfo {author} {\bibfnamefont {Y.}~\bibnamefont
  {Wang}}, \bibinfo {author} {\bibfnamefont {J.}~\bibnamefont {Lv}}, \bibinfo
  {author} {\bibfnamefont {L.}~\bibnamefont {Zhu}}, \ and\ \bibinfo {author}
  {\bibfnamefont {Y.}~\bibnamefont {Ma}},\ }\href
  {https://doi.org/10.1103/PhysRevB.82.094116} {\bibfield  {journal} {\bibinfo
  {journal} {Phys. Rev. B}\ }\textbf {\bibinfo {volume} {82}},\ \bibinfo
  {pages} {094116} (\bibinfo {year} {2010})}\BibitemShut {NoStop}%
\bibitem [{\citenamefont {Oganov}, \citenamefont {Lyakhov},\ and\ \citenamefont
  {Valle}(2011)}]{oganov2011evolutionary}%
  \BibitemOpen
  \bibfield  {author} {\bibinfo {author} {\bibfnamefont {A.~R.}\ \bibnamefont
  {Oganov}}, \bibinfo {author} {\bibfnamefont {A.~O.}\ \bibnamefont {Lyakhov}},
  \ and\ \bibinfo {author} {\bibfnamefont {M.}~\bibnamefont {Valle}},\ }\href
  {https://doi.org/10.1021/ar1001318} {\bibfield  {journal} {\bibinfo
  {journal} {Acc. Chem. Res.}\ }\textbf {\bibinfo {volume} {44}},\ \bibinfo
  {pages} {227} (\bibinfo {year} {2011})}\BibitemShut {NoStop}%
\bibitem [{\citenamefont {Hautier}\ \emph {et~al.}(2011)\citenamefont
  {Hautier}, \citenamefont {Fischer}, \citenamefont {Ehrlacher}, \citenamefont
  {Jain},\ and\ \citenamefont {Ceder}}]{hautier2011data}%
  \BibitemOpen
  \bibfield  {author} {\bibinfo {author} {\bibfnamefont {G.}~\bibnamefont
  {Hautier}}, \bibinfo {author} {\bibfnamefont {C.}~\bibnamefont {Fischer}},
  \bibinfo {author} {\bibfnamefont {V.}~\bibnamefont {Ehrlacher}}, \bibinfo
  {author} {\bibfnamefont {A.}~\bibnamefont {Jain}}, \ and\ \bibinfo {author}
  {\bibfnamefont {G.}~\bibnamefont {Ceder}},\ }\href
  {https://doi.org/10.1021/ic102031h} {\bibfield  {journal} {\bibinfo
  {journal} {Inorg. Chem.}\ }\textbf {\bibinfo {volume} {50}},\ \bibinfo
  {pages} {656} (\bibinfo {year} {2011})}\BibitemShut {NoStop}%
\bibitem [{\citenamefont {Collins}\ \emph {et~al.}(2017)\citenamefont
  {Collins}, \citenamefont {Dyer}, \citenamefont {Pitcher}, \citenamefont
  {Whitehead}, \citenamefont {Zanella}, \citenamefont {Mandal}, \citenamefont
  {Claridge}, \citenamefont {Darling},\ and\ \citenamefont
  {Rosseinsky}}]{Collins-2017-MCEMMA}%
  \BibitemOpen
  \bibfield  {author} {\bibinfo {author} {\bibfnamefont {C.}~\bibnamefont
  {Collins}}, \bibinfo {author} {\bibfnamefont {M.~S.}\ \bibnamefont {Dyer}},
  \bibinfo {author} {\bibfnamefont {M.~J.}\ \bibnamefont {Pitcher}}, \bibinfo
  {author} {\bibfnamefont {G.~F.~S.}\ \bibnamefont {Whitehead}}, \bibinfo
  {author} {\bibfnamefont {M.}~\bibnamefont {Zanella}}, \bibinfo {author}
  {\bibfnamefont {P.}~\bibnamefont {Mandal}}, \bibinfo {author} {\bibfnamefont
  {J.~B.}\ \bibnamefont {Claridge}}, \bibinfo {author} {\bibfnamefont {G.~R.}\
  \bibnamefont {Darling}}, \ and\ \bibinfo {author} {\bibfnamefont {M.~J.}\
  \bibnamefont {Rosseinsky}},\ }\href {\doibase 10.1038/nature22374} {\bibfield
   {journal} {\bibinfo  {journal} {Nature}\ }\textbf {\bibinfo {volume}
  {546}},\ \bibinfo {pages} {280} (\bibinfo {year} {2017})}\BibitemShut
  {NoStop}%
\bibitem [{\citenamefont {Grimvall}\ \emph {et~al.}(2012)\citenamefont
  {Grimvall}, \citenamefont {Magyari-K{\"o}pe}, \citenamefont
  {Ozoli{\c{n}}{\v{s}}},\ and\ \citenamefont {Persson}}]{grimvall2012lattice}%
  \BibitemOpen
  \bibfield  {author} {\bibinfo {author} {\bibfnamefont {G.}~\bibnamefont
  {Grimvall}}, \bibinfo {author} {\bibfnamefont {B.}~\bibnamefont
  {Magyari-K{\"o}pe}}, \bibinfo {author} {\bibfnamefont {V.}~\bibnamefont
  {Ozoli{\c{n}}{\v{s}}}}, \ and\ \bibinfo {author} {\bibfnamefont {K.~A.}\
  \bibnamefont {Persson}},\ }\href {https://doi.org/10.1103/RevModPhys.84.945}
  {\bibfield  {journal} {\bibinfo  {journal} {Rev. Mod. Phys.}\ }\textbf
  {\bibinfo {volume} {84}},\ \bibinfo {pages} {945} (\bibinfo {year}
  {2012})}\BibitemShut {NoStop}%
\bibitem [{\citenamefont {Sikka}, \citenamefont {Vohra},\ and\ \citenamefont
  {Chidambaram}(1982)}]{sikka1982omega}%
  \BibitemOpen
  \bibfield  {author} {\bibinfo {author} {\bibfnamefont {S.}~\bibnamefont
  {Sikka}}, \bibinfo {author} {\bibfnamefont {Y.}~\bibnamefont {Vohra}}, \ and\
  \bibinfo {author} {\bibfnamefont {R.}~\bibnamefont {Chidambaram}},\ }\href
  {https://doi.org/10.1016/0079-6425(82)90002-0} {\bibfield  {journal}
  {\bibinfo  {journal} {Prog. Mater. Sci.}\ }\textbf {\bibinfo {volume} {27}},\
  \bibinfo {pages} {245} (\bibinfo {year} {1982})}\BibitemShut {NoStop}%
\bibitem [{\citenamefont {Rahim}, \citenamefont {Skelton},\ and\ \citenamefont
  {Scanlon}(2020)}]{Rahim-2020-aBi2Sn2O7}%
  \BibitemOpen
  \bibfield  {author} {\bibinfo {author} {\bibfnamefont {W.}~\bibnamefont
  {Rahim}}, \bibinfo {author} {\bibfnamefont {J.~M.}\ \bibnamefont {Skelton}},
  \ and\ \bibinfo {author} {\bibfnamefont {D.~O.}\ \bibnamefont {Scanlon}},\
  }\href {\doibase 10.1039/D0TA03945D} {\bibfield  {journal} {\bibinfo
  {journal} {J. Mater. Chem. A}\ }\textbf {\bibinfo {volume} {8}},\ \bibinfo
  {pages} {16405} (\bibinfo {year} {2020})}\BibitemShut {NoStop}%
\bibitem [{\citenamefont {Kayastha}\ and\ \citenamefont
  {Ramakrishnan}(2021)}]{kayastha2021high}%
  \BibitemOpen
  \bibfield  {author} {\bibinfo {author} {\bibfnamefont {P.}~\bibnamefont
  {Kayastha}}\ and\ \bibinfo {author} {\bibfnamefont {R.}~\bibnamefont
  {Ramakrishnan}},\ }\href {https://doi.org/10.1063/5.0041717} {\bibfield
  {journal} {\bibinfo  {journal} {J. Chem. Phys.}\ }\textbf {\bibinfo {volume}
  {154}},\ \bibinfo {pages} {061102} (\bibinfo {year} {2021})}\BibitemShut
  {NoStop}%
\bibitem [{\citenamefont {Stillinger}(1999)}]{stillinger1999exponential}%
  \BibitemOpen
  \bibfield  {author} {\bibinfo {author} {\bibfnamefont {F.~H.}\ \bibnamefont
  {Stillinger}},\ }\href {https://doi.org/10.1103/PhysRevE.59.48} {\bibfield
  {journal} {\bibinfo  {journal} {Phys. Rev. E}\ }\textbf {\bibinfo {volume}
  {59}},\ \bibinfo {pages} {48} (\bibinfo {year} {1999})}\BibitemShut {NoStop}%
\bibitem [{\citenamefont {Adams}\ \emph {et~al.}(2021)\citenamefont {Adams},
  \citenamefont {Wang}, \citenamefont {Steinle-Neumann}, \citenamefont
  {Passerone},\ and\ \citenamefont {Churakov}}]{adams2021anharmonic}%
  \BibitemOpen
  \bibfield  {author} {\bibinfo {author} {\bibfnamefont {D.~J.}\ \bibnamefont
  {Adams}}, \bibinfo {author} {\bibfnamefont {L.}~\bibnamefont {Wang}},
  \bibinfo {author} {\bibfnamefont {G.}~\bibnamefont {Steinle-Neumann}},
  \bibinfo {author} {\bibfnamefont {D.}~\bibnamefont {Passerone}}, \ and\
  \bibinfo {author} {\bibfnamefont {S.~V.}\ \bibnamefont {Churakov}},\ }\href
  {\doibase 10.1088/1361-648x/abc972} {\bibfield  {journal} {\bibinfo
  {journal} {J. Condens. Matter Phys.}\ }\textbf {\bibinfo {volume} {33}},\
  \bibinfo {pages} {175501} (\bibinfo {year} {2021})}\BibitemShut {NoStop}%
\end{thebibliography}%

\end{document}